\documentclass[a4paper,onecolumn,accepted=2023-02-28]{quantumarticle}
\pdfoutput=1

\usepackage[T1]{fontenc}
\usepackage[numbers]{natbib}
\usepackage{graphicx}
\usepackage{amsfonts}
\usepackage{amssymb}
\usepackage{amsmath}
\usepackage{thmtools}
\usepackage{bbm}
\usepackage{dsfont}
\usepackage{hyperref}
\usepackage{color}
\usepackage{tikz}
\usepackage{tikz-cd}
\hypersetup{
	colorlinks,
	linkcolor={black!30!blue},
	citecolor={red},
	urlcolor={blue}
}
\usepackage[margin=1cm]{caption}  
\usepackage{multirow}

\newtheorem{thm}{Theorem}

\newtheorem{lem}[thm]{Lemma}
\newtheorem{prop}[thm]{Proposition}

\newtheorem{sumry}[thm]{Summary}
\newenvironment{proof}[1][]{\par\vskip12pt\noindent%
     {\it Proof\ \ifx#1==\else{#1\/}\newline\fi\ }\bgroup}
     {\egroup\quad\strut\hfill$\blacksquare$\vskip12pt}

\def\proofsin#1{($\rightarrow$#1)} 
\def\proofin#1{\proofsin{\reff{#1}}} 

\let\reff\autoref

\def\qtno#1{#1}
\def\qtno#1{\ifx#11cx\else\ifx#12ce\else\ifx#13nx\else\ifx#14qx\else\ifx#15ei\else\ifx#16in\else\ifx#1nnn\else\ifx#1mmm\fi\fi\fi\fi\fi\fi\fi\fi}
\def\Qt#1{($\Qm$\qtno#1)} 
\def\Qtc#1{[$\Qm$\qtno#1]} 
\def\Qb#1#2{($\Qm$\qtno#1,\qtno#2)} 
\def\Qset#1{\Qm_{(#1)}} 

\def\Rl{{\mathbb R}}\def\Cx{{\mathbb C}}

\def\HH{{\mathcal H}}   
\def\Hh{{\mathfrak h}}
\def\BB{{\mathcal B}}
\def\KK{{\mathcal K}}  
\def\Alg{{\mathcal A}} 

\def\marge{{\!\tt|\!}}

\def\Qm{{\mathcal Q}} 
\def\Qmc{{\mathcal Q}_{(e)}} 
\def\Qmst{\Qm_{\mathrm{st}}}
\def\Qmobs{\Qm_{\mathrm{obs}}}
\def\QmObs{\mathsf O(2)}
\def\Cl{{\mathcal C}} 
\def\Ns{{\mathcal N}} 

\def\ncyc{{\mathbf N}} 
\def\gauge{\gamma}

\def\abs#1{\lvert#1\rvert}
\def\tr{{\rm tr}}

\def\idty{{\leavevmode\rm 1\mkern -5.4mu I}}
\def\idty{{\mathds1}}
\def\ketbra#1#2{\vert #1\rangle \langle #2\vert}
\def\kettbra#1{\ketbra{#1}{#1}}              
\def\braket#1#2{\langle#1\vert#2\rangle}
\def\ket#1{\lvert{#1}\rangle}

\def\rank{\mathop{\mathrm{rank}}\nolimits}

\def\re{\Re e}

\def\stt{\mid}

\def\norm#1{\Vert{#1}\Vert}

\def\ext{\partial_{\mathrm e}} 
\def\conv{\mathop{\mathrm{co}}}
\def\fivepmatrix#1\\#2\\#3\\#4\\#5\\{\left(\begin{array}{c|cc|cc}#1\\\hline #2\\#3\\\hline#4\\#5\relax \end{array}\right)}

\def\Sin{{\mathop{\mathbf{sin}}\nolimits}}
\def\Cos{{\mathop{\mathbf{cos}}\nolimits}}
\def\sign{{\mathop{\mathrm{sign}}\nolimits}}
\def\diag{\mathop{\mathrm{diag}}\nolimits}
\def\polar{^\circ}
\def\bipolar{^{\circ\circ}}

\def\dufa{\sharp} 
\def\dufa{\perp}
\def\dufak#1{^{\dufa_{#1}}}

\def\evev{\varepsilon} 

\begin{document}

\title{Quantum Correlations in the Minimal Scenario}
\author{Thinh P. Le}
\affiliation{Institute for Quantum Optics and Quantum Information Vienna, Boltzmanngasse 3 1090 Vienna, Austria}
\email{Thinh.Le@oeaw.ac.at}
\author{Chiara Meroni}
\affiliation{Institute for Computational and Experimental Research in Mathematics, 121 South Main Street Providence RI 02903, USA}
\email{chiara\_meroni@brown.edu}
\author{Bernd Sturmfels}
\affiliation{Max Planck Institute for Mathematics in the Sciences Leipzig, Inselstrasse 22 04103 Leipzig, Germany}
\affiliation{Department of Mathematics, University of California, Berkeley, 970 Evans Hall \#3840 Berkeley CA 94720-3840, USA}
\email{bernd@mis.mpg.de}
\author{Reinhard F. Werner}
\affiliation{Insitute f\"ur Theoretische Physik, Leibniz Universit\"at Hannover, Appelstrasse 2 30167 Hannover, Germany}
\email{reinhard.werner@itp.uni-hannover.de}
\author{Timo Ziegler}
\affiliation{Insitute f\"ur Theoretische Physik, Leibniz Universit\"at Hannover, Appelstrasse 2 30167 Hannover, Germany}
\email{timo.ziegler@itp.uni-hannover.de}
\maketitle

\begin{center}
{\Large Dedicated to the memory of Boris Tsirelson}
\end{center}

\begin{abstract}
In the minimal scenario of quantum correlations, two parties can choose from two observables with two possible outcomes each. Probabilities are specified by four marginals and four correlations. The resulting four-dimensional convex body of correlations, denoted $\mathcal{Q}$, is fundamental for quantum information theory. We review and systematize what is known about $\Qm$, and add many details, visualizations, and complete proofs. In particular, we provide a detailed description of the boundary, which consists of three-dimensional faces isomorphic to elliptopes and sextic algebraic manifolds of exposed extreme points.  These patches are separated by cubic surfaces of non-exposed extreme points. We provide a trigonometric parametrization of all extreme points, along with their exposing Tsirelson inequalities and quantum models. All non-classical extreme points (exposed or not) are self-testing, i.e., realized by an essentially unique quantum model.

Two principles, which are specific to the minimal scenario, allow a quick and complete overview: The first is the pushout transformation, i.e., the application of the sine function to each coordinate. This transforms the classical correlation polytope exactly into the correlation body $\mathcal{Q}$, also identifying the boundary structures. The second principle, self-duality, is an isomorphism between $\Qm$ and its polar dual, i.e., the set of affine inequalities satisfied by all quantum correlations (``Tsirelson inequalities''). The same isomorphism links the polytope of classical correlations contained in $\Qm$ to the polytope of no-signalling correlations, which contains $\Qm$.

We also discuss the sets of correlations achieved with fixed Hilbert space dimension, fixed state or fixed observables, and establish a new non-linear inequality for $\Qm$ involving the determinant of the correlation matrix.
\end{abstract}

\begin{center}
\includegraphics[trim=0.9cm 0.8cm 1.2cm 1.3cm,clip,scale=0.7]{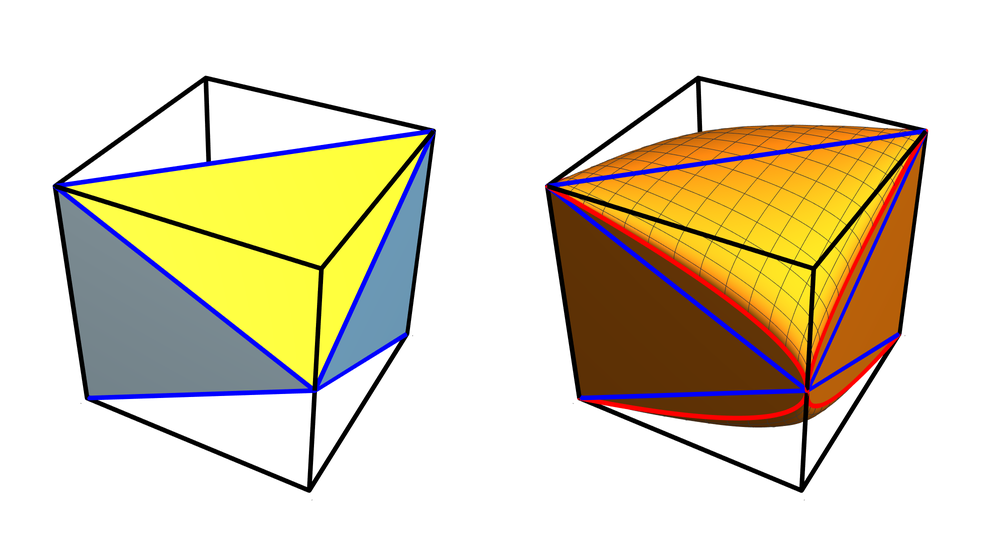}
\end{center}

\newpage
\tableofcontents

\section{Introduction}

The quantum correlation set is a fundamental object in quantum information theory. The key point is that some correlations predicted by quantum theory cannot be modeled within classical probability, more precisely under the constraints of ``local realism''.
This far-reaching insight can be gained from just a single example, the
singlet state of two quantum bits with a particular measurement setup. The correlations predicted by quantum theory were demonstrated experimentally \cite{Aspect} to high precision, and recently also while carefully closing some loopholes that persisted for decades \cite{lopholefree1,loopholefree2}.
The non-classical nature is certified by the violation of the CHSH Bell inequality
\cite{Bell,CHSH}, up to a maximum value of $2\sqrt{2}$ allowed by quantum theory. The correlation achieving $2\sqrt{2}$ also has the property of ``self-testing'', which means that its quantum realization is essentially unique. Just from the correlation measurement, one can infer not only the quantum state, but also
the action of the measurement devices.

In this situation it is natural to ask: Where exactly lies the boundary between the classical and the quantum, between the quantum and the post-quantum correlations?
What other correlations or quantum states exhibit the same features? In a correlation setting,
$N$ parties share some quantum state, so that each party can choose from $M$ different measurements, each of which can have $K$ different outcomes. So what exactly is the set of correlation data that can arise from either classical probability or quantum? This is a hard problem even for fairly small $N,M,K$, as seen
 on the open problems  website \cite[Problems\,1,26,27,32,33,34]{problemSite}. The scenario mentioned in the first paragraph corresponds to $N=M=K=2$, the smallest nontrivial scenario. Here, characterizations have been known since Tsirelson's seminal work \cite{Tsi85},
particularly in the ``zero-marginals'' case 222\marge0, defined by the property that each outcome by itself, without considering the results of other parties, is equidistributed. This is the scenario indicated by the adjective ``minimal'' in our title. We provide an overview of the literature below in \autoref{sec:review}. Known
results are scattered. The connections between different characterizations are
rarely given, the overall structure of the boundary is not analyzed, and no attempts at a full geometric understanding or visualization are made. Moreover, the self-duality of the body seems to have escaped notice of the wider community. All this will be provided in the present article, along with self-contained proofs of all assertions.

Our paper arose from a project of T.P.L.~and R.F.W.~aimed at a better understanding of self-testing.
T.Z.~joined as a Masters student.
When we realized that even the 222-case was not clear in reasonable generality, we focused on that and could win the help of C.M.\ and B.S. from the mathematical side.
The 222\marge0 case was to serve as the ``well-understood example''.
Only that it was not understood at the desired level of detail.
Furthermore, most of the available techniques for 222\marge0 do not apply to the full-marginal 222 case, from Tsirelson's correlation matrices to the cosine parametrization and the pushout principle to the semidefinite hierarchy. Doing justice to these techniques would have been a distraction in the full 222 context, so  we decided to separate it, and organize the material into this 222\marge0  review with a geometric flavor.
As work progressed, we realized that there was more in the works of Boris Tsirelson than we had recollections of. Tsirelson was writing at a time when the relevant community was very small (and included one of us, R.F.W., who should have remembered more).  Back then long proofs of exotic material were hard to publish, which may be why he often chose to state his results without proof. But he was clearly a pioneer, coming more than a decade before the surge of interest with the rise of Quantum Information. Boris Tsirelson died last year, so we felt it was fitting to dedicate this paper to him, and include the proofs he left out. We like to think that he would have enjoyed our presentation.

This article is very much a two-communities paper. We ask experts from the quantum side to bear with us when we cover standard material, just as we ask patience from geometers when we explain basic notions. Aiming for completeness on a well-researched subject means that it is largely a review.
But we hope that even those experienced with quantum correlations will find new connections, just as we have.

Our presentation
 is organized as follows: We first set the scene with a brief introduction to quantum correlations.
This is followed by a mathematical discussion which states main results in a concise form.
We then also give a brief summary of previous work. \autoref{sec:describe}
offers a more extended description of
the correlation body $\Qm$, from basic visualization and overall properties to a detailed classification of boundary points.  In \autoref{sec:dualscribe} we focus on the dual body and its connection to $\Qm$. These geometric and algebraic features are related back to quantum issues in \autoref{sec:desquribe}. We also discuss constrained sets in which some of the quantum data are fixed, namely Hilbert space dimension (\reff{sec:dimH}), state (\reff{sec:fixedState}), or the observables  (\reff{sec:fixedObs}).
In these descriptive sections we give no proofs. Proofs are collected in \autoref{sec:proofs}. Every statement of a proposition or theorem begins with a clickable pointer such as \proofin{sec:proofs} to the subsection containing the proof. An exception to this rule are statements that are clear from the context, and merely summarize a narrative just given.
The proof section is organized in logical order, and should be readable from beginning to end without forward references. Naturally, this order differs from
the narrative in \autoref{sec:describe}-\autoref{sec:desquribe}, and also from the theorems in \autoref{sec:introMath}.

\subsection{Background from Physics: Quantum Correlations}\label{sec:introPhys}

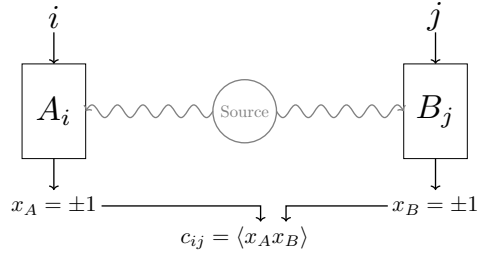
\begin{figure}\centering
\resizebox{0.4\textwidth}{!}{
\begin{tikzpicture}
\draw[gray, very thick] (0,0) circle (10mm) node[anchor=center,scale=1.5]  {Source};
\draw[black, thick] (-7,-1.5) rectangle (-5,1.5);
\path (-6,3) node[scale=3] (i) {$i$};
\path (6,3) node[scale=3] (j) {$j$};
\draw[->, very thick] (-6,2.5) -- (-6,1.5);
\draw[->, very thick] (6,2.5) -- (6,1.5);
\draw[black, thick] (7,-1.5) rectangle (5,1.5);
\path (-6,0) node[scale=3] {$A_i$};
\path	(6,0) node[scale=3] {$B_j$};
\draw[->, gray, very thick] (-1,0) sin (-1.2,-0.2) cos (-1.4,0) sin (-1.6,0.2) cos (-1.8,0) sin (-2,-0.2) cos (-2.2,0) sin (-2.4,0.2) cos (-2.6,0) sin (-2.8,-0.2) cos (-3,0) sin (-3.2,0.2) cos (-3.4,0) sin (-3.6,-0.2) cos (-3.8,0) sin (-4,0.2) cos (-4.2,0) sin (-4.4,-0.2) cos (-4.6,0) sin (-4.8,0.2) cos(-5,0);
\draw[->, gray, very thick] (1,0) sin (1.2,-0.2) cos (1.4,0) sin (1.6,0.2) cos (1.8,0) sin (2,-0.2) cos (2.2,0) sin (2.4,0.2) cos (2.6,0) sin (2.8,-0.2) cos (3,0) sin (3.2,0.2) cos (3.4,0) sin (3.6,-0.2) cos (3.8,0) sin (4,0.2) cos (4.2,0) sin (4.4,-0.2) cos (4.6,0) sin (4.8,0.2) cos(5,0);
\draw[->, very thick] (-6,-1.5) -- (-6,-2.5);
\draw[->, very thick] (6,-1.5) -- (6,-2.5);
\path (-6,-3) node[scale=2] (XA) {$x_A=\pm 1$};
\path (6,-3) node[scale=2] (XB) {$x_B=\pm 1$};
\draw[->, very thick] (-4.5,-3) -- (0.5,-3) -- (0.5,-3.5);
\draw[->, very thick] (4.5,-3) -- (1.3,-3) -- (1.3,-3.5);
\path (0,-4) node[scale=2] (correlate) {$c_{ij}=\langle x_Ax_B\rangle$};
\end{tikzpicture}}
\caption{A correlation experiment: Alice chooses setting $i$ and Bob chooses setting $j$.
  The outcomes of $A_i$ and $B_j$ are $\pm 1$.
One measures the correlation $c_{ij}$ of these outcomes. }
\label{fig:corrscheme}
\end{figure}

In a correlation experiment, several parties carry out measurements on a shared quantum system.
We consider $N=2$ causally disconnected parties, conventionally called Alice and Bob. Each of them chooses from $M=2$ possible measurements, labeled $i,j=1,2$, with $K=2$ possible outcomes,
labeled $\pm1$ (see \reff{fig:corrscheme}). Thus
there are four experiments, labeled by the pairs $(i,j)$
of choices for Alice and Bob. The correlation $c_{ij}\in[-1,1]$
is the probability of equal outcomes minus that of different outcomes. Equivalently, $c_{ij}$ is the
expectation of the product of the outcomes, when these are labeled as $\pm1$.
The $c_{ij}$ are not sufficient to reconstruct the full statistics.
That would give an $8$-dimensional convex body, whose coordinates are
the marginals for single outcomes, plus one correlation for every pair $(i,j)$. This count
incorporates the no-signalling condition, namely that the marginals do not depend
on the setting chosen at the other site. Restricting to the $4$-tuples $c=(c_{11},c_{12},c_{21},c_{22})$
corresponds to a {\it projection} of the $8$-dimensional body. We can realize this projection geometrically by taking the equal weight mixture of the given model with one in which all outcomes are flipped to their negatives. This operation changes the sign of
the marginals, but not of the correlations $c_{ij}$. Therefore, we can alternatively think of the $4$-dimensional body as that {\it section} of the $8$-dimensional body, in which the marginals are set equal to zero.  This explains why we call our scenario in the $4$-dimensional $c$-space the zero marginals case.

We are interested in the set $\Qm$ of correlations
$c=(c_{11},c_{12},c_{21},c_{22})$ that are consistent with quantum theory.
Quantum systems are described in some separable Hilbert space $\HH$ over $\Cx$. The source is given by a
 positive Hermitian operator  $\rho$  acting on $\HH$. It satisfies $\tr(\rho)=1$ and is called
   the \textit{density operator}. The measurements are characterized by Hermitian operators
   $A_1,A_2,B_1,B_2$  on  $\HH$ that satisfy the  hypotheses
   \begin{equation}\label{eq:ABhypotheses}
  [A_i,B_j]=0   \quad {\rm and} \quad
        -\idty \le A_i,B_j\le\idty
     \qquad
   \hbox{for} \, 1 \leq i, j \leq 2.
\end{equation}

If $\HH = \Cx^m$ then  $A_i$ and $B_j$ are Hermitian $m \times m$ matrices
and $\idty$ is the identity matrix. Condition
\eqref{eq:ABhypotheses} says  that  $A_i$ commutes with $B_j$ and
that the eigenvalues of all these matrices are in $[-1,1]$.
The commutation condition represents the hypothesis that the two parties are causally disconnected, i.e., all
measurements by Alice can be executed jointly with those of Bob.
In contrast, the commutators $[A_1,A_2]$ and $[B_1,B_2]$ are usually nonzero, i.e., the
two measurement choices of each party individually are not commensurate.

We remark that in the quantum information community, the commuting condition is often made in the stronger form, namely that the separated labs of Alice and Bob are rendered by a Hilbert space tensor product $\HH=\HH_A\otimes\HH_B$, so that $A_i=A_i'\otimes\idty_B$ and $B_j=\idty_A\otimes B_j'$ with $A_1',A_2'$ acting on $\HH_A$, and $B_1',B_2'$ acting on $\HH_B$. In principle, this might lead to a smaller quantum correlation set than the one defined above. This is a hard problem in general, discussed as {\it Tsirelson's problem} at the end of this section. But, as we will also show below, in the minimal scenario there is no difference, so there is no possible harm in starting from the more inclusive condition \eqref{eq:ABhypotheses}.

The correlations are computed from the operators above by taking traces:
\begin{equation} \label{eq:getCfromAB}
 c_{ij}\,=\,\tr(\rho A_iB_j)
   \qquad
   \hbox{for} \quad 1 \leq i, j \leq 2.
\end{equation}
The {\em correlation body} $\Qm$ consists of all points $c$ in the cube
$ [-1,1]^4$ that admit such a representation.

There is an analogous set $\Cl$ in classical probability theory, where
the $A_i$ and $B_j$ are $\pm1$-valued random variables, with  joint
probability distribution $\rho$.
Writing angle brackets for expectations, the formula is
\begin{equation} \label{eq:cClassical}
 c_{ij}\,=\,\bigl\langle A_iB_j\bigr\rangle
   \qquad
   \hbox{for} \quad 1 \leq i, j \leq 2.
\end{equation}
The classical set $\Cl$ consists of all points $c$ in the cube
$ [-1,1]^4$ that admit such a representation.

We note that \eqref{eq:cClassical} is the special case of
\eqref{eq:getCfromAB} when all $A_i$ and $B_j$ commute.
All matrices can then be taken to be diagonal, and the diagonal entries of $\rho$ form a probability distribution.
(Analogous statements hold in infinite dimensional Hilbert spaces, where the $A_i,B_j$ may have continuous spectra). Hence $\Cl \subset \Qm$.

The whole point of our correlation body is that the reverse inclusion is false.
A prominent example is
\begin{equation}\label{cStandard}
  c \,=\, \begin{matrix} \frac{1}{\sqrt2}(1,1,1,-1) \end{matrix} \, \in  \Qm \setminus \Cl.
\end{equation}
It is easy to build a quantum representation, and this has been realized experimentally to very high precision. So the realizability of this $c$ is a well-confirmed experimental fact.
On the other hand, $c$ cannot be classical, because
 the Clauser-Horne-Shimony-Holt version of John Bell's inequality
holds for all $c\in\Cl$:
\begin{equation}\label{CHSH}
  {\rm CHSH}(c):\qquad\qquad \begin{matrix}
\frac{1}{2}(c_{11}+c_{12}+c_{21}-c_{22}) \end{matrix} \leq 1. \qquad
\qquad
\end{equation}
The point $c$ in  \eqref{cStandard} is not classical because
the  left hand side of \eqref{CHSH} equals  $\sqrt2$, and this exceeds $1$.
This is a remarkable result, the basis of an experimentum crucis ruling out a whole mode of describing Nature.

The experiments put quantum theory to a sharp test: The value $\sqrt2$ is an upper bound for all quantum correlations. If a value significantly larger than $\sqrt2$ had been found, then this would refute the quantum way of describing Nature, in just the same way as classical theories are excluded by a violation of Bell's CHSH inequality. This inequality and all linear inequalities bounding $\Qm$ are called {\em Tsirelson inequalities}.

Tsirelson's bound has led to speculations about super-quantum correlations in families of theories (``generalized probabilistic theories''), and to the desire to view the quantum case in a larger context. The only constraint then would be that Alice choosing a measurement device  makes no detectable difference for the probabilities of outcomes seen by Bob alone, i.e., without comparing outcomes with Alice.
This is the {\em no signalling set} of correlations, denoted by $\Ns$.
It satisfies  $\Cl\subset\Qm\subset\Ns$. Since in this paper we ignore marginals seen by only
one partner, the remaining constraint is that $c$ lies in the cube,
so  $\Ns=[-1,1]^4$.

A crucial property for quantum key distribution is that a maximal violation of the CHSH inequality can be achieved in an essentially unique way. Thus, by just verifying such correlations, without any knowledge  about the construction of the devices, one can reconstruct $\rho,A_i,B_j$ up to trivial enlargements. This property is called {\it self-testing}. It  implies that any further system will be uncorrelated, so an eavesdropping third party could never learn anything about the data collected by Alice and Bob. We extend this property to all non-classical extreme points of $\Qm$ in \reff{prop:selftest}, and explain the cryptographic background in \reff{sec:qkd}.

When the numbers $N,M,K$ of parties, settings and outcomes are larger than $2,2,2$, computing optimal bounds for $\Qm$ is a hard problem, and necessary and sufficient conditions describing $\Qm$ are rare. Basically, one needs special algebraic properties to characterize the measurement operators, that are needed to show self-testing. In Tsirelson's work \cite{Tsi85}, Clifford algebra techniques allow statements about $NMK=2M2$, and the C*-algebra generated by two projections allows at least a parametrization of the extreme points of $NMK=N22$ \cite{WW01a}. Thus, these cases can effectively be reduced to a finite dimensional Hilbert space. In general this is not possible: It is known \cite{Slofstra} that restricting the dimension to be finite, in general, gives a set that is not closed, so some limiting correlations require infinite dimension.

Related is the problem of distinguishing the sets of correlations, in which the operators are just required to commute, as in \eqref{eq:ABhypotheses}, versus the more restrictive tensor product form described after that equation. In finite dimension the two are readily shown to be equivalent (see, e.g., \cite{Tsir_Scholz}), which may have led Tsirelson to claim in \cite{TS93} that the two were equivalent. The problem was first clearly noticed in the project \cite{navascues2008} chararacterizing quantum correlations by semidefinite matrix hierarchies, where the commutation condition is easy to implement, but tensor products are more elusive. The authors of that study contacted Tsirelson, who (``to his crying shame'' according to his own website) could not make work the argument that he had had in mind in \cite{TS93}. This became {\it Tsirelson's problem}. This problem turned out to be equivalent to a celebrated conjecture in operator algebras \cite{Tproblem_Werner,Fritz}, and was finally solved $14$ years later in the negative \cite{Tsirelson_solved}. So we now know that in general there is a difference. From the detailed sharp characterizations of the minimal $\Qm$ collected in this paper, including explicit finite dimensional tensor product models for all correlations, we see that the problem does not arise here. But such detailed knowledge is rare. This is one main motivation for the detailed geometric study that is to follow.

\subsection{A View from Mathematics: Convex Algebraic Geometry}\label{sec:introMath}

The theory of polytopes is a mature subject in mathematics \cite{Ziegler}.
Based on linear algebra, it leads to rich geometric objects classified by
discrete combinatorial structures and a beautiful duality theory.
In its guise as {\em linear programming}, i.e, optimization over a polytope with
linear objective, it has found many applications. The natural next step in complexity is
to go from linear algebra to {\em nonlinear algebra} \cite{Michalek}. If we replace
convex sets and objectives described by linear inequalities with
those described by polynomials,
the basic geometric appeal and duality theory are still there. But
we now enter into the world of algebraic geometry.
 Much richer ways of combining sets, partial descriptions and geometric
constraints have to be considered. Examples play an important role in exploring these possibilities.
Our study of quantum correlations serves as a showcase for nonlinear phenomena that occur
in {\em convex algebraic geometry}~\cite{Blekherman, Uhler}.

The correlation body $\Qm$ is compact and  convex.
Compactness is not obvious but follows from  \reff{thm:main}.
For convexity, let $c , \tilde c \in \Qm$
have realizations  \eqref{eq:getCfromAB} by matrices
of size $m$ and $\tilde m$.
Any convex combination  $\lambda c + (1-\lambda) \tilde c$
is realized by block matrices of size $m + \tilde m$, namely
$\lambda \rho \oplus (1-\lambda) \tilde \rho$,
$\,A_i \oplus \tilde A_i\,$ and $\,B_j \oplus \tilde B_j$.
By contrast, if we were to fix $m$ then
compactness is easy to see but convexity generally fails.
For instance, fixing $m=1$, the image under  \eqref{eq:getCfromAB}
is the set of $2 \times 2 $ matrices of rank $\leq 1$
with each entry in $[-1,+1]$.

The body $\Qm$ lies between two polytopes. First, $\Qm$ is contained in
the $4$-cube $\Ns =[-1,1]^4$, which has $16$ vertices,
$32$ edges, $24$ ridges and $8$ facets. Second, $\Qm$ contains the classical
set $\Cl$, which is also a polytope. It has two equivalent descriptions, both of which are helpful for our project.
On one hand, it is a {\em demicube},
which is defined to be the convex hull of the eight even vertices of $\Ns$, i.e., the vertices for which the product of the four $\pm1$-coordinates is $+1$. This describes the inclusion $\Cl\subset\Ns$.
On the other hand, we can think of $\Cl$ as the {\em cross polytope} \cite[Example 0.4]{Ziegler}, combinatorially dual to $\Ns$, whose extreme points lie on the Cartesian coordinate axes at unit distance from the origin.
Hence $\Cl$ has $8$ vertices, $24$ edges, $32$ ridges and $16$ facets, the reversed sequence of that for $\Ns$. This census of the faces of $\Cl$ is
of direct relevance for our description of the boundary of the convex body $\Qm$, to be given in \autoref{prop:boundary}.

The correlation body $\Qm$ is semialgebraic: It can be described by a
Boolean combination of polynomial inequalities. Here a phenomenon
arises that is unfamiliar from polytope theory. It is not sufficient to use a
conjunction of polynomial inequalities. In other words, $\Qm$ is
not a basic semialgebraic set. Moreover, while both polynomials
$g$ and $h$ in \autoref{thm:main} (d) are needed, only $h$ is determined by
$\Qm$, as the unique algebraic description of a part of the boundary,
while there is some freedom of choice for $g$.

A main source of convex semialgebraic sets is the cone of positive semidefinite matrices.
Quantum theory is entirely based on this cone. Its states,
observables, and channels are all defined in terms of it.
The intersection of the semidefinite matrix cone with an affine-linear space
is called a {\it spectrahedron}. The set $\Qm$ arises from a spectrahedron by projection, and
it is thus in the class of {\em spectrahedral shadows} \cite{Scheiderer}.

Each of the themes described in the previous paragraphs can be used
to characterize the set $\Qm$. This leads to six descriptions that
look different at first glance. We summarize these in the following theorem.

\begin{thm} \label{thm:main}\proofin{sec:pround}
The following six items all describe the same subset $\Qm$ in $\,\Rl^4$:
\begin{itemize}
\item[(a)] The set of quantum correlations $c$, as  defined in \autoref{sec:introPhys}, i.e.,
  the $c_{ij}$ from \eqref{eq:getCfromAB} satisfying \eqref{eq:ABhypotheses}.
\item[(b)] The convex hull of the  hypersurface
$ \bigl\{  (\cos\alpha,\cos\beta,\cos\gamma,\cos\delta) \in \Rl^4 \stt
  \alpha+\beta+\gamma+\delta \equiv 0 \mod2\pi   \bigr\}$.
\item[(c)] The image of the demicube $\,\Cl$ under the homeomorphism
$\,\Sin : \Ns \rightarrow \Ns ,\,
c \, \mapsto \,
   \sin\bigl(\frac\pi2 c_{ij}\bigr)_{1 \leq i,j \leq 2}$.
\item[(d)] The semialgebraic set $\,\bigl\{\, c \in \Ns \stt   g(c) \geq 0\,\,\, {\rm or}\,\,\,
h(c) \geq 0 \,\bigr\}$, where
\begin{eqnarray}
  g(c) &=& 2-(c_{11}^2+c_{12}^2+c_{21}^2+c_{22}^2)+2 c_{11} c_{12} c_{21} c_{22} \label{gpoly}\\
  h(c) &=& 4(1-c_{11}^2)(1-c_{12}^2)(1-c_{21}^2)(1-c_{22}^2)\,-\,g(c)^2\ . \label{hpoly}
\end{eqnarray}
\item[(e)] The spectrahedral shadow consisting of all points $(c_{11},c_{12},c_{21}, c_{22} )\in\Rl^4$
such that the matrix
\begin{equation}\label{eq:specshadow0}
C\,=\, \left(\begin{matrix}
    1 & u & c_{11} & c_{12} \\
    {u} & 1 & c_{21} & c_{22} \\
    c_{11} & c_{21} & 1 & v \\
    c_{12} & c_{22} & {v} & 1\end{matrix}\right)
\end{equation}
is positive semidefinite for some choice of $u,v\in\Rl$.
\item[(f)] The scalar products of pairs of unit vectors $a_i,b_j$ in some Euclidean space:
$c_{ij}=a_i\cdot b_j$ for $i,j=1,2$.
\end{itemize}
\end{thm}

\noindent
One way to describe a convex body is by the maxima of all linear functionals.
This is the {\em support function}
\begin{equation}\label{supportFct}
  \phi(f)=\sup\bigl\{ f\cdot c \stt  c\in\Qm \bigr\}.
\end{equation}
Here $f\in\Rl^4$ and ``$\cdot$'' denotes the scalar product in $\Rl^4$.
The following theorem gives an explicit formula.

\begin{thm} \label{thm:supp}\proofin{sec:supproof}
Consider the following expressions in four variables $f = (f_{11},f_{12},f_{21},f_{22})$:
{\def\suck{\hskip-8pt}
\begin{eqnarray}
 k(f) \suck&=& (f_{11} f_{22}{-}f_{12} f_{21}) (f_{11} f_{12}{-}f_{21}f_{22})(f_{11} f_{21}{-}f_{12} f_{22}) \label{kpoly}\\
 p(f) \suck&=& f_{11}f_{12} f_{21} f_{22}\label{ppoly} \\
 \phi_\Cl(f)
      \suck&=& \max\{|f_{11}{+}f_{12}{+}f_{21}{+}f_{22}|,|f_{11}{+}f_{12}{-}f_{21}{-}f_{22}|,
      |f_{11}{-}f_{12}{+}f_{21}{-}f_{22}|,|f_{11}{-}f_{12}{-}f_{21}{+}f_{22}|\}
      \label{phiCearly} \\
 m(f) \suck&=& \Bigl(\min_{i,j} |f_{ij}| \Bigr)\Bigl( \sum_{i,j} |f_{ij}|^{-1} \Bigr).
\end{eqnarray}
The support function of the correlation body $\Qm$ equals
\begin{equation}\label{phicases}
  \phi(f)=\left\lbrace\begin{array}{ll}
            \sqrt{\frac{k(f)}{p(f)}} & \mbox{if} \ \ p(f)<0 \ \ \mbox{and} \ \ m(f) >2 ,\\[10pt]
            \phi_{\Cl}(f)& \mbox{otherwise.}
            \end{array}\right.
\end{equation}}
\end{thm}

The case distinction in \eqref{phicases} is between the ``classical'' case and the  ``quantum'' case. Indeed, as the notation suggests,
the piecewise linear expression $\phi_\Cl$ in \eqref{phiCearly} is the support function of the cross polytope $\Cl$.
Hence $\phi_\Cl$ represents inequalities for
classical correlations. On the other hand, in first case
of \eqref{phicases},  the maximizers are
non-classical correlations, which share all essential properties with CHSH: For fixed $f$ the maximizer is unique (see \reff{prop:hom}) and given by a unique quantum model (see \reff{sec:desquribe}).

Let $\KK$ be a convex body that contains the origin in its interior.
Then its dual or polar is a convex body $\KK\polar$ that
represents the linear inequalities satisfied by $\KK$.  In symbols, we have
 $\KK\polar=\{f\stt f{\cdot}c\leq1,\ c\in\KK\}$.
If $\KK$ is a polytope then so is $\KK\polar$, and the face numbers of
$\KK\polar$ are the reversal of the face numbers of $\KK$.

Consider the sequence of inclusions  $\Cl\subset\Qm\subset\Ns$ we discussed above. Then we have
$\Ns\polar \subset \Qm\polar \subset \Cl\polar$. Here $\Cl$ and $\Ns\polar$ are cross polytopes,
while $\Ns$ and $\Cl\polar$ are $4$-cubes. However, even stronger statements are true:
$\Cl$ is affinely isomorphic to $\Ns\polar$, and  $\Ns$ is affinely isomorphic to $\Cl\polar$.
Moreover, these isomorphisms extend to the middle term in the inclusion
$\Cl\subset\Qm\subset\Ns$, i.e., the correlation body $\Qm$ is {\it self-dual}:

\begin{thm} \label{thm:selfdual}\proofin{sec:self-dual}
There is an orthogonal transformation $H$ on $\Rl^4$ such that
$$\Cl\polar=\frac12H\Ns, \quad \Ns\polar=\frac12H\Cl \quad {\rm and} \quad \Qm\polar=\frac12H\Qm . $$
\end{thm}

The proofs of the three theorems are presented in \autoref{sec:proofs}.
While proving that they agree, we write
 $ \Qm_{(a)},  \Qm_{(b)},  \Qm_{(c)},  \Qm_{(d)},  \Qm_{(e)},  \Qm_{(f)}$
for the six sets in \autoref{thm:main}. All
objects and assertions  are explained in detail in \autoref{sec:describe}.
along with lots of additional information.
For instance, \autoref{prop:boundary} describes
the stratification of the boundary of $\Qm$ into various patches.
Readers might start with \autoref{fig:symAB}, \autoref{fig:horicut}, and \autoref{fig:diacut}.

\subsection{Short review of previous work and preview of new results}\label{sec:review}

The correlation body $\Qm$ first came into focus in Tsirelson's work \cite{Tsi80}. That
paper gives no proofs, but some of them were supplied in \cite{Tsi85}. This includes the
characterization $\Qm_{(a)}=\Qm_{(f)}$ in the more general 2M2\marge0 case (\cite[Thm.\,1]{Tsi80}=\cite[Thm.\,2.1]{Tsi85}). Thereby the study of $\Qm$, whose definition
also allows infinite dimensional Hilbert spaces, is reduced to a finite dimensional problem. A  semialgebraic
description for the 222\marge0 case is given in \cite[Thm.\,2.2]{Tsi85}, along with an expression for the support
function \cite[Thm.\,2.2]{Tsi85}. This is our \autoref{thm:supp}.
Tsirelson calls these results `elementary' consequences of (f), and does not provide a proof.
He also thought about issues not covered in our review, like the full 222 case,
multipartite scenarios ($N>2$) \cite[Sect.~5]{Tsi85}, and violations of CHSH inequalities by position
and momentum (\cite{Tsi85}, see also \cite{Kiukas}).

The spectrahedral shadow $\Qm_{(e)}$ first appeared in Landau's work~\cite{Lan88} as a relaxation of the correlation body.
That the relaxation is tight follows from Tsirelson's theorem~\cite{Tsi80}. Landau also gave a nearly semialgebraic characterization of $\Qm$ (see \eqref{LandauPoly} below), which only misses the semialgebraic standard form by containing a square root. He almost achieved the description (c).

The pushout (c) was found by Masanes~\cite{Masanes03}, who stated that it identifies $\Cl$ and $\Qm$. He also considered the cosine-parametrized manifold of correlations in (b). The pushout was used implicitly in \cite{navascues2008,singletST}, in the form of  a characterization of $\Qm$ by linear inequalities applied to the inverse pushout. However, it was not pointed out that the linear inequalities just characterize $\Cl$.
Of course, spectrahedral shadows are used as outer approximants to $\Qm$ in the semidefinite
hierarchies \cite{doherty2008, navascues2008}. This is an important technique for higher NMK, even
though one gets the convex body exactly only in the minimal case.

We thought that self-duality (\reff{thm:selfdual}) was entirely our own idea, until we found it  mentioned in \cite[Sect.4]{N22duality} with only a private communication by Palazuelos and Villanueva (2011) as a reference. We could not trace a published proof.

The uniqueness of quantum models, now known as self-testing~\cite{MayersYao}, was also
studied by Tsirelson~\cite{Tsi85}. Independently, \cite{SW1} obtained the self-testing property for the CHSH inequality. A covering of the set of
extreme points $\partial_e\Qm$, which results in the cosine parametrization (b) and its analog in the N22\marge0-case,
were found in \cite{WW01a}, see also \cite{Masanes05}. The exact identification of the set of extreme points was found much later in~\cite{Le18, singletST}.
The minimal case is an important example in many applications such as quantum nonlocality~\cite{bellnonlocality,Goh}, self-testing~\cite{reviewST}, and quantum cryptography~\cite{schwonnek2020qkd,tan2020qkd}.

Our contribution is, first, a systematic account with proofs of all these results. The condensation of all the existing characterizations of $\Qm$ is \autoref{thm:main}. We give the first published proofs of support function (\reff{thm:supp}) and self-duality (\reff{thm:selfdual}). Our second contribution is the detailed geometric understanding of $\Qm$, and particularly of its boundary. This includes visualizations wherever possible, the classification of types of boundary points, i.e., exposed and non-exposed extreme points, edges and $3$-dimensional facets (\autoref{prop:boundary}) and the analysis of how these parts fit together in $4$ dimensions (\reff{fig:projective}). Moreover, we provide semialgebraic descriptions of all boundary pieces, and analyze the symmetry group of $\Qm$ (a bit larger than trivially expected).  Third, we analyze the relation between $\Qm$ and its polar dual $\Qm\polar$, i.e., the set of linear inequalities satisfied by $\Qm$. By self-duality, this is affinely isomorphic to $\Qm$, so much of the analysis can be taken over. But we also analyze how points on the boundary of $\Qm$ map to their supporting hyperplanes in $\Qm\polar$ (\reff{fig:T2T}). This is a part of the so-called normal cycle, which we studied in \reff{sec:convNC}. Fourth, we consider the quantum models for all $c\in\Qm$. This is again a traditional subject, going back at least to Bell, and to Tsirelson for the uniqueness question (see above). Our review is condensed into a theorem of equivalences (\reff{prop:selftest}) including geometric conditions, and two kinds of cryptographic security. We also address correlation bodies under constraints such as bounded Hilbert space dimension, fixed state, or fixed observables. The last question suggested an apparently new and surprisingly simple non-linear inequality for $c\in\Qm$, namely $\abs{\det{c}}\leq1$, with sharp condition for equality (\reff{prop:det}).

\section{Description of the Correlation Body}
\label{sec:describe}

A convex body is a compact convex set with nonempty interior. The convex body $\Qm$ has dimension four.  Our aim is to describe it in every detail.
Naturally, the geometric description will strain our $3$-dimensional intuition. As always, the solution is to build the geometric intuitions (German ``Anschauung'', visualization) on analytic notions, such as sections, projections, affine submanifolds, extreme points, and faces, which have clear definitions, but also on low-dimensional instances on which intuitions can be grounded. In the case at hand, the dimension gap is not too large, and we will see that some three-dimensional sections faithfully display important features of the four-dimensional body. We will also
point out where this becomes too misleading. One general cautionary remark is that extreme points of a section usually fail to be extremal in the higher-dimensional body.

This section is divided into subsections, which are organized by geometric features, from overall properties to the classification of boundary points and their explicit description, to the dual inequalities, and finally the quantum realizations.  This is different from the logical ordering in a proof. A complete set of proofs will be given only later in \reff{sec:proofs},
which is accordingly organized in logical progression.

\subsection{Gallery}
We visualize the $4$-dimensional body $\Qm$ by showing $3$-dimensional sections. Some of these will be at the same time projections.
For example, the zero marginal case arises from the full marginal case either by ignoring the
marginals (a projection) or by taking the subset with zero marginals (a section by a linear subspace).
Such sections/projections often arise by averaging over a symmetry group \cite{VW}.

\begin{figure}\centering
	\includegraphics[width=13cm]{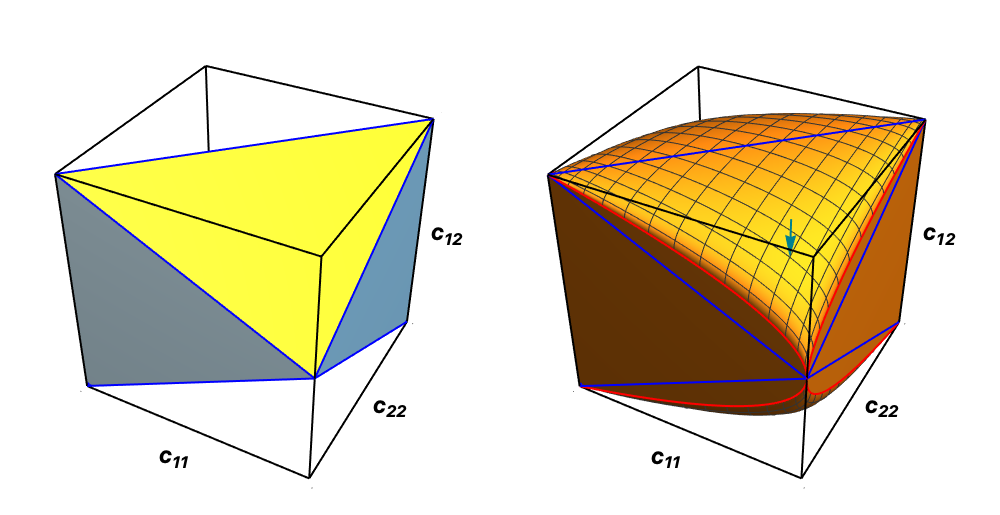} \vspace{-0.2in}
	\caption{The Alice-Bob symmetric subsets of correlations $(c_{12} =c_{21})$, both a projection and a section of their $4$-dimensional counterparts.
    Left: the polytopes: The cube $\Ns$ has black edges, and the octahedron $\Cl$ has
 blue edges, with two kinds of facets distinguished: CHSH-facets (yellow) defined by saturation of a CHSH-Bell inequality and ``$\Ns$-facets'' extending to facets of $\Ns$ (gray).
    Right: the correlation body $\Qm$. Its boundary consists of a strictly convex surface of exposed extreme points, together with
     $\Ns$-faces extending those of $\Cl$, whose boundaries are outlined in red. The arrow points to the CHSH point $c=(1,1,1,-1)/\sqrt{2}$. }
\label{fig:symAB}
\end{figure}

We begin in \reff{fig:symAB} with correlations that are
symmetric under exchanging Alice and Bob, i.e., $c_{12}=c_{21}$. The corresponding
projection is represented by
$(c_{11},c_{12},c_{21},c_{22})\mapsto  (c_{11},c_{22},(c_{12}+c_{21})/2)$.
The corresponding sections of the polytopes $\Ns$ and $\Cl$ are a cube and an octahedron, respectively. The extreme points of the section of $\Cl$ are not all extremal in four dimensions:
The point $(1,1,0)$, where two parabolas meet in \reff{fig:symAB}, has $c_{12}=c_{21}=0$.
It is not extremal in $\Cl$. But it is the midpoint between the extreme points $(1,1,-1,1)$ and $(1,-1,1,1)$ of $\Cl$, which lie outside the section shown.
The nonlinear boundary in \reff{fig:symAB} (right) is a quartic surface, obtained by setting $c_{12}=c_{21}$ in the sextic $h(c)$ and cancelling a factor $(c_{11}-c_{22})^2$.
These geometric features have $4$-dimensional counterparts, to be described later.

\begin{figure}\centering
\includegraphics[width=16.1cm]{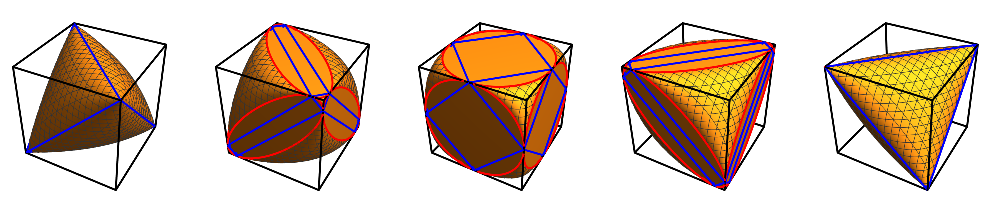}
	\caption{Parallel sections $c_{11}=t$ through
the body $\Qm$, for
$t=-1,-0.8, \,0, \,0.8,\,1$.
}
\label{fig:horicut}
\end{figure}

In \reff{fig:horicut} we show the sections parallel to the facets of the cube $\Ns$.
These are obtained by fixing the value of one coordinate, say $c_{11}$, at a number
$t$ in the interval $[-1,1]$.
This family of pictures gives a full description of $\Qm$. The corresponding projections are
non-informative: They are equal to the full $3$-cube.
The special sections $c_{11}=\pm1$ are facets of $\Qm$.
This $3$-dimensional shape is known as the {\em elliptope}.

\begin{figure}[ht]\centering
	\includegraphics[width=11cm]{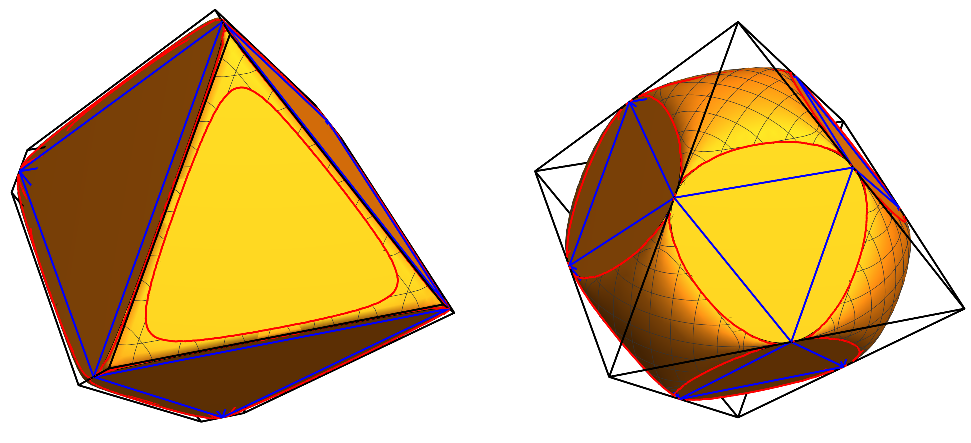} \vspace{-0.1in}
	\caption{Sections through $\Qm$ perpendicular to a long diagonal of
the cube $\Ns$. There are two
symmetry classes of diagonals: those connecting two classical correlations (left),
  and those connecting two PR-boxes (right). The cut on the right goes through the origin.
On the left we would get the full octahedron. Our cut is thus
slightly off-center. }
\label{fig:diacut}
\end{figure}

Other cutting directions, which can be expected to have an interesting symmetry are  sections orthogonal to the main diagonals of $\Ns$. Like the vertices, they come in two kinds, either connecting two classical correlations or connecting two PR-boxes. In \reff{fig:diacut} we show, on the left, a cut very close to
the hyperplane $c_{11}+c_{12}+c_{21}+c_{22}= 0$, which is orthogonal
to the diagonal connecting the classical point $(1,1,1,1)$ and its antipode. On the
right in \reff{fig:diacut}, we see the cut given by $c_{11}+c_{12}+c_{21}-c_{22}= 0$.

\subsection{Enclosing and enclosed polytopes}\label{sec:polytopes}

The cube $\Ns$ consists of the tuples $c=(c_{11},c_{12},c_{21},c_{22})$ with $-1\leq c_{ij}\leq1$.
It has $2^4=16$ vertices, namely the points in $\{-1,1\}^4$.
The classical distributions are the mixtures of uncorrelated $c$, i.e.,
$c_{ij}=a_ib_j$ for some $a_i,b_j\in[-1,1]$.
A classical correlation is extremal when $a_i,b_j=\pm1$. Hence, the extreme points of $\Cl$ are also extreme points of $\Ns$, but which?
This is decided by a sign: For a classical extreme point we have  $c_{11}c_{12}c_{21}c_{22}=a_1^2b_1^2a_2^2b_2^2=+1$. So
the classical extreme points of $\Cl$ are just the $8$ even vertices of $\Ns$. The odd vertices are the so-called Popescu-Rohrlich (PR-)box correlations.
They have neither classical nor quantum realizations: In \eqref{CHSH} they satisfy $\mathrm{CHSH}(c)=2$, which clearly exceeds the quantum maximum.

The $16$ facets of the demicube $\Cl$ come in two classes. These are distinguished by how they sit inside the  cube $\Ns$. A
facet in the first class extends to a facet of $\Ns$.
It is the intersection of $\Cl$ with a hyperplane like $c_{11}=1$. There are
eight such facets, which we call {\em $\Ns$-facets} of $\Cl$.
The more interesting kind is called a {\em CHSH-facet}, because it saturates a Clauser-Horne-Shimony-Holt inequality \eqref{CHSH}. There are
eight such inequalities: Any odd number of minus signs can multiply the four correlations.

All of this can be seen also in the $3$-dimensional cut along the hyperplane
$ c_{12}=c_{21}$ shown on the left in  \reff{fig:symAB}.
The two polytopes are the intersection of $\Cl$ and $\Ns$ with this hyperplane.
The CHSH-facets and the $\Ns$-facets are marked in different colors.
Other slices of $\Cl$ (blue frame) and $\Ns$ (black frame) are shown in Figures \ref{fig:horicut} and \ref{fig:diacut}. The cut through the origin, orthogonal to the long diagonal connecting classical vertices (approximately as in the left panel in \reff{fig:diacut}) has the intersections with $\Ns$ and $\Cl$ equal to the same octahedron. Since $\Cl\subset\Qm\subset\Ns$, this cut also makes
our convex body $\Qm$ look like a polytope.

Of course, in every cut we see the inclusions $\Cl\subset\Qm\subset\Ns$. More precisely, the $\Ns$-facets of $\Cl$ extend to facets of $\Qm$, which we also call $\Ns$-facets. This is typically a strict extension. The CHSH-facets are no longer faces of $\Qm$. Instead, they become the basis of a bulging part of $\Qm$, above which we find a single vertex of $\Ns$. In fact, this is a feature of any body between $\Cl$ and $\Ns$. We record this fact in the following proposition (also a corollary of~\cite{Bierhorst_2016}), which helps with keeping track of the parts of $\Qm$.

\begin{prop}\label{prop:1chsh}\proofin{sec:uniqueChsh}
Every non-classical correlation $c\in\Ns\backslash\Cl$ violates exactly one of the
eight CHSH-inequalities.
\end{prop}

A feature which will play a major role later, and is
characteristic of the minimal case,
is the {\it duality} between the polytopes $\Cl$ and $\Ns$. We saw this in
\autoref{thm:selfdual}, and it
 will be considered in detail in \reff{sec:dualscribe}.

\subsection{Pushout}\label{sec:push}
The connection between the boundary structures of $\Qm$ and $\Cl$ can be raised from a qualitative
observation to a precise mathematical statement. There is a natural
homeomorphism between the convex bodies. To this end, we define a
transformation $\Sin:\Ns\to\Ns$ of the cube, which we call the {\bf pushout} operation:
\begin{equation}\label{defpush}
   \bigl(\Sin\, c\bigr)_{ij} \,=\, \sin\bigl(\frac\pi2 c_{ij}\bigr).
\end{equation}
This is the coordinatewise application of a suitably scaled sine function. Since the sine maps
the interval $[-\pi/2,\pi/2]$ bijectively and continuously onto $[-1,1]$, we see that $\Sin$ is bijective, continuous and has a continuous inverse. This is relevant because of the
following astonishingly simple characterization of~$\Qm$.

\begin{prop}\label{prop:push}\proofin{sec:pushout}
  \qquad $\Cl$ is homeomorphic to $\Qm$ under $\Sin$, in particular $\Sin\,\Cl=\Qm$.
\end{prop}

The fact that $\Sin\,\Cl=\Qm$ was known to L. Masanes~\cite{Masanes03}. This fact was also often represented as $\Cos(\pi\mathrm{MET}(K_{2,2}))$ in literature on matrix completion (see also~\autoref{sec:pushout}). Here we strengthen the statement to a homeomorphism.

\begin{figure}[ht]\centering
	\includegraphics[width=10cm]{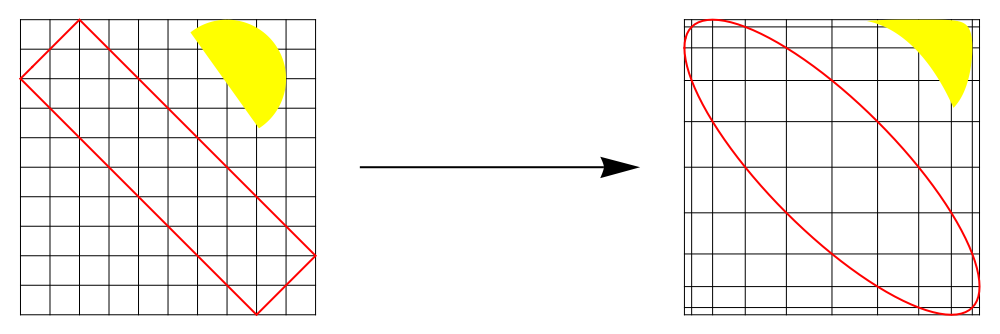}
	\caption{The $2$-dimensional version of the pushout map in \eqref{defpush}.
The rectangle on the left gets mapped to the ellipse on the right.
To verify this compare the scaled grid lines. The yellow semicircle shows that
  convexity is  generally not  preserved under the pushout map. }
\label{fig:push2}
\end{figure}

As a connection between convex bodies this is quite strange: The sine is neither convex nor concave on $[-\pi/2,\pi/2]$, so  the $\Sin$  transformation applied to a convex set normally does not give a convex set. Moreover, the $\Sin$ function is transcendental but
both sets have an algebraic description. So why does this work?
What is the general principle? The pushout property is inherited by the sections in \reff{fig:symAB} and by the $\Ns$-facets because the hyperplanes $\{c_{12}=c_{21}\}$ and $\{c_{11}=-1\}$ are invariant under $\Sin$. It also connects $\Cl$-sections and $\Qm$-sections in \autoref{fig:horicut} when the fixed $c_{11}$-coordinate is appropriately transformed. The closest we can come to a general principle is related to the fact that the pushout of the tetrahedron is the elliptope (see also~\cite{L97}). This fact also underlines the
cosine parametrization in \reff{sec:Dcosines}.
The threefolds of extreme points are parametrized by angles satisfying a linear constraint.
A similar connection arises between two families of curves inside the $3$-cube, on one hand Lissajous knots, i.e., cosine-parametrized closed rational curves, and, on the other, billiard knots, i.e., closed piecewise linear curves bouncing from the boundaries by specular reflection;
see \cite{Lissknot}, \cite[Fig.~3]{soccer}.
Another feature can be understood from the pushout characterization: At the edges of a polytope extending all the way to the boundary we get a rounded surface with continuous tangents. This is
explained in \reff{fig:push2}.

\subsection{Symmetry}\label{sec:symm}

The symmetry group of the regular $4$-cube $\Ns = [-1,1]^4$ is the
{\em hyperoctahedral group} $B_4$. This group has order $384 = 2^4 \times 4!$.
All of these symmetries are given by rotations and reflections in $\Rl^4$.
Each symmetry either preserves the parity of the vertices of $\Ns$,
or it swaps the eight even vertices and the eight odd vertices.
The symmetry group $G$ of the demicube $\Cl$ is an index two subgroup of $B_4$, meaning $|B_4|/|G|=2$. It
consists of symmetries of $\Ns$ that preserve the parity. It follows from
\autoref{prop:push}, and the fact that the pushout map commutes with coordinate
permutations and sign changes,  that $\Qm$ has the same symmetry group $G$ as $\Cl$.

\begin{prop}\label{prop:symmetry}
The common symmetry group of $\Qm$ and $\Cl$ has order
$192$. It is the semidirect product $ G = (\mathbb{Z}/2 \mathbb{Z})^3 \rtimes S_4$,
where $S_4$ is the symmetric group on four elements.
The first factor swaps~labels.
\end{prop}

\begin{figure}[ht]\centering
	\includegraphics[width=10cm]{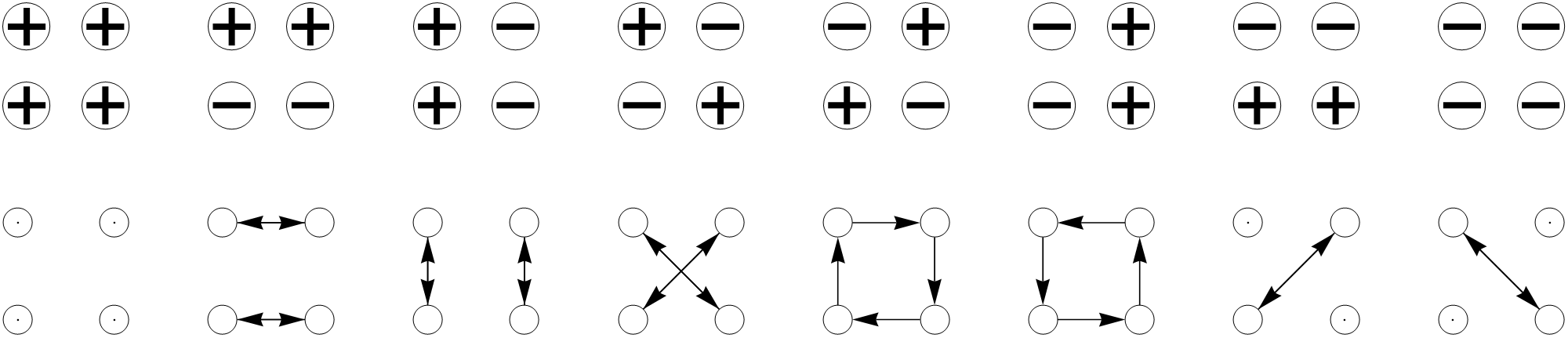}
	\caption{Symmetries that preserve the definition of quantum correlations.
First row: sign patterns.  Second row: the eight
permutations arising from a swap of partners or relabeling.
The symmetry group $G$ of $\Qm$, as described in \reff{prop:symmetry}, is larger:
 It allows {\it all} $24$ permutations.}
\label{fig:groups}
\end{figure}

It is noteworthy that this group is larger than one would expect from the definition of $\Qm$. Indeed, some obvious symmetries arise from changing the conventions for describing the correlations:
Which party is called Alice, which is Bob? Which outcome is $+1$ or $-1$, and which settings get the labels $1$ or $2$? Changing any of these conventions defines a transformation that clearly leaves each of the correlation sets invariant. The resulting group is visualized in \autoref{fig:groups}, and acts on the tuples $(c_{11},c_{12},c_{21},c_{22})$ by sign changes (first row) and permutations (second row) giving only $64$ transformations. Not all sign changes can be obtained, only even ones, and the permutations cannot break pairs of diagonally opposing pairs in the square form in which we arranged the $c_{ij}$.
As the additional ``non-trivial'' transformation needed to generate the group given in
\reff{prop:symmetry}, one can take the swap $(c_{11},c_{12},c_{21},c_{22})\mapsto(c_{12},c_{11},c_{21},c_{22})$.

A group that played a key role in analyzing the classical polytope in the N22\marge0 case in \cite{WW01a} will also play a special role below for establishing self-duality. It is familiar to quantum physicists as a discrete version of phase spaces, describing a quantum or classical system in terms of ``position'' and ``momentum'', or ``multiply'' and ``shift'' operators, like $\sigma_x$ and $\sigma_z$ for qubits \cite{wootters,psuncert}.  To mathematicians it is familiar as the standard case of symplectic self-duality \cite{Prasad}, where the position variables are given by a locally compact abelian group $X$, and the momenta by elements of the dual group $\widehat X$. The Hilbert space of the system is $\Hh=L^2(X)$, on which $X$ acts by transformation of the arguments, and $\widehat X$ by multiplication. This determines a projective representation of $X\times\widehat X$ on $\Hh$ by the Weyl operators.
The case of mechanics has $X=\Rl^f\cong\widehat X$, where $f$ is called the number of degrees of freedom. For a single qubit, $X$ is the two element group, and the Weyl operators are the Pauli matrices as above. For our problem we obtain the Klein four-group: a product of two copies of the two element group. The elements of this group are the indices of $c_{ij}$, with the convention that $1$ is the identity and $2{\cdot}2=1$. Thus the correlation body is a subset of $\Hh$. The 16 Weyl operators are the products of the first four elements of the each of the two lines in \reff{fig:groups}, the top row giving the momentum translations and the bottom row the position translations. A standard automorphism of this structure is the Fourier transform that takes functions on $X$ to functions on $\widehat X$. This is not a symmetry of $\Qm$, but, as we will see, takes $\Qm$ to its dual convex body.

\subsection{Boundary, faces, and extreme points }

A {\it face} $F$ of $\Qm$ is a subset such that a convex decomposition of a point in $F$
with non-zero weights must have all components in $F$. We use the term {\it facet} for faces of codimension $1$. An {\it extreme point}
of $\Qm$ is a point $c$ in $\Qm$ such that $\{c\}$ is a face. A face is called {\it exposed} if it is the zero set
of an affine-linear function which is non-negative on $\Qm$.
Since the pushout is a homeomorphism, it identifies the boundary of $\Qm$ with the boundary of
$\Cl$. The facets of the demicube $\Cl$ thus provide a partition of the boundary of
the correlation body $\Qm$. However, the image of a facet may now become curved, making
 interior points of the facet extremal in $\Qm$. This happens exactly for the CHSH-facets of $\Cl$, as the following proposition shows. In the following we always list the interiors of faces
because the boundaries already belong to another face. Here by
 ``interior'' we mean the relative interior, i.e., the interior in the affine span of the~face.

\begin{prop}\label{prop:boundary}\proofin{sec:normals}
The convex body $\Qm$ is the disjoint union of the following semialgebraic~sets:
\begin{itemize}
\item[\Qt1] $8$ {\bf classical exposed points}  $c\in\ext\Cl$.
\item[\Qt2] $24$ {\bf exposed edges}, i.e., the interiors of line segments that connect
pairs of classical exposed points.
\item[\Qt3] $32$ surfaces of {\bf non-exposed extreme points}.\\
     Each is the pushout of a triangle in  $\partial \Cl$,
which is the intersection of an $\Ns$-facet and a CHSH-facet.
\item[\Qt4] $8$ threefolds of {\bf exposed extreme points}.\\
      Each threefold is the curved pushout of the interior of a CHSH-facet of $\Cl$,
      which is a tetrahedron.
\item[\Qt5] $8$ {\bf elliptopes}' interiors, i.e., the interiors of the facets
that arise from $\Ns$-facets of $\Cl$, as in \autoref{fig:symAB}.
\item[\Qt6] The interior of $\Qm$.
\end{itemize}
\end{prop}

\begin{figure}[ht]\centering
	\includegraphics[width=5.6cm]{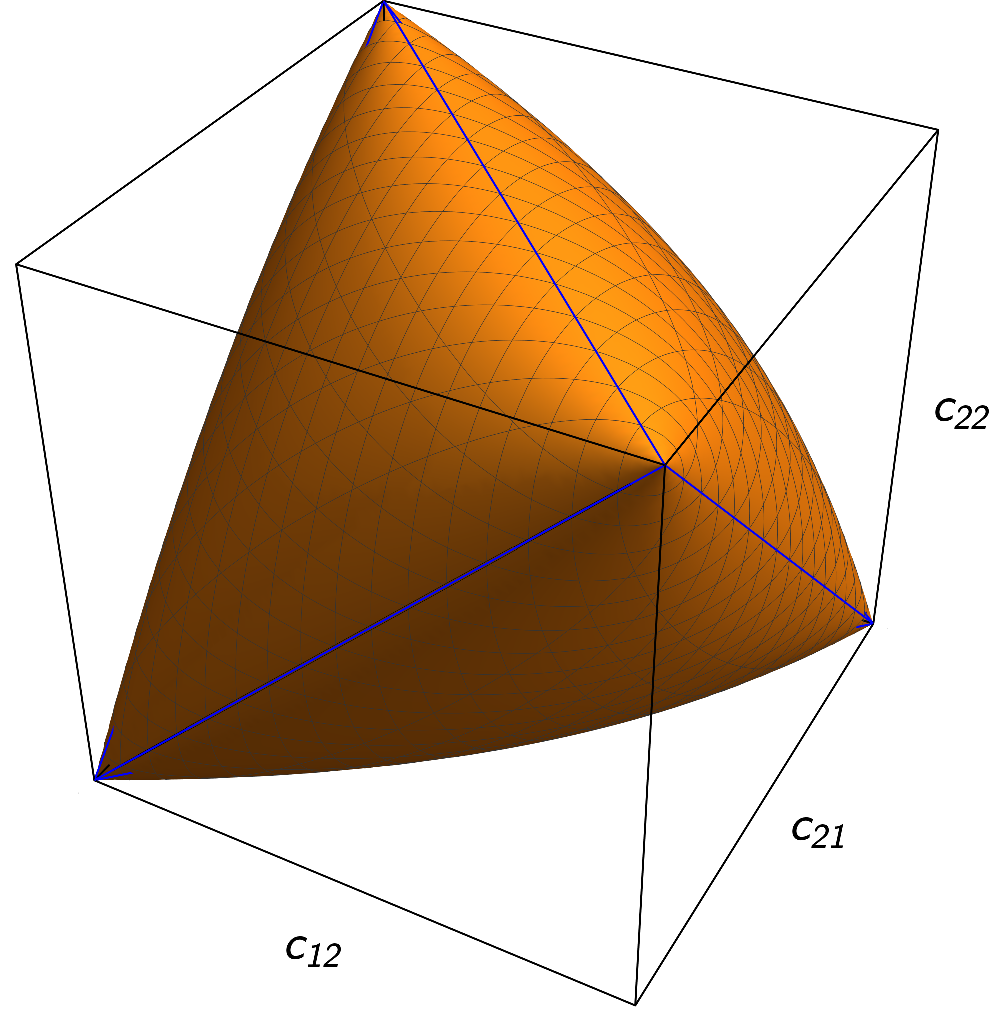}
	\caption{Every $3$-dimensional face of $\Qm$ is an elliptope.
Points of type \Qt1 are vertices of the blue tetrahedron,  \Qt2 give blue edges,
\Qt3 is the orange surface. Extreme points of the elliptope are exposed in
$3$ dimensions, but not in $4$. Type \Qt5 comprise the interior. }
\label{fig:elliptope}
\end{figure}

The stratification of $\Qm$ shown in \reff{prop:boundary}  mirrors the
stratification of the demicube $\Cl$ into relatively open faces. Indeed,
the numbers seen in \reff{prop:boundary} count the faces of various dimension
of $\Cl$. Namely, $\Cl$ has $8$ vertices, $24$ edges, $32$ two-dimensional faces, etc.
In short, the {\em f-vector} of $\Cl$ equals $(8,24,32,8{+}8,1)$.

The closures of sets of type \Qt x are said to be {\em of type} \Qtc x.
Types \Qtc4 and \Qtc5 suffice to cover the boundary.
In the literature in convex algebraic geometry \cite{ciripoi2018computing,Plaumann},
boundary strata (cf. \autoref{sec:convNC} for a rough explanation) of codimension one are now called {\it patches}.
Thus, our body $\Qm$ has $16$ patches, eight of type \Qtc4 and eight of type~\Qtc5.

The difference between the types \Qt4 and \Qt5 will be important for what follows.
It also relates to the two types of {\it maximal non-trivial faces} of $\Qm$. The only two types are the singletons
 in \Qt4, and the $\Ns$-facets \Qtc5. These maximal faces are also exposed.
Exposedness properties will be discussed later in their natural context: duality.
An intuitive geometric understanding of the non-exposedness of type \Qt3  can be gained from considering the pushout mechanism: This converts the junction between a CHSH-facet and an $\Ns$-facet of $\Cl$ to a junction with matching tangents.
The prototype for this is shown in \reff{fig:push2}. The red ellipse is tangent to the boundary of the
square at the four points of intersection. For \Qt3-points the same happens in higher dimension.

Since the \Qt4 extreme points will be discussed in more detail, let us briefly describe the \Qtc5 facets.
These elliptopes are visualized in \reff{fig:elliptope}, an enlarged version of the first panel in
\reff{fig:horicut}.
Analytically, the elliptope, say the subset of correlations $c=(-1,x,y,z)\in\Qm$,
is described by the single inequality $1-x^2-y^2-z^2+2xyz\geq0$ in the cube; cf.~\cite[eqn (1.1)]{Michalek}.
Since the pushout map restricts to $\Ns$-facets, it is the pushout of the $\Cl$-like tetrahedron that forms its skeleton. Indeed, the edges of $\Cl$ are marked by two of the $c_{ij}$ being $\pm1$, and the other two equal up to a sign. This condition is invariant under pushout, so the edges of $\Cl$ are invariant as sets, but not pointwise fixed, and become exactly the \Qtc2-edges of $\Qm$.

How do all these pieces fit together? Once again this is readily answered by looking at the pushout identification of $\Qm$ with $\Cl$.
Consider the bicoloring of the facets of the cross polytope $\Cl$.
The two colors distinguish CHSH and $\Ns$ types.  Only tetrahedra of different types
intersect in a surface, and each  surface \Qtc3 is the intersection of a \Qtc4 and a \Qtc5-set. The
triangular  surfaces  \Qt3 are thus faithfully portrayed in \reff{fig:elliptope}. The intersection of two \Qtc4-surfaces is lower dimensional: If they are not disjoint opposites, they intersect in a straight edge \Qtc2.

\begin{figure}[h]
\centering
	\includegraphics[width=14cm]{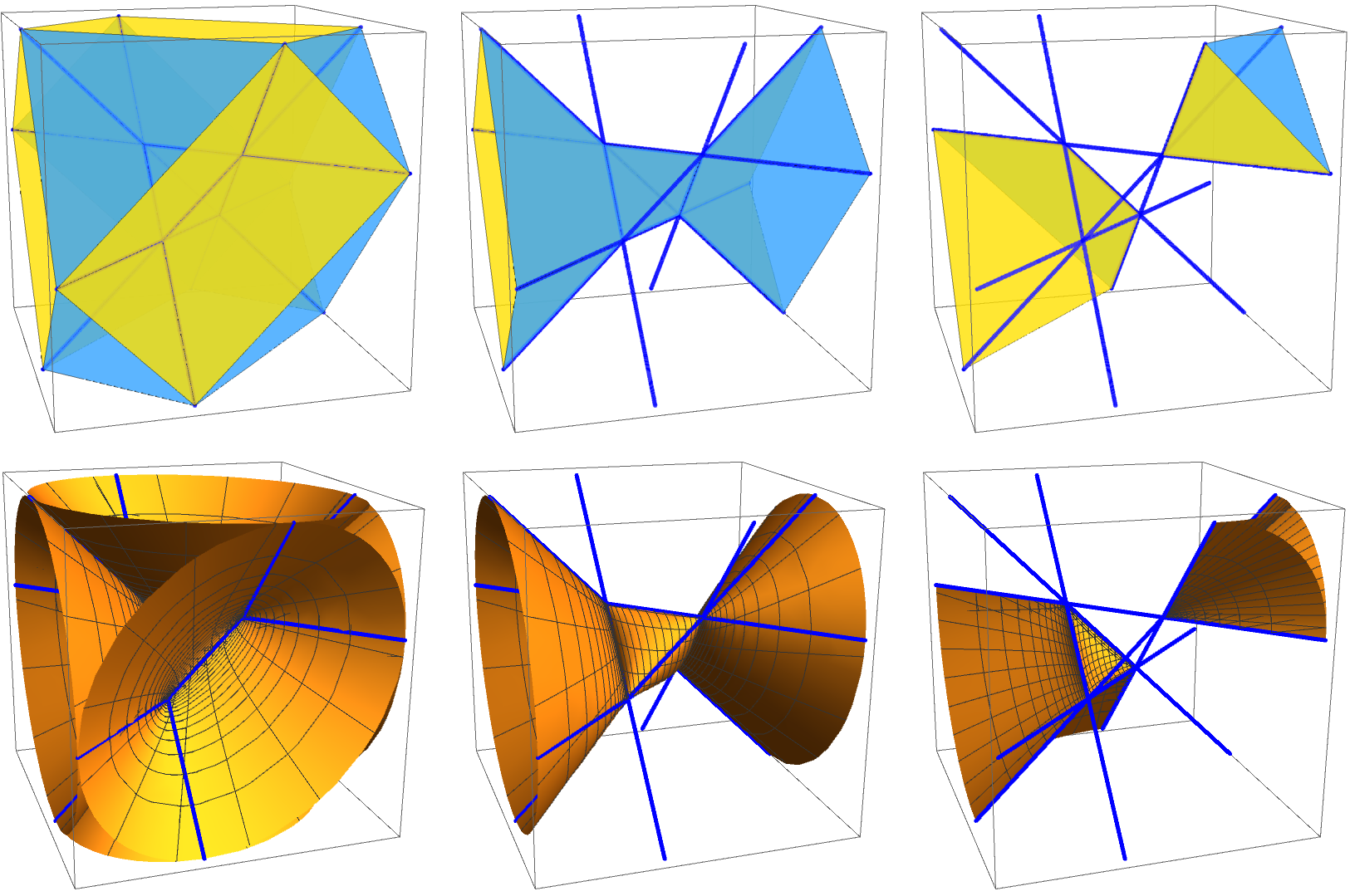}
	\caption{Stereographic projection of the boundaries of $\Cl$ (top row) and $\Qm$ (bottom row).
     The common edges are shown as blue lines. The surfaces separate $\Ns$-facets from CHSH-facets. Their coloring for $\Cl$ corresponds to \reff{fig:symAB}.
     For $\Qm$ the surfaces are the boundary sets of type \Qt3.
     Left panels: all boundaries. Center panels: The $\Ns$-face at the center, together with another $\Ns$-face. The latter is displayed in two pieces, connected via the plane at infinity.
     Right panels: a CHSH-face, resp.\ a \Qt4 curved tetrahedron. }
\label{fig:projective}
\end{figure}

In order to visualize the boundary structure we can use a {\it stereographic projection} from the origin.
Each boundary point is first mapped to the line through the origin it generates, and then
to the intersection of that line with a fixed reference hyperplane, to be identified with $\Rl^3$.
This defines a rational map from the boundary of our convex body onto $\Rl^3$.
It is $2$-to-$1$, because it identifies points $\pm c$, but this hardly matters because this is a symmetry of $\Qm$ and $\Cl$.
If the reference hyperplane supports the body, then  this will produce a local image of the boundary.
Stereographic projection takes line segments to line segments, so the boundary of a polytope like $\Cl$ induces a subdivision of $\Rl^3$ into  polyhedra. This is  shown in the top row of \reff{fig:projective}, where the reference plane has been chosen as $\{c_{11}=1\}$. Points with $c_{11}=0$ are mapped to infinity, so facets containing such points are represented in two parts, connected via some infinite points, but forming a single polyhedron in projective geometry. This partitions $\Rl^3$ into $8$ projective tetrahedra, all of which have the same four vertices. This reminds us that the ``line segment between two given points'' is intrinsically ambiguous in projective geometry. The term ``projective polyhedron'' is therefore often used for the whole tessellation, and not just for ``connected region in projective space bounded by finitely many hyperplanes'' as in this paragraph.

The resulting pictures are inside-out versions of {\em Schlegel diagrams} \cite[Chapter 5]{Ziegler}, which represent $4$-dimensional polytopes
by stereographic projection from a point just outside a reference facet. When the projection point is close enough to that facet, the whole boundary is mapped to a polyhedral subdivision of the facet, thus avoiding infinite points and the identification of antipodes. The construction easily generalizes to other convex bodies. In this generalized setting convexity still guarantees that the ray from the projection point will cut the boundary in at most two points. But it may happen that one cannot arrange for one of these to be in the plane of the reference facet. Our body $\Qm$ demonstrates this point. The only facets are $\Ns$ facets, and these are bounded by non-exposed extreme points, whose unique tangent plane is the plane of the reference facet. Therefore,  no matter how close the projection point is chosen to the reference facet, the cone generated by the facet does not contain the whole body $\Qm$. So the Schlegel map does not map $\partial\Qm$ into the facet, and  has non-trivial double images. Therefore we opted for the projection from the center, which is shown in the second row of \reff{fig:projective}.

Since the boundaries $\partial\Qm$ and $\partial\Cl$ are identified via pushout, their stereographic projections are topologically equivalent partitions. Stereographic projection suggests another identification of $\partial\Cl$ and $\partial\Qm$, namely by identifying points on the same ray through the origin. Such points are simply the same in \reff{fig:projective}, and that the stereographic projections of the boundaries are distinct shows the non-linearity of the pushout map. Indeed there are points $c$ in a CHSH face, so that the multiple $\lambda c\in\partial\Qm$ lies in an $\Ns$-face. Such $c$ are readily found already in \reff{fig:symAB}.

\subsection{Curved tetrahedra} \label{sec:Dcosines}

\reff{prop:boundary} allows us to turn an affine parametrization of a CHSH-face into a trigonometric parametrization of a curved tetrahedron \Qt4. For later purposes, we find cosines a bit more convenient than sines.

\begin{prop}\label{prop:paramExt}\proofin{sec:cos_param}
The threefolds \Qt4 of exposed extreme points on  $\Qm$ are parametrized~by
\begin{eqnarray}
  c=(c_{11},c_{12},c_{21},c_{22}) &=& (\cos\alpha,\cos\beta,\cos\gamma,\cos\delta)\qquad\mbox{where} \nonumber\\
  \alpha+\beta+\gamma+\delta &=& 0\ \mod2\pi   \qquad\mbox{and} \label{eqn:cosacosb} \\
  \Delta = \sin\alpha\cdot \sin\beta\cdot\sin\gamma\cdot\sin\delta & < & 0.
\nonumber
\end{eqnarray}
\end{prop}

The angle parameters can be taken as the triples $(\alpha,\beta,\gamma)$, with $\delta=-(\alpha+\beta+\gamma)$.
In this $3$-space, the sign of $\Delta$ marks a partition into two kinds of subsets: On the one hand,
we have the curved tetrahedra with $\Delta<0$ considered in \reff{prop:paramExt}.
A prototype which contains the CHSH-point $(\pi/4,\pi/4,\pi/4)$ is given
by the conditions $\alpha,\beta,\gamma,\alpha+\beta+\gamma \in(0,\pi)$.

\begin{figure}[h] \centering
	\includegraphics[width=6cm]{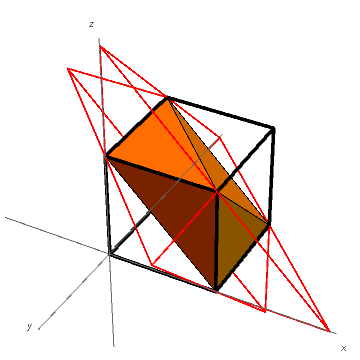} \qquad \includegraphics[width=6cm]{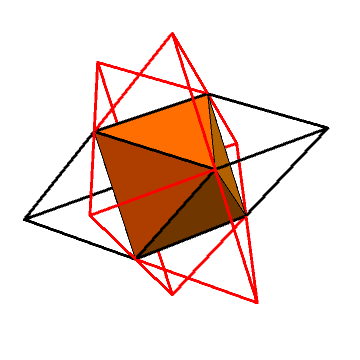}\vspace{-0.2in}
	\caption{Left: the cube $(0,\pi)^3$ containing the octahedron given by $\alpha+\beta+\gamma \in(\pi,2\pi)$, on which $\Delta>0$. The (red and black) tetrahedra built on its faces have $\Delta<0$,
    and are hence mapped to the curved tetrahedra \Qt4 according to  \reff{prop:paramExt}.
             Right: an affine transformation of the left panel making the symmetry of permuting angles and the regularity of the tetrahedra and octahedra more evident.  }
\label{fig:tetrock}
\end{figure}

On the other hand,  consider the angles with $\Delta>0$. Adding multiples of $\pi$ to any of $\alpha,\beta,\gamma$ (and hence implicitly to $\delta$) corresponds to an even sign change on the $c_{ij}$, and hence a symmetry of $\Qm$. Therefore, it suffices to consider the cube $(0,\pi)^3$. In this cube only $\sin\delta$ can be negative, so $\Delta>0$ means $\alpha+\beta+\gamma\in(\pi,2\pi)$.
This is an octahedron, as shown in \reff{fig:tetrock}. By taking cosines, these points end up in the interior of $\Qm$.

Finally, note that the angle parameters $(\alpha,\beta,\gamma)$ with $\Delta=0$ correspond to further boundary elements from \reff{prop:boundary}, as follows.
The symmetries also help to reduce each of the classes in the following boundary version of the parametrization to a single case, which is readily checked.

\begin{prop}\label{prop:paramExtbd}\proofin{sec:facerank}
In the cosine parametrization of \reff{prop:paramExt}, taking $\Delta=0$  parametrizes further boundary strata
of lower dimension.
 Specifically, we get points of type
\begin{itemize}
\item[\Qt1] if and only if $\alpha,\beta,\gamma,\delta$ are all multiples of $\pi$,
\item[\Qt2] if and only if  exactly two of these angles are multiples of $\pi$, and
\item[\Qt3] if and only if  exactly one of these angles is a multiple of $\pi$.
\end{itemize}
\end{prop}

\subsection{Semialgebraic description}\label{subsec26}

One possibility to decide quickly, for a given $c\in\Rl^4$, whether $c\in\Qm$, is to apply the inverse pushout map, and to check whether the result lies in $\Cl$.
This involves a transcendental function, and requires the checking of  $16$ linear inequalities. Here we consider an alternative,
which only requires checking the positivity of two polynomials in $c$. In other words,
we will describe our body $\Qm$ as a semialgebraic set.

There is a standard method to obtain relevant polynomials from the parametrization given in \reff{prop:paramExt}. Namely,
one represents each angle variable $\eta$ by the point $(\cos\eta,\sin\eta)$ on the unit circle, i.e., one introduces new variables $s_{ij}$ with $c_{ij}^2+s_{ij}^2=1$, and expresses the constraint on the sum of angles by trigonometrically expanding. Then one eliminates the $s_{ij}$-variables.
Computer algebra systems, such as {\tt Mathematica}
or {\tt Macaulay2} \cite{M2} handle such tasks routinely and, in this instance, in no time. The result is the identity $h(c)=0$ with $h$
the sextic polynomial in
 \eqref{polyh} below. But also the condition $\Delta<0$ has to be transcribed, for which we use the following polynomial $g$, which on \Qt4 satisfies $g(c)=\Delta$.
\begin{eqnarray}
  g(c) &=&  2-(c_{11}^2+c_{12}^2+c_{21}^2+c_{22}^2)+2 c_{11} c_{12} c_{21} c_{22}.\label{polyg}\\
  h(c) &=&  4(1-c_{11}^2)(1-c_{12}^2)(1-c_{21}^2)(1-c_{22}^2)-g(c)^2\label{polyh1}\\
       &=&  4 (c_{11} c_{22}-c_{12} c_{21}) (c_{11} c_{21} - c_{12} c_{22}) (c_{11} c_{12} - c_{21} c_{22})-\label{polyh}\\
       &&   \qquad -(c_{11} {+} c_{12} {-} c_{21} {-} c_{22}) (c_{11} {-} c_{12} {+} c_{21} {-} c_{22}) (c_{11} {-} c_{12} {-} c_{21} {+} c_{22}) (c_{11} {+} c_{12} {+} c_{21} {+} c_{22}). \nonumber
\end{eqnarray}
We wrote two formulas for $h$, because \eqref{polyh1} shows that $h$ is invariant
under the full symmetry group in \reff{sec:symm}, and \eqref{polyh} clarifies that the degree of $h$ is six and not eight, as suggested by \eqref{polyh1}. Then we have:

\begin{prop}\label{prop:semialg}\proofin{sec:parab}
A point $c$ in the cube $ \Ns$ lies in $\Qm$ if and only if
 $\, h(c)\geq0$ or $g(c)\geq0$.
\end{prop}

\begin{figure}[ht]
\centering
	\includegraphics[width=6.7cm]{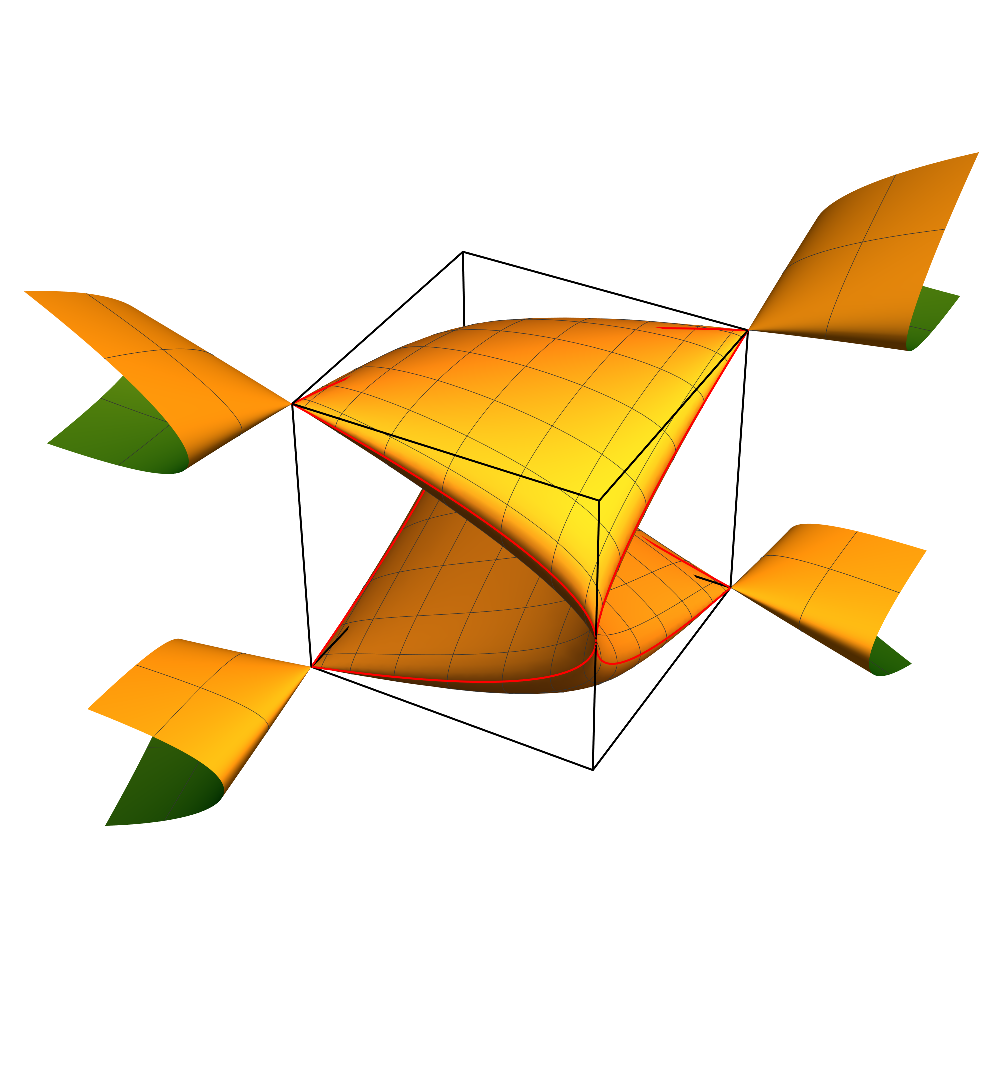}\qquad \quad
	\includegraphics[width=6.7cm]{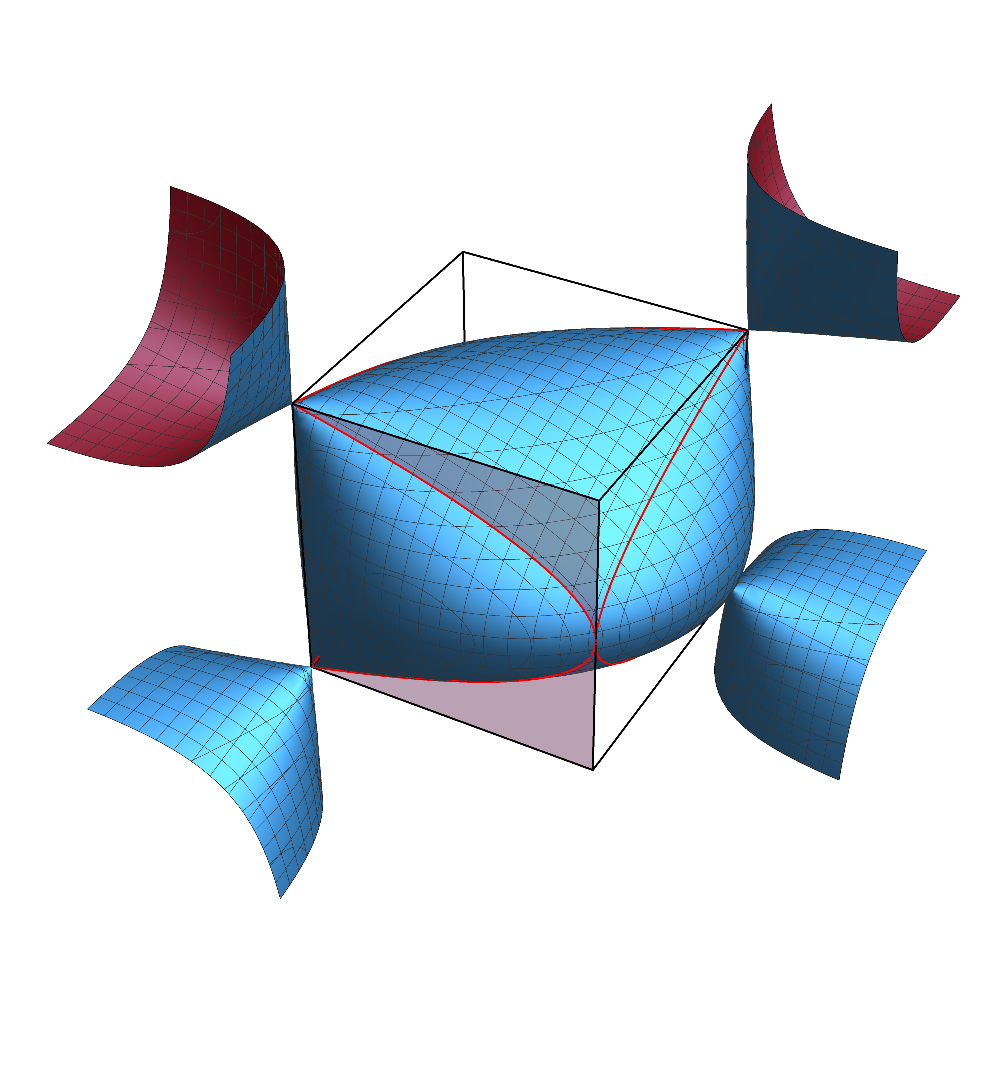} \vspace{-0.5in}
	\caption{Inside the hyperplane $\{c_{12}=c_{21}\}$,  we consider
	the two quartic surfaces $\{g=0\}$ (right)	 and
	  $\{(c_{11}-c_{22})^{-2}h=0\}$ (left).	
The two surfaces intersect transversally
in the red curves that are seen in the
boundary of the cube.}
\label{fig:Zariski}
\end{figure}

While the polynomial $h$ is an intrinsic
feature of $\Qm$, there is some freedom in the choice of $g$.
Indeed, knowing a small piece of the boundary suffices to determine $h$. This is expressed by saying that $h$,
together with the linear polynomials $1\pm c_{ij}$ describe the {\it algebraic boundary} of
$\Qm$. At each boundary point of $\Qm$ one of these polynomials vanishes.
In  algebraic geometry language  \cite[Chapter 2]{Michalek},
 the  threefold $\{h(c) = 0\}$ is the {\it Zariski closure} of the
curved tetrahedra in \Qt4, or pieces thereof. This threefold
has unbounded pieces outside the cube $\Ns$, but taking its convex hull
after the intersection with $\Ns$ gives exactly $\Qm$.
This is visualized in \reff{fig:Zariski} (left) which shows the surface $\{h(c) = 0\}$ in the $3$-space $ \{c_{12}-c_{21} = 0\}$.
The unbounded pieces arise because the algebraic elimination process works just as well in the complex domain. So the circle $c^2+s^2=1$, as a complex variety, also contains real points with imaginary $s$, corresponding to angles $\alpha=ir$ or $\alpha=\pi+ir$ with $r\in\Rl$,  and hence $\cos\alpha=\pm\cosh r$ in \reff{prop:paramExt}.

\begin{figure}[ht]\centering
	\includegraphics[width=4.5cm]{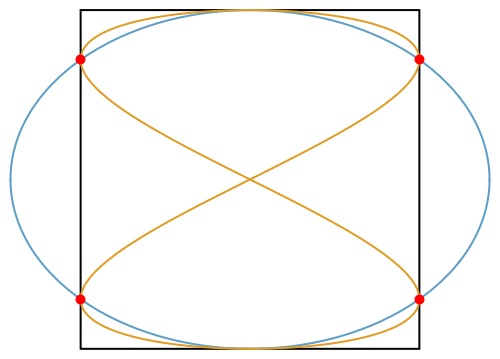}
\caption{A slice that illustrates how the two surfaces in  \reff{fig:Zariski} intersect.
The convex hull of the orange curve represents $\Qm$.
The disjunction {\em $\, h(c)\geq0$ or $g(c)\geq0$} describes  $\Qm$.
Note that $\Qm$  is not a basic semialgebraic set, i.e., not a conjunction of
polynomial inequalities.
 }
\label{fig:Zarisect}
\end{figure}

What is the role of the second polynomial $g$?  This polynomial is needed to make \reff{prop:semialg}  true.
\reff{fig:Zariski} (left) shows
 the zero set of $h$, so $h$ is negative on the outside of the yellow surface.  This extends well into the cube, where $c\in\Qm$, and even $c\in\Cl$, e.g., near
the origin. Hence,  ``$c\in\Ns$ and $h(c)\geq0$'' would produce many false negatives.
The disjunction with $g(c) \geq 0$ captures the convex hull of the
threefold inside the cube~$\Ns$.
The surface defined by $g$ in the hyperplane $\{c_{12}-c_{21} = 0\}$
is shown in \reff{fig:Zariski} (right).
 \reff{fig:Zarisect} is a two-dimensional representation that shows the geometry of the relevant intersections.

We remark that more compact forms than the characterization by two polynomials can be given, if we allow the use of absolute values or roots. Such conditions can be converted to polynomial expressions. However, they typically generate a case distinction, hence some overhead in the logical part of a semialgebraic description. For example, by taking a root in \eqref{polyh1} and combining with \reff{prop:semialg}, we can see that $c\in\Qm$ is equivalent to $c\in\Ns$ together with the single inequality
\begin{equation}\label{TimoPoly}
    g(c)\,\geq \, -2\sqrt{(1-c_{11}^2)(1-c_{12}^2)(1-c_{21}^2)(1-c_{22}^2)}.
\end{equation}
The following reformulation, due to Landau \cite{Lan88},
is not invariant under the full symmetry group of $\Qm$:
\begin{equation}\label{LandauPoly}
  \sqrt{(1-c_{11}^2)(1-c_{12}^2)}+\sqrt{(1-c_{21}^2)(1-c_{22}^2)}
   \geq \abs{c_{11}c_{12}-c_{21}c_{22}}.
\end{equation}
Note that these inequalities must be combined with the hypothesis $c\in\Ns$. This
excludes unbounded solutions with a product of two negative factors under the square root.

\subsection{Spectrahedral shadow}
\label{sec:specshadow}

For any quantum correlations,
given by a density operator $\rho$ and observables $A_1,A_2,B_1,B_2$, and complex numbers $\xi_1,\ldots,\xi_4$, we consider the operator
$X=\xi_1A_1+\xi_2A_2+ \xi_3B_1+\xi_4B_2$. Since $X^*X$ is positive semidefinite, we
conclude that $\tr(\rho X^*X) =\sum_{\nu,\mu=1}^4\overline{\xi_\nu}C_{\nu\mu}\xi_\mu\geq0$,
where we introduced the  $ 4 \times 4$  matrix
\begin{equation}\label{bigC}
    C=  (C_{\nu \mu})=
\begin{pmatrix}
                d_1&u&c_{11}&c_{12}\\
                \overline u&d_2&c_{21}&c_{22}\\
                c_{11}&c_{21}&d_3&v\\
                c_{12}&c_{22}&\overline v&d_4\end{pmatrix}.
\end{equation}
The entries other than $c_{ij}$ are
\begin{equation}\label{uvFromrho}
  u=\tr(\rho A_1A_2),   \qquad v=\tr(\rho B_1B_2), \qquad \mbox{and}\quad
d_1 =\tr(\rho A_1^2)\geq 0,\, \mbox{etc.}
\end{equation}
The positivity stated above is $C\geq0$, our notation for $C$ being positive semidefinite.
The existence of $u,v$ and $d_i$ with $d_i^2\leq1$ making $C\geq0$ is thus a necessary condition
for $c\in\Qm$. This is the bottom level of the semidefinite hierarchy \cite{navascues2008}
for quantum correlations. In the case at hand, the necessary condition is also sufficient, known since the time of Tsirelson~\cite{Tsi85}.
We can further assume that  $u$ and $v$ are real and that
the diagonal entries are all $1$.

\begin{prop}\label{prop:Cpos}\proofin{sec:Peasy}
A point $c$ lies in  the convex  body
$\Qm$ if and only if there exist numbers $u,v\in\Rl$ such that $C\geq0$
in~\eqref{bigC} with $d_1=d_2=d_3=d_4=1$.
\end{prop}

Hence, $\Qm$ is characterized by a semidefinite matrix completion problem. This is essentially the completion problem for the $4$-cycle,
as discussed in \cite[Example 12.16]{Michalek}. Our  boundary polynomial $h(c)$ is obtained from the degree eight polynomial given
there by setting the diagonal entries to be $1$.
The matrix inequality $C \geq 0$ defines a spectrahedron in $\Rl^6$, i.e., an intersection of the semidefinite matrix cone with an affine-linear space.
Deleting the matrix entries $u$ and $v$ specifies a projection $\Rl^6 \rightarrow \Rl^4$.
The correlation body $\Qm$ is the image of the spectrahedron $\{C \geq 0\}$ under this projection.
Thus, \autoref{prop:Cpos} furnishes an explicit realization of $\Qm$ as a
{\em spectrahedral shadow} \cite{Scheiderer}.

\begin{figure}[ht]\centering
  	\includegraphics[width=5cm]{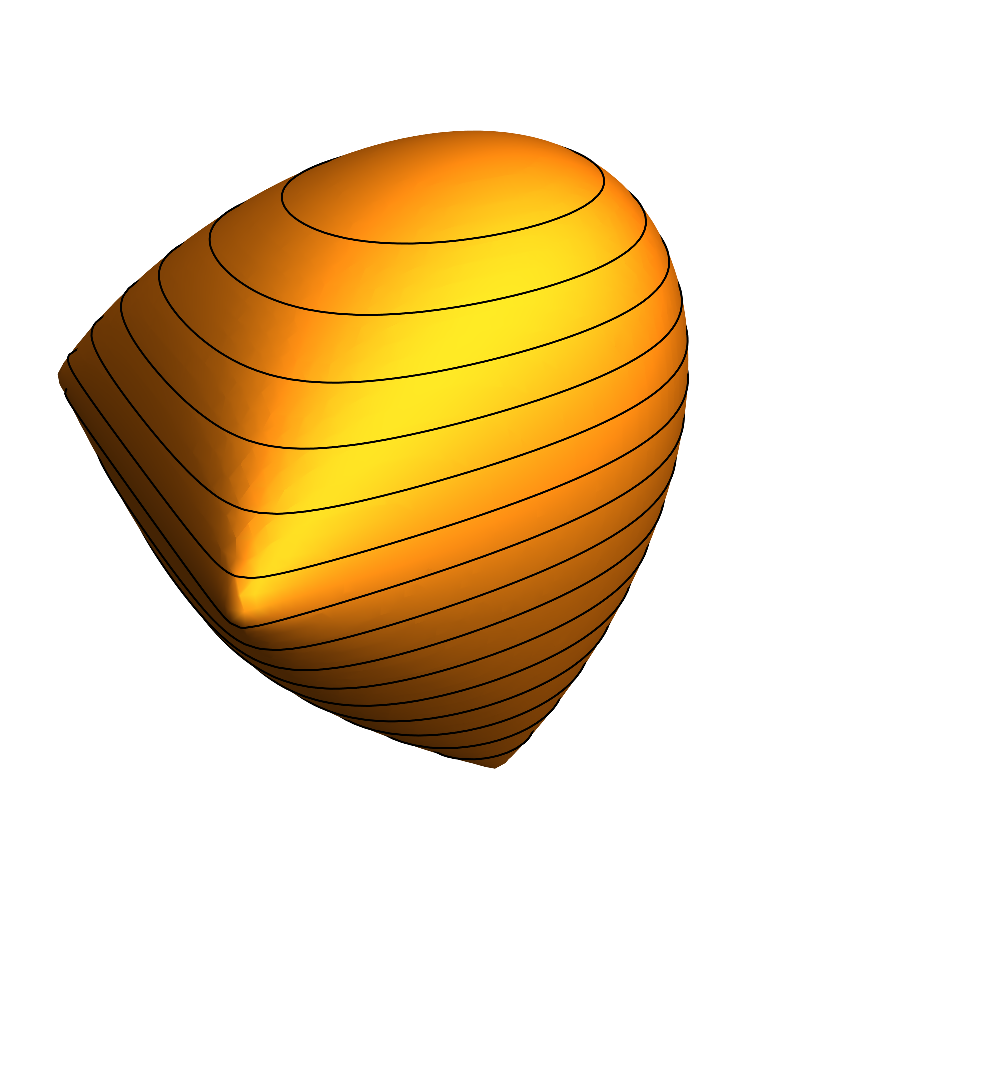} \vspace{-0.1in}
  	\caption{Nonuniqueness of matrix completion for a family of correlations $c=tc_\text{CHSH}+(1-t)c_\text{center}$ with $c_\text{center}=(-1,0,0,0)$ along the line segment from
 the center of an elliptope facet (top) to the CHSH-point (bottom).
        The horizontal cuts for fixed $t$ (black meshes) represent the pairs $(u,v)$
that make \eqref{bigC} positive semidefinite. These slices are $2$-dimensional
convex bodies that shrink to a point at both boundaries.
        The figure is a {\em quartic spectrahedron}~\cite{Vinzant}.}
          \label{fig:uvcut}
\end{figure}

A natural question to ask here is: How many completions can a correlation $c$ have? Geometrically, this corresponds to studying the inverse images (``fibers'') of the projection, that takes a $C\geq0$ with unit diagonal to its upper right $2\times2$-submatrix $c\in\Qm$. By definition, the set in the $(u,v)$-plane making $C\geq0$ for given $c$ is the intersection of an affine manifold with a positive definite cone: a spectrahedron. For example, \autoref{fig:uvcut} shows the union of these $2$-dimensional fibers over a line that cuts through $\Qm$. The following proposition establishes that the uniqueness of the completion problem (a one-point fiber) characterizes the boundary. Thus the rank of $C$ can be seen as a property of $c\in\partial\Qm$, and this turns out to distinguish different boundary elements.

\begin{prop}\label{prop:Cposb}\proofsin{\reff{lem:Qmc} and \reff{sec:facerank}}
\begin{itemize}
\item[(1)] The matrix completion problem has a unique solution $(u,v)$ if and only if
            $c\in \partial \Qm$.
\item[(2)] The resulting unique matrix $C$ has rank $1$ if $c$ is of type \Qt1, it has rank $2$ for types \Qt2, \Qt3 and \Qt4,
and it has rank $3$ for type \Qt5.
\item[(3)] The point $c$ is in the interior \Qt6 if and only if there is some completion with $\rank C=4$.
\end{itemize}
\end{prop}

We close with one remark regarding item (3). It is not true that \emph{all} the completions $C$ of a given interior point $c\in$ \Qt6 have rank $4$.
For instance, the origin $c=0\in$\Qt6 has among its completions the completion $C=\idty$ of rank 4, but also the completion $C$ with $u=v=1$ of rank 2 (direct verification).

\subsection{Volume}\label{sec:Vol}

The volume of $\Qm$ is a fundamental geometric invariant. It could be understood as the probability of quantum correlations in an ensemble, for which $c\in\Ns$ is distributed according to Lebesgue measure.
This ensemble of generalized probabilistic theories has no operational meaning, so the volume has no direct physical relevance. However, the probabilistic interpretation suggests a way to compute it stochastically: The semialgebraic description gives us a fast way to decide
whether $c\in\Qm$, for points $c$ in a random ensemble with each $c_{ij}$ independent and equidistributed
in $[-1,1]$. A run of $10^6$ samples
led to
\begin{equation}\label{volumeStoch}
  \frac{V(\Qm)}{V(\Ns)} \,\,\approx \,\, 0.925898.
\end{equation}
On account of $\sqrt N$-fluctuations, this can be expected to be accurate to within three digits.

On the other hand, we can compute the volume exactly, by integrating the pushout over $\Cl$. Since the pushout acts coordinatewise, the Jacobi matrix is diagonal and the functional determinant is readily determined. The resulting trigonometric integrals can be solved, giving the overall result
\begin{equation}\label{volume}
  \frac{V(\Qm)}{V(\Ns)} \,=\, \frac{3\pi^2}{2} \cdot \frac{1}{16}\quad \approx\,\, 0.9252754126.
\end{equation}

Similarly, the ratio $V(\Cl)/V(\Ns)$ can be estimated stochastically to be $0.666$. Using the linear transformation $2H$ in~\eqref{Hadamard}, $V(\Cl)$ is $\sqrt{\det((2H)^*(2H))}=16$ times the volume of the 4-dimensional cross-polytope, defined as the convex hull of 8 points $(\pm1,0,0,0), (0,\pm1,0,0), (0,0,\pm1,0), (0,0,0,\pm1)$ in $\Rl^4$. In general, the volume of the $n$-dimensional cross-polytope is $2^n/n!$, shown by induction from the observation that it equals $2^n$ times the volume of the standard $n$-simplex (the convex hull of the origin and standard basis of $\Rl^n$). Thus, $V(\Cl)=2^5/3$ and we get exactly $V(\Cl)/V(\Ns)=2/3$ and
\begin{equation}
  \frac{V(\Qm)}{V(\Cl)} \,=\, \frac{3\pi^2}{2} \cdot \frac{3}{2^5}\quad \approx\,\, 1.3879131189.
\end{equation}

Our results agree with that of~\cite{volumes}. Also, these volumes give the probability $\approx0.258609$ on the task of sampling nonlocal quantum correlations $c\in\Qm\setminus\Cl$, if one were to use the translationally invariant Lebegues measure to obtain a random point in $\Ns$. For this task, a better approach is clearly to change the measure to the unitarily invariant Haar measure, as studied in~\cite{sampling_nonlocal}.

For the surface area, we get the volume of the $\Ns$-faces as the well-known elliptope volume $\pi^2/2$ \cite{volumeElliptope}. For the volume of the curved tetrahedra (that is the one of the elliptope face, the pushout of the three dimensional tetrahedron) we did not find a closed expression. Testing the numerical value for being a simple fraction times a low power of $\pi$ by continued fraction expansion also did not seem to give a simple expression. The \Qt3 boundaries are directly expressed by the
surface area of the elliptope. This area is known to be $5\pi$, by a direct calculation found on
{\tt math.stackexchange.com}.

\section{Description of the Dual Body}\label{sec:dualscribe}
The polar $\KK\polar=\{f\stt f{\cdot}c\leq1,\ c\in\KK\}$ of a convex body $\KK$ provides the
description of $\KK$ by the affine inequalities it satisfies. Since $\KK\bipolar=\KK$, by \cite[Thm.\,IV.1.5]{Schaefer}, this is a symmetric relation. We could ask all the questions  we studied so far about
the three bodies $\Cl\subset\Qm\subset\Ns$ also about their polars.
 For the polars the inclusion is reversed, so $\Ns\polar\subset\Qm\polar\subset\Cl\polar$. The big surprise, which is a rather special feature of the minimal case, is that this dual chain of inclusions is essentially
{\it the same}  as the original chain. We will spell this out in detail.
For now it just means the good news that much of the work is already done.

\subsection{The duality transform}\label{sec:Hadamard}

We first noticed the duality in the polytopes $\Cl\subset\Ns$ from the observation that their face
counts by dimension (the f-vectors) are reversals of each other. Strengthening this
to an isomorphism $\Cl\cong\Ns\polar$ goes as follows.
We start from the inequalities describing $\Ns$.
These come from the  $8$ vertices of $\Ns\polar$, which are
\begin{equation}\label{Nspextreme}
  \pm(1,0,0,0),\quad \pm(0,1,0,0),\quad \pm(0,0,1,0),\quad \pm(0,0,0,1).
\end{equation}
These have to be identified with the vertices of $\Cl$, i.e., the $8$ even vertices of $\Ns$ itself.
They are
\begin{equation}\label{Nspextreme2}
  \pm(1,1,1,1),\quad \pm(1,-1,1,-1),\quad \pm(1,1,-1,-1),\quad \pm(1,-1,-1,1).
\end{equation}
The following transformation $H$ maps the points in \eqref{Nspextreme} to
those in \eqref{Nspextreme2}. So we get $\Cl=2H\Ns\polar$ with
\begin{equation}\label{Hadamard}
\qquad  H \,=\frac12 \begin{small} \begin{pmatrix}
\, 1 & \phantom{-}1 &  \phantom{-}1  &  \phantom{-}1 \, \\
\, 1 &  -1 & \phantom{-}1 & - 1 \, \\
\, 1 & \phantom{-}1 &   -1 &  -1  \, \\
\, 1 & -1 & -1 &  \phantom{-}1 \, \end{pmatrix}\end{small}.
\end{equation}
Here we included the factor $1/2$ so that $H^2=\idty$. Since $H^*=H$, this matrix is then also unitary, i.e., a Hadamard matrix \cite{KarolHadamard} (note different conventions for the normalization of such matrices though). Recall two basic properties of polar duality: $(s\mathcal{K})\polar=\mathcal{K}\polar/s$ for $s\neq0$ and $(U\mathcal{K})\polar=U\mathcal{K}\polar$ for unitary $U$. This makes it easy to write down
the consequences of $\Cl=2H\Ns\polar$ from dualization or multiplication with $H$:
\begin{equation}\label{polytopeDuals}
  \textstyle \Ns\polar=\frac12H\Cl\qquad\mbox{and }\qquad \Cl\polar=\frac12H\Ns.
\end{equation}

\begin{figure}[h]\centering
	\includegraphics[width=12cm]{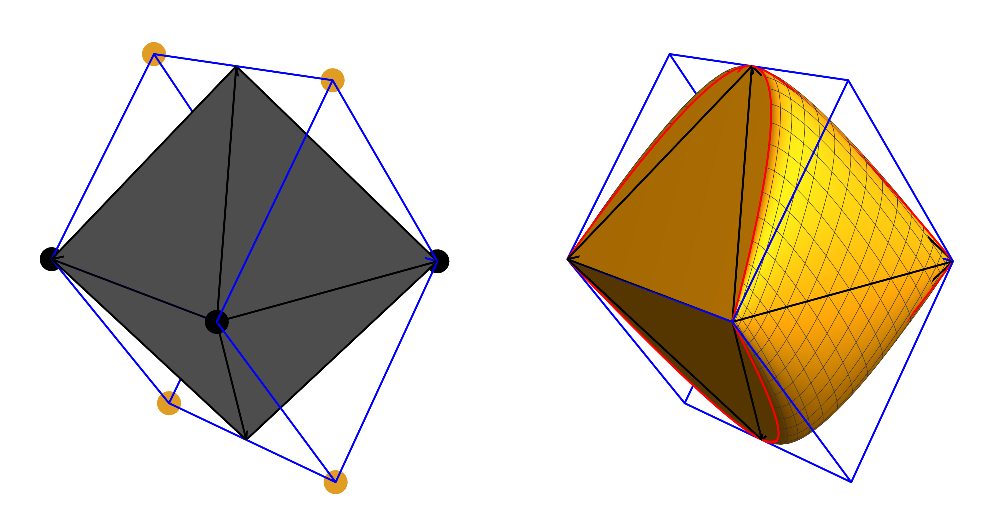} \vspace{-0.2in}
	\caption{Polar duals for the Alice-Bob symmetric sections shown in \reff{fig:symAB}. The marked vertices of $\Cl\polar$ (blue cube frame) correspond to the facets of $\Cl$ in \reff{fig:symAB}.
    The figure of $\Qm\polar$ (right panel) is an affine transformation of $\Qm$ in \reff{fig:symAB}.
    This expresses the self-duality.}
\label{fig:ABpolar}
\end{figure}

To visualize this duality, consider the left panel of \reff{fig:ABpolar}.
Since the section with the hyperplane $\{c_{12}=c_{21}\}$ is also a projection, and duality
swaps projection and intersection operations, this $3$-dimensional picture faithfully represents
the $4$-dimensional polarity relations. The outer blue frame represents $\Cl\polar$, the dual of the inner frame in \reff{fig:symAB} (Left). It is a cube whose vertices correspond to the facets of $\Cl$. They can thus correspond either to a CHSH-facet (marked in yellow) or to a  $\Ns$-facet (marked in black). Overall, the figure looks like a rotation of its counterpart in \reff{fig:symAB}. In $4$ dimensions the required $H$ is exactly an orthogonal reflection. In the $3$-dimensional section it is still a submatrix of a rotation, and looks like a rotation because human 3D perception is highly capable of ignoring uneven affine stretching.

The right panel in \reff{fig:ABpolar} shows the pertinent section of
the dual body $\Qm\polar$. We can draw it using the parametrized points $f$
from \reff{prop:paramExtf}. This mimics the right panel of \reff{fig:symAB}, by the same transformation as the one used for the polytopes. Indeed, this is true exactly, also for the full $4$-dimensional body:

\begin{prop}\label{prop:duality}\proofin{sec:self-dual}
 $\Qm\polar=\frac12H\Qm$. \end{prop}

This is \reff{thm:selfdual}, but with the matrix $H$ spelled out.
The linear transformation $H$ is far from unique. Indeed, if $S_1,S_2$ are matrices representing any of the symmetries from \reff{sec:symm}, then so are their transposes. Hence $H'=S_1HS_2$, incidentally always again a Hadamard matrix,  also maps the inclusion chain to its dual. Note that  $S\mapsto S'=HSH^{-1}$ is an automorphism of the symmetry group which changes the semidirect product decomposition, so multiplication by an even number of signs can become a permutation and conversely. In fact, $H$ is just the Fourier transform in the discrete phase space representation mentioned in \reff{sec:symm}, and in this picture it is the swap between position shifts and momentum shifts.

In the terminology of axiomatic quantum mechanics,
 this form of self-duality of a convex set is called ``weak self-duality'' \cite{Howard}, as opposed to stronger forms with a canonical isomorphism between the set and its dual, that is characteristic for Jordan algebra state spaces and in particular the quantum state space.
Can one strengthen \reff{prop:duality} in that direction?
Indeed, we already have a canonical mapping taking each point $c$ of a \Qt4 patch to the unique maximizer $f$, via \reff{prop:paramExt} and \reff{prop:paramExtf}. By multiplying $H$ with a symmetry
of $\Qm$ we can achieve that it takes the patch of $c$ to the patch of $f$, but can we do it for all patches simultaneously? The answer is no: A map with that property would have to commute with all symmetries.
Since our representation of the group $G$ on $\Rl^4$ is irreducible, we would conclude that $H$ is a multiple of the identity.
Another context of interest are normed spaces: Since $-\Qm=\Qm$, our convex body $\Qm$ is the unit ball of a norm in $\Rl^4$ and the dual normed space has unit ball $\Qm\polar$. So we have an example of a normed space that is isomorphic to its dual, a subject studied more generally, e.g., in \cite{yannakakis2010}.

After the completion of this work, we became aware of T. Fritz's results~\cite{N22duality} which extend the duality of $\Cl$ and $\Ns$ to many parties and shows the impossibility of this duality for $M,K>2$, i.e. scenarios with more than two measurements or two outcomes. There it was stated that the self-duality of $\Qm$ was also known to C. Palazuelos and I. Villanueva (in a private communication), and that self-duality fails for the analogously defined three-partite body, but leaving the possibility for $N>3$.

\subsection{Parametrized extreme points of $\Qm\polar$}
The boundary patches \Qt4 of exposed extreme points have a unique maximizing functional $f\in\Qm\polar$. The parametrization by angles can be taken over directly.

\begin{prop}\label{prop:paramExtf}\proofin{sec:fnondeg}
Let $\alpha,\beta,\gamma,\delta$ and $c=(\cos\alpha,\cos\beta,\cos\gamma,\cos\delta)$ satisfy the conditions of \reff{prop:paramExt}.
Define
\begin{equation}\label{exposingf}
  f\,=\,(f_{11},f_{12},f_{21},f_{22})
 \,=\, \frac1K\left(\frac1{\sin\alpha},\frac1{\sin\beta},\frac1{\sin\gamma},\frac1{\sin\delta}\right),
\end{equation}
where $K= \cot\alpha+\cot\beta+\cot\gamma+\cot\delta$.
Then $f \cdot c'\leq1$ for all $c'\in\Qm$ with equality if and only if $c'=c$.
Moreover, the vector $f$ is uniquely determined by this property.
\end{prop}

Applying the duality transform to the point $f$ in \eqref{exposingf}  gives again a point $2Hf\in\Qm$ of type \Qt4, which in turn can be parametrized by angles.
The resulting map $\Phi$ from one tetrahedron of angles to another expresses the duality of boundary points.

\begin{figure}[h]\centering
	\includegraphics[width=13cm]{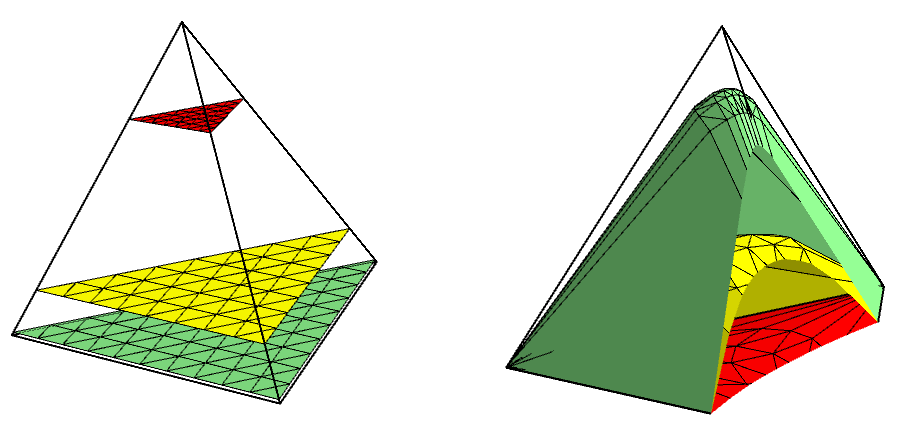} \vspace{-0.1in}
	\caption{The map $\Phi$ from the tetrahedron $T$ to itself:
		 The triangular meshes on the left are mapped to the surfaces on the right.
         The figure on the right has been truncated so that the image surfaces can be seen.
         Since $\Phi$ commutes with vertex permutations, analogous surfaces can be drawn between any tetrahedron face and the opposing vertex.
         Since $\Phi^2$ is the identity, the diagrams can also be read from right to left. }
\label{fig:T2T}
\end{figure}

For a concrete computation let $T$ be the tetrahedron defined by $\alpha,\beta,\gamma,-\delta=\alpha+\beta+\gamma\in(0,\pi)$.
Denote $\theta=(\alpha,\beta,\gamma,\delta)$ and write $c(\theta)$ and $f(\theta)$ for the images in
\reff{prop:paramExt} and  \reff{prop:paramExtf}. We obtain a self-map $\Phi:T\to T$ with the property
\begin{equation}\label{PhiEq}
  2H\,f(\theta)=c(\Phi(\theta)).
\end{equation}
By definition, this map will commute with the permutations of vertices of $T$, which extend to symmetries of $\Qm$. By self-duality it is also its own inverse.  This map is visualized in \reff{fig:T2T}.

By symmetry, these properties uniquely fix the map also for other \Qt4-patches. When doing this concretely, one has to observe that whereas the association of $f$ with the unique maximizer $c$ is a property of $\Qm$, the concrete parametrization of the tetrahedra involves a convention, and depends on the choice of self-duality operator $H$. Therefore in solving \eqref{PhiEq} for $\Phi(\theta)$, on any of the tetrahedra in \reff{fig:tetrock}, one has to carefully pick the branches of the $\arccos$ function.

The map $\Phi$ is continuous and even analytic on the open tetrahedron. But it does not have a continuous extension to the closure of the tetrahedron. Indeed, a glance at \eqref{exposingf} shows that when one of the angles in $\theta$ goes to zero, and the others to values not in $\{0,\pi\}$, the image $\Phi(\theta)$ approaches the opposite vertex. Hence the whole open part of the bottom face on the left goes to a single point. This is evident from \reff{fig:T2T} in the form that most of the triangles in the evenly spaced triangulation close to the base triangle end up close to the top vertex. Similarly, when $\theta$ approaches a point on the edge, the limit of $\Phi(\theta)$ depends on the direction from which the edge is approached.

\subsection{Dualized descriptions}
We can now apply the duality transform to each of the previous subsections to get analogues for $\Qm\polar$ of all the statements made about $\Qm$. There is no simple analogue of the pushout. The cosine parametrization of the curved tetrahedra was already given an analogue in \reff{prop:paramExtf} (not via duality transform).

Consider the polynomials \eqref{polyg} and \eqref{polyh} of the semialgebraic description. Since $f=(f_{11},f_{12},f_{21},f_{22})\in\Qm\polar$ is equivalent to $2Hf\in\Qm$, we need to consider these polynomials after substituting $c\mapsto 2Hf$. Note that such a substitution takes polynomials which are invariant under the symmetry group to polynomials with the same property. Moreover, by linearity of the substitution, homogeneous polynomials of some degree go to homogeneous polynomials of the same degree. This constrains the number of polynomials we need to consider. The invariant polynomials of degree two are proportional to $\abs c^2=c_{11}^2+c_{12}^2+c_{21}^2+c_{22}^2$, and since $H$ is orthogonal, this goes to itself under substitution. Among the quartics we need to consider the product $p(c)=c_{11}c_{12}c_{21}c_{22}$ that already appeared in \reff{thm:supp}. Under substitution it becomes
\begin{equation}\label{polyq}
  q(f)=p(2 H f)=(f_{11}+f_{12}+f_{21}+f_{22})(f_{11}-f_{12}+f_{21}-f_{22})(f_{11}+f_{12}-f_{21}-f_{22})(f_{11}-f_{12}-f_{21}+f_{22}).
\end{equation}
This is the quartic part in  \eqref{polyh}. The sextic part of $h$ is the polynomial $k$
from \reff{thm:supp}. This is self-dual in the sense that $k(2Hf)=64 k(f)$.
With these building blocks, the polynomials describing $\Qm$ and $\Qm\polar$~are
\begin{equation}\label{polypolar}
\begin{array}{rlrll}
   h(c)&=4k(c)-q(c),     &\hskip60pt  h\polar(f) &:= \frac1{256}\,h(2Hf)&=
k(f)-p(f),    \\[10pt]
   g(c)&=2-\abs c^2 +2 p(c),  &     g\polar(f) &:= \,\frac12\,g(2Hf)&= 1-2 \abs f^2 +q(f).
   \end{array}
\end{equation}
We conclude that $\,f\in\Qm\polar \,$ if and only if $\,2Hf\in\Ns\,$ and
(\,$g\polar(f)\geq0\,$ or $\,h\polar(f)\geq0$\,).

Describing $\Qm\polar$ as a spectrahedral shadow is done in general by the dual semidefinite
hierarchy \cite{doherty2008}. Indeed, the dual of a spectrahedral shadow is again of this form
\cite[Remark 5.43]{Blekherman}. This holds because the operations
of linear section and projection change roles under dualization, and the semidefinite cone is anyhow self-dual. In terms of matrix completion problems, where the primal matrix has unspecified entries (a subspace constraint), the dual matrix has zeros (resulting from the dual projection).
In the case at hand, the unspecified entries $u,v$ in $C$ become zeros in the dual matrix, which we call
$F$. The condition that the diagonal entries of $C$ are $1$ dualizes to
constraint on the trace of $F$. This gives:

\begin{prop}\label{prop:dualMatComp}\proofin{sec:self-dual}
A point $f=(f_{11},f_{12},f_{21},f_{22})$ lies in the
dual body $\Qm\polar$ if and only if there exist nonnegative
real numbers $p_1,p_2,p_3,p_4 $ with $\sum_{i=1}^4 p_i=2$ such that
\begin{equation}\label{Fmatrix}
  F=\begin{pmatrix}
      p_1 & 0 & -f_{11} & -f_{12} \\
      0 & p_2 & -f_{21} & -f_{22} \\
      -f_{11} & -f_{21} & p_3 & 0 \\
      -f_{12} & -f_{22} & 0 & p_4
    \end{pmatrix} \, \geq \, 0.
\end{equation}
Moreover,  there exists a completion satisfying the above constraints, and additionally $p_1+p_2=p_3+p_4=1$. This condition is automatically satisfied for all boundary points.
\end{prop}

Indeed, one checks that $\,\tr\, CF= 2-2 f\cdot c \,$ holds for $C$ from \eqref{bigC},
so $\tr \,CF\geq 0$ is equivalent to~$f\cdot c\leq1$.

\subsection{The normal cycle}\label{sec:convNC}

The definition of the polar suggests to look at the {\it incidence relation} between points $c\in\Qm$ and $f\in\Qm\polar$ for which the inequality $c\cdot f\leq1$ is tight. For a fixed $f$, this naturally answers the question about what correlation maximizes this (Bell) inequality, and for a fixed $c$, the question which (Bell) inequality contain this as a maximizer. The resulting set is called the {\em normal cycle} \cite{ciripoi2018computing,Plaumann}.

Given any  convex body $\KK$ in $\Rl^d$, which contains the origin in its interior, the normal cycle is defined~as
\begin{equation}\label{ncyle}
  \ncyc(\KK)  \,=\,  \bigl\{\,(c,f) \in \partial \KK \times \partial \KK\polar \stt c\cdot f = 1 \,\bigr\}.
\end{equation}
This integrates $\KK$ and $\KK\polar$ into a single structure, but it is useful far beyond this role. Importantly for our project, it gives a unified description of curved manifolds and polyhedral kinks, both of which are features of the correlation body $\Qm$.
It appears that this structure is not so well known in the quantum community. Therefore we begin with a brief description for an arbitrary $\KK$
before specializing to $\Qm$.

Fixing $c$ to range over a subset $A\subset \partial\KK$ leaves $f$ to range over a closed face of $\KK\polar$. We denote by $A^\dufa=\{f\,\in\KK\polar\stt \,\forall c\in A\,: \,c{\cdot}f=1\,\}$ the set incident to $A$. We use the orthogonality symbol, and call $A^\dufa$ the face orthogonal to $A$, although the relation is not given by the vanishing of a scalar product,
unless we work in projective geometry, with homogenous coordinates and cones instead of convex sets (see \cite{Plaumann}). The relation behaves in
many ways like an orthogonality relation: Since $A\subset B$ implies $B^\dufa\subset A^\dufa$, and $A\subset B^\dufa\,\Leftrightarrow\,B\subset A^\dufa$, the map $A\mapsto (A^\dufa)^\dufa$ is a closure operation, taking $A$ to the smallest face of the form $B^\dufa$ containing it.
This implies $A^{\dufa\dufa\dufa}=A^\dufa$. An {\it exposed face} is by definition a set of the
form~$\{x\}^\dufa$.

Let us look at the two extreme cases of convex sets: First, we have strictly convex sets with smooth boundary whose only faces are exposed singletons $\{c\}$.
Here, $\ncyc(\KK)$ is the graph of a homeomorphic identification of $\partial\KK$ with
$\partial\KK\polar$. At each point we can introduce coordinates so that $\partial\KK$ is the graph of a convex function. To second order it is given near $c$ by a quadratic form, containing the curvature information. Dually, $\partial\KK\polar$ is given near $f$ by the Legendre transform, hence the inverse quadratic form. Physicists are familiar with this reciprocity from the kinetic energy forms of Lagrange and Hamilton mechanics. In the limit, where curvature goes to zero in some directions, the dual curvature form diverges, corresponding to kinks of the dual manifold. Carrying this further, one gets to a situation where all curvature is concentrated on lower dimensional manifolds. This happens for polytopes.

For a polytope $\KK$, the lattice of closed faces is neatly ordered by dimension,
and $A\mapsto A^\dufa$ is an order reversing bijection of face lattices. Although the normal cycle is far from being a convex set,
given that $c\cdot f=1$ is nonlinear,
for every $k$-dimensional face $F$,  the set
 $F\times F^\dufa \subset \ncyc(\KK)$ is a polytope of dimension $k+(d-k-1)=d-1$.
Consider the  $4$-cube  $ \Ns$
and the demicube $ \Cl=\Ns\polar$. Their common normal cycle
$\ncyc(\Ns) = \ncyc(\Cl) $ consists of $80$ strata,
one for each of the $8+24+32+16$ pairs $(F,F^\dufa)$,
where $F \subset \partial \Cl$ and $F^\dufa \subset \partial \Ns$ are faces.
For instance, if ${\rm dim}(F) = 2$ then ${\rm dim}(F^\dufa)
 = 1$ and $F \times F^\dufa$ is a  triangular prism.

Let us clarify the word \emph{strata}. Consider an arbitrary semialgebraic set $S$. As a consequence of the Cylindrical Algebraic Decomposition (see, e.g., \cite[Theorem 2.3.6]{bcr:realag}), the set $S$ can be decomposed as a finite disjoint union of strata $S = \cup_{i=1}^N C_i$ with $\dim C_i = d_i$.
Each stratum $C_i$ is semialgebraically homeomorphic to $(0,1)^{d_i}$. Moreover, the closure of $C_i$ in $S$ is given by the union of $C_i$ with some other $C_j$ such that $d_j<d_i$.
Note that the decomposition of $S$ into these strata is far from unique.

The power of the normal cycle lies in the uniform treatment covering the whole range of convex bodies from polytopes to smooth. Indeed, for any compact convex $d$-dimensional set $\KK$ containing the origin in its interior, $\ncyc(\KK)$ is always a $(d-1)$-dimensional (Lipschitz Legendrian) submanifold of $\Rl^{2d}$ \cite{fu2014algebraic}. Moreover, the map $\KK\mapsto\ncyc(\KK)$ is continuous in the Hausdorff metric for sets. This makes the normal cycle a remarkably stable structure under approximations either by smooth manifolds or by polytopes.

The normal cycle plays the role of the normal bundle for more general geometric objects.
It was defined by Federer \cite{federer1959curvature} for sets of positive reach.
These include convex bodies but also much crazier sets. It is an important tool from geometric measure theory, used for defining curvature
measures \cite{wintgen1982normal, zahle1986integral}. This is related to the classic result of Steiner, who noted that the volume of the smooth approximation to a body, which one gets as the union of all balls of small radius $r$ centered on the body, is a polynomial in $r$, whose coefficients relate to curvature of different dimensionality. On the other hand, stable polyhedral approximations are needed  in visualization, computer graphics and computational anatomy (see \cite{cohen2003restricted, roussillon2017surface, su2019curvature}).
Recently, the normal cycle has also emerged as a key player in convex algebraic geometry \cite{ciripoi2018computing, Plaumann}.

If a convex body $\KK \subset \Rl^d$ is semialgebraic, then its normal cycle $\ncyc(\KK)$ is semialgebraic
as well, by \cite[Theorem 1.7]{Plaumann}. Moreover, $\ncyc(\KK)$
 admits a finite semialgebraic stratification whose top-dimensional strata are
$(d-1)$-dimensional.
To recognize the nonlinear nature of the normal cycle, we also introduce the
{\em algebraic normal cycle} $\overline{\ncyc(\KK)}$.
This is, by definition, the Zariski closure~\cite{CLO} of the normal cycle. It is a
 $(d-1)$-dimensional subvariety~\cite{CLO} of the complex space $\Cx^{2d}$, modeled by conormal varieties as in \cite[Section~2]{Plaumann}.
The radical ideal~\cite{CLO} of the algebraic normal cycle is the intersection of the
prime ideals~\cite{CLO} associated to the various strata. If $\KK$ is a polytope
then each such prime ideal is generated by the linear polynomials
in $c$ that vanish on $F$, the linear polynomials in $f$ that vanish on $F^\dufa$,
plus one bilinear polynomial $c \cdot f - 1$.

Our goal is to describe the normal cycle $\ncyc(\Qm)$ of the correlation body $\Qm$. We describe a stratification of $\ncyc(\Qm)$ that mirrors the stratification of $\Qm$ in \reff{prop:boundary}.
The symbol \Qb nm\  will represent the set of points $(c,f)$ in $\ncyc(\Qm)$ such that $c\in$\Qt n and $f\in$\Qt m, for the families from \reff{prop:boundary}.

\begin{prop}\label{prop:normalcycle}
The normal cycle $\ncyc(\Qm)$ is
divided into $24=8+8+8$ strata of dimension~$3$ whose~types are \Qb44, \Qb51, \Qb15.
These threefolds are separated by $88=32+32+24$ surfaces  \Qb31, \Qb13, \Qb22.
The eight strata of type \Qb44 belong to the same irreducible component
of the algebraic normal cycle $\overline{\ncyc(\Qm)}$, so
 the radical ideal of $\overline{\ncyc(\Qm)}$ is the intersection of $17 = 1+8+8$ prime~ideals.
\end{prop}

All these features are made more explicit in \reff{table:normalcycle}.
The labels refer to the strata of $\partial \Qm$ (cf.~\reff{prop:boundary}).

\begin{table}[h]
\centering
\begin{tabular}{|r|ccccc|}\hline
  $c$ is a point of type               & \Qt4  &  \Qt1    & \Qt5  & \Qt2   & \Qt3     \\\hline
  $\{c\}^\dufa$ is a set of type        & $\{f\}$, $f\in$\Qt4  &  \Qtc5   & \Qt1    & \Qtc2  & \Qt1  \\
  $\{c\}^{\dufa\dufa}$ is a set of type & $\{c\}$ &  $\{c\}$  & \Qtc5  & \Qtc2  & \Qtc5  \\
   $\dim\{c\}^{\dufa\dufa}$          &  0    &    0    &  3     & 1     &   3       \\
    \hline
  $\dim$ of face generated by $c$           &  0    &    0  &  3     & 1     &   0       \\
  $\dim\{c\}^\dufa$                     &  0    &    3     &  0      & 1     &  0     \\
  $\dim$ of manifold of such faces      & 3     &   0      &  0      & 0     &   2       \\
  sum of these dimensions                & 3     &   3      &  3      & 2     &   2    \\\hline
  number of open strata in $\ncyc(\Qm)$                 & 8     &   8      &  8    & 24    &   32     \\\hline
  irreducible components in $\Bigl.\Bigr.\overline{\ncyc(\Qm)}$ & 1     &   8     &  8    & 0   &   0    \\\hline
\end{tabular}
\caption{
Classification of boundary points
$c\in\partial\Qm$, in the notation of \reff{prop:boundary}, and what they correspond to in the
normal cycle $\ncyc(\Qm)$.
The first three columns give the full-dimensional strata
of $\ncyc(\Qm)$. The last two columns are two of the three strata of dimension $2$.
The fact that the stratum \Qb13 does not appear here reflects the non-exposed nature of \Qt3.}
\label{table:normalcycle}
\end{table}

The extreme points \Qt4 and their duals are both zero dimensional, hence maximally violate the typical counts for dual faces of polytopes. This is made up for by having a manifold of such points, so the local dimension of $\ncyc(\Qm)$ is still $d-1=3$. It is also clear, from the pushout transformation, that a tradeoff between manifold dimension and the dimensions of individual face pairs must be possible.

We now briefly discuss the strata \Qt2 and \Qt3 in the fourth and fifth column of \reff{table:normalcycle}. These~are surfaces  in the threefold $\ncyc(\Qm)$. Points $c$ on the surfaces \Qt3 are supported only by the linear functional that supports the entire elliptope, which is $\{c\}^{\dufa\dufa}$.
Hence there are $32$ surfaces \Qb31 in $\ncyc(\Qm)$. Dually, we also get $32$ surfaces of type \Qb13. These do not appear in \reff{table:normalcycle} because the faces of type \Qt3 are not exposed and thus cannot be realized as $\{f\}^\dufa$. This feature is
highlighted by the jump
in the dimension of $\{c\}^{\dufa\dufa}$ for $c \in$\Qt3. All $64$ surfaces arise as the intersection of the closure of \Qb44 with the closure of a stratum \Qb51 or \Qb15. Each intersection produces four such surfaces.

The exposed edges \Qt2 of $\Qm$ are supported by one-dimensional families of normal directions. These are the exposed edges \Qt2 on $\Qm\polar$. We therefore have $24$ squares of type  \Qb22 in  $\ncyc(\Qm)$. Each square separates two $3$-dimensional strata of type \Qb51 and \Qb15. In addition to the $64$ curved triangles of type \Qb31 or \Qb13, this accounts for all $2$-dimensional cells in our stratification of $\ncyc(\Qm)$.

The $3$-dimensional strata of $\ncyc(\Qm)$ give the irreducible components of the
algebraic normal cycle $\overline{\ncyc(\Qm)}$.
 We begin with the most nonlinear stratum, denoted \Qb44.
This stratum is characterized by \reff{prop:paramExt} and \reff{prop:paramExtf}.
 It is parametrized by angles $\alpha,\beta,\gamma,\delta$ that
add up to $0$ modulo $2\pi$, and it consists of pairs $(c,f)$ where $c$
satisfies \eqref{eqn:cosacosb} and $f$ satisfies \eqref{exposingf}.
Its  Zariski closure is an irreducible threefold in $\Cx^8$.
Its prime ideal is generated by $17$ polynomials.
The first three of these $17$ generators are familiar:
$$ \begin{matrix} \ell & = & c_{11} f_{11} + c_{12} f_{12} + c_{21} f_{21} + c_{22} f_{22} - 1 ,
\qquad \qquad \qquad \qquad \qquad \qquad \qquad \quad  \\
 h &= &  4 (c_{11} c_{22}-c_{12} c_{21}) (c_{11} c_{21} - c_{12} c_{22}) (c_{11} c_{12}  - c_{21} c_{22})
 \qquad \qquad  \qquad \qquad  \qquad  \\ & & \qquad \qquad
       -(c_{11} {+} c_{12} {-} c_{21} {-} c_{22}) (c_{11} {-} c_{12} {+} c_{21} {-} c_{22}) (c_{11} {-} c_{12} {-} c_{21} {+} c_{22}) (c_{11} {+} c_{12} {+} c_{21} {+} c_{22}) , \\
  h\polar & = & f_{11}f_{12}f_{21}f_{22} \,-\,(f_{11} f_{22}{-}f_{12} f_{21})
 (f_{11} f_{21}{-}f_{12} f_{22})(f_{11} f_{12}{-}f_{21}f_{22}) .\qquad \qquad
       \end{matrix} $$
The remaining $14$ generators of our prime ideal are the following polynomials:
       $$ \begin{small} \begin{matrix}
   c_{11}^2 f_{11}^2-c_{22}^2 f_{22}^2-f_{11}^2+f_{22}^2, \\
      c_{21} f_{11} f_{12} f_{21}+c_{22} f_{11} f_{12} f_{22}+c_{11} f_{11} f_{21} f_{22}+c_{12} f_{12} f_{21} f_{22}, \\
      c_{11}^2 f_{11} f_{12}-c_{21}^2 f_{21} f_{22}-c_{12} c_{21} f_{12} f_{22}+c_{11} c_{22} f_{12} f_{22}-f_{11} f_{12}+f_{21} f_{22}, \\
      c_{11}^2 f_{11} f_{21}-c_{12}^2 f_{12} f_{22}-c_{12} c_{21} f_{21} f_{22}+c_{11} c_{22} f_{21} f_{22}-f_{11} f_{21}+f_{12} f_{22}, \\
      c_{12}^2 f_{11} f_{12}-c_{21}^2 f_{21} f_{22}-c_{11} c_{21} f_{11} f_{22}+c_{12} c_{22} f_{11} f_{22}-f_{11} f_{12}+f_{21} f_{22}, \\
      c_{12}^2 f_{12} f_{21}-c_{11}^2 f_{11} f_{22}-c_{11} c_{21} f_{21} f_{22}+c_{12} c_{22} f_{21} f_{22}-f_{12} f_{21}+f_{11} f_{22}, \\
      c_{21}^2 f_{11} f_{21}-c_{12}^2 f_{12} f_{22}-c_{11} c_{12} f_{11} f_{22}+c_{21} c_{22} f_{11} f_{22}-f_{11} f_{21}+f_{12} f_{22}, \\
       c_{21}^2 f_{12} f_{21}-c_{11}^2 f_{11} f_{22}-c_{11} c_{12} f_{12} f_{22}+c_{21} c_{22} f_{12} f_{22}-f_{12} f_{21}+f_{11} f_{22}, \\
    (c_{11} c_{12}^2+c_{11} c_{21}^2+c_{11} c_{22}^2-2 c_{12} c_{21} c_{22}) f_{11} + c_{12}^3 f_{12} + c_{21}^3 f_{21} + c_{22}^3 f_{22} - 1, \\
          c_{11} c_{12} f_{12} f_{21}-c_{21} c_{22} f_{12} f_{21}+c_{12} c_{21} f_{21} f_{22}-c_{11} c_{22} f_{21} f_{22}+c_{12}^2 f_{12} f_{22}-c_{22}^2 f_{12} f_{22},   \\
      c_{12} c_{21} f_{11} f_{21}-c_{11} c_{22} f_{11} f_{21}+c_{11} c_{21} f_{11} f_{22}-c_{12} c_{22} f_{11} f_{22}+c_{21}^2 f_{21} f_{22}-c_{22}^2 f_{21} f_{22}, \\
       c_{12} c_{21} f_{11} f_{12}-c_{11} c_{22} f_{11} f_{12}+c_{11} c_{12} f_{11} f_{22}-c_{21} c_{22} f_{11} f_{22}+c_{12}^2 f_{12} f_{22}-c_{22}^2 f_{12} f_{22}, \\
      (c_{12}^3{-}c_{11}^2 c_{12}{-}c_{12} c_{21}^2{-}c_{12} c_{22}^2{+}2 c_{11} c_{21} c_{22}) f_{12}
      - (c_{22}^3{-}c_{11}^2 c_{22}{-}c_{12}^2 c_{22}{-}c_{21}^2 c_{22}{+}2 c_{11} c_{12} c_{21}) f_{22}, \\
      (c_{21}^3{-}c_{11}^2 c_{21}{-}c_{12}^2 c_{21}{-}c_{21} c_{22}^2{+}2 c_{11} c_{12} c_{22}) f_{21}
      - (c_{22}^3{-}c_{11}^2 c_{22}{-}c_{12}^2 c_{22}{-}c_{21}^2 c_{22} {+}2 c_{11} c_{12} c_{21}) f_{22}.
      \end{matrix} \end{small}
$$
This list was generated by computer algebra as follows.
We start from a list of polynomials that cuts out
$\overline{\ncyc(\Qm)}$ as a subset of $\Cx^8$.
That list consists of $\ell,h,h\polar$
and the twelve $2 \times 2$ minors of the two matrices
$$ \begin{pmatrix}
c_{11} & c_{12} & c_{21} & c_{22} \\
\partial h\polar / \partial f_{11} & \partial h\polar / \partial f_{11} &
\partial h\polar / \partial f_{21} & \partial h\polar / \partial f_{21} \\
\end{pmatrix}
\quad {\rm and} \quad  \begin{pmatrix}
f_{11} & f_{12} & f_{21} & f_{22} \\
\partial h / \partial c_{11} & \partial h / \partial c_{11} &  \partial h / \partial c_{21} & \partial h / \partial c_{21} \\
\end{pmatrix}.
$$
The respective rows are linearly dependent for any pair
$(c,f)$ of supporting linear functions. The resulting ideal has
the desired prime ideal as its radical, by Hilbert's Nullstellensatz.
We computed that radical.

We now describe the other  components of the variety $\overline{\ncyc(\Qm)}$.
The eight strata  \Qb51 consist of
points $(c,f)$ where one of the entries of $f$ is $\pm 1$ and the others are $0$.
This linear functional $f$ exposes one of the elliptopes in $\partial\Qm$. For example, consider $f = (1,0,0,0)$.
The prime ideal of this component is
$$\,\langle f_{11} -1, f_{12}, f_{21}, f_{22},c_{11} -1 \rangle . $$
The corresponding stratum in the semialgebraic set $\ncyc(\Qm)$
satisfies the additional cubic inequality
\begin{equation}
\label{eq:cubicboundary}
 c_{12}^2 + c_{21}^2 + c_{22}^2 - 2c_{12}c_{21}c_{22}  \leq 1 .
 \end{equation}
The boundary of the elliptope, seen inside $\ncyc(\Qm)$, separates
 \Qb51 from the nonlinear stratum  \Qb44.

By duality, there are also eight strata \Qb15 in the normal cycle $\ncyc(\Qm)$.
Now, the elliptopes appear in the $f$-coordinates and $c$ is one of the $8$ classical extreme points \Qt1.
These are obtained by exchanging the roles of $f$ and $c$, using the linear transformation
$H$. The ideal of one of the components  \Qb15~is
\begin{equation}\nonumber
\langle c_{11} -1, c_{12} -1, c_{21} -1, c_{22} -1, f_{11} + f_{12} + f_{21} + f_{22} -1 \rangle .
\end{equation}
The semialgebraic description of this stratum is obtained by setting  $c = 2Hf$
in the cubic inequality \eqref{eq:cubicboundary}.

\subsection{Support function and gauge function}\label{sec:suppfct}

Two real valued functions are commonly used to describe a convex set $\KK$.
The first is the {\it gauge function}  $\gauge_\KK$
of the set \cite{Schaefer}.  It measures how far we have to go along a ray in some direction
until we leave the set.
 The best known example is a norm as determined from its unit ball.
The gauge function $\gauge_\KK$ is the reciprocal of another well-known function in convex geometry: the {\it radial function} of the body.

The other standard measure is the maximum of a given linear functional over the given set. This is called the {\it support function} $\phi_\KK$. In short, for any convex body $\KK$ with $0$ in its interior,
we consider:
\begin{eqnarray}
  \gauge_\KK(x) &=& \inf \, \{\lambda\geq0\stt x\in\lambda\KK\}, \label{thirtyfour} \\
  \phi_\KK(f) &=& \sup \, \{f\cdot x\stt x\in\KK\}.
\end{eqnarray}
Both functions are homogeneous of degree $1$. From \eqref{thirtyfour} we recover
$\KK=\{x\stt \gauge_\KK(x)\leq 1\}$, and dually
$\KK\polar=\{f\stt\phi_\KK (f)\leq 1\}$. Hence, the support function of $\KK$ is
 the gauge functional of $\KK\polar$.
By self-duality, the two functions are closely related for $\KK = \Qm$. In the sequel we study
 $\phi = \phi_\Qm$, and we drop the~subscript.

Since the set $\Qm$ has two markedly different kinds of boundary points, the support function requires a binary distinction concerning a functional $f$:
{\it Will $\max_{c\in\Qm}f\cdot c$ be attained at a classical point or at an exposed extreme point?}
The answer is the same for all multiples of $f$, including $-f$, so we are really asking about the projective geometry of the boundary points of $\Qm\polar$, which is visualized for $\Qm$ in \reff{fig:projective}. The separating surfaces are the  images of the \Qtc3 surfaces. Note that (after the identification of $f$ with $-f$)  there are four such surfaces. We can therefore not expect a simple algebraic relation to mark the distinction: Each of the elliptope boundaries has a Zariski completion with multiple additional irrelevant components overlapping the division \reff{fig:projective}, so some careful sifting of inequalities is needed.

Once we know the nature of the maximizer, however, the computation of $\phi$ is straightforward.
In the first case, it is the maximum of an affine functional over the $8$ vertices of
 $\Cl$. Namely, $\phi(f)=\phi_\Cl(f)$ with
\begin{eqnarray}\label{phiC}
  \phi_\Cl(f)&=&\max\{f\cdot c\stt c\in\partial_e\Cl\}\nonumber\\
             &=&\max\{|f_{11}{+}f_{12}{+}f_{21}{+}f_{22}|,|f_{11}{+}f_{12}{-}f_{21}{-}f_{22}|,|f_{11}{-}f_{12}{+}f_{21}{-}f_{22}|,|f_{11}{-}f_{12}{-}f_{21}{+}f_{22}|\} \nonumber\\
             &=& \norm{2Hf}_\infty,
\end{eqnarray}
In the second line, we grouped the maxima over a pair of antipodal classical extreme points
by writing an absolute value. In the third line,
$\norm\cdot_\infty$ denotes the maximum norm on $\Rl^4$. This case
applies exactly when the ray $\Rl\,f$ intersects the boundary of $\Qm\polar$ in an $\Ns$-facet.

The other possibility is that the ray intersects the boundary in a curved tetrahedron.
In this case we think of $\phi$ as the gauge function of $\Qm\polar$.
We need to determine the intersection point of the ray $\{\lambda f\}$ with
 $\partial \Qm\polar$.
This point is in the zero set of $h\polar$.
The equation $h\polar(\lambda f)=0$ in \eqref{polypolar} is readily solved for $\lambda$,
using the splitting of $h\polar$ into homogeneous parts.
We get $\lambda^6k(f)-\lambda^4p(f)=0$, so in this case $\lambda=\phi(f)$ is
\begin{equation}\label{phiQ}
  \widetilde\phi(f)=\sqrt{\frac{k(f)}{p(f)}}.
\end{equation}
This function is homogeneous of degree $1$, as required.  The above heuristics is made precise
in the following proposition, which extends \reff{thm:supp}:
\begin{prop}\label{prop:hom}\proofin{sec:supproof}
For any $f\in\Rl^4 \backslash \{0\}$, the following conditions are equivalent:
\begin{itemize}
\item[(1)] The ray $\{\lambda f\}$ intersects the boundary of $\Qm\polar$ in a point of type \Qt4.
\item[(2)] Some point $c\in\Qm$ maximizing $f\cdot c$ is of type \Qt4.
\item[(3)] We have \,$p(f)<0\,$ and
\begin{equation}\label{Tsirelson_condition}
  m(f) \,= \min_{i,j = 1,2} |f_{ij}| \Bigl( \sum_{i,j = 1,2} |f_{ij}|^{-1} \Bigr) \,>\, 2.
\end{equation}
\item[(4)] We have \,$p(f)<0\,$ and
{\def\tightffff#1#2#3{\bigl(\textstyle\frac1{f_{11}}{#1}\frac1{f_{12}}{#2}\frac1{f_{21}}{#3}\frac1{f_{22}} \bigr)}
\begin{equation}
  \tilde m(f) \,= \tightffff+++ \tightffff+-- \tightffff-+- \tightffff--+ \,<\, 0. \nonumber
\end{equation}}
\item[(5)] Perform a symmetry transformation (even number of sign changes) so that the maximum \eqref{phiC} of $f\cdot c$ over $\Cl$ is attained at the extreme point $c=(1,1,1,1)$, i.e., $\phi_\Cl(f)=f_{11}{+}f_{12}{+}f_{21}{+}f_{22}$. Then
\begin{equation}\label{cubicbdary}
  f_{11}f_{12}f_{21}+f_{11}f_{12}f_{22}+f_{11}f_{21}f_{22}+f_{12}f_{21}f_{22}<0.
\end{equation}
\end{itemize}
In this case the support function of $\Qm$ is  $\phi(f)=\widetilde\phi(f)$ from \eqref{phiQ}, otherwise $\phi(f)=\phi_\Cl(f)$ from~\eqref{phiC}.
\end{prop}

In the domain described by  \reff{prop:hom},
the maximizer of $\phi(f)$ is a unique exposed point $c^*\in\Qm$. An explicit formula $f\mapsto c^*$ would correspond to solving for $c$ given $f$ in the system of 17 polynomials defining the stratum \Qb44 in $\overline{\ncyc(\Qm)}$. In terms of the angle parametrization of boundary pieces, it is essentially the self-dual counterpart of the map $\Phi$ sketched in~\reff{fig:T2T}.

We also numerically checked that our formula for the support function agrees with Tsirelson's formula~\cite{Tsi85}
\begin{equation}\label{phiTsirelson}
  \phi_T(f)=\left\lbrace\begin{array}{ll}
            \sum_{ij}\abs{f_{ij}} & \mbox{if} \ \ p(f)\geq0 \\[10pt]
            \sum_{ij}\abs{f_{ij}}-2\min_{ij}\abs{f_{ij}}& \mbox{if} \ \ p(f)<0 \ \ \mbox{and} \ \ m(f) \leq2 ,\\[10pt]
            \sqrt{\left(\sum_{ij} f_{ij}^2\right)+\abs{p(f)}\left(\sum_{ij} f_{ij}^{-2}\right)}& \mbox{if} \ \ p(f)<0 \ \ \mbox{and} \ \ m(f) \geq2 ,
            \end{array}\right.
\end{equation}
but we were not able to reconstruct his proof.

\section{Quantum Connections}\label{sec:desquribe}

We now return to \reff{sec:introPhys}, and we discuss what
our geometric findings on $\Qm$ mean for quantum theory. We
begin by exhibiting explicit quantum models for the extremal points $c$ of type \Qt4.
In \autoref{sec:qkd}
we examine what geometric features of some point $c\in\Qm$ make it suitable for quantum key
distribution. Actually, the extremal correlations have much stronger uniqueness properties,
known as self-testing. We state these in \autoref{sec:selftest}.
This is followed by a short reflection on the historical role of the elliptope case.

\subsection{Quantum models}
\label{sec:quantummodels}

We introduced the convex body $\Qm$ as the set of quantum correlations. We then mostly studied necessary conditions, i.e., consequences of the quantum form. But in order to show the sufficiency of any such condition,
we must construct explicit quantum models. In this section we write out a parametrized family of quantum models, that realizes all points described in \reff{prop:paramExt}.
That such a simple family already exhausts the extreme points of $\Qm$ is, of course, a special feature of the minimal 222 case.
In a sense which we will describe in \reff{sec:selftest}, this family is actually {\it the only one} realizing the non-classical extreme points of $\Qm$.

Using the notation introduced in \reff{sec:introPhys}, we set $m=4$ and $\mathcal{H} = \Cx^4$.
Algebraically, a quantum state is a positive semidefinite $4 \times 4$ matrix $\rho$ with trace $1$.
For the density operator, we fix it as the rank one matrix with entries $\rho_{\alpha\beta} = \Psi_\alpha\overline{\Psi_\beta}$,
defined by the unit vector
\begin{equation}\label{psi0marge}
\Psi =  \biggl( \,0\,, \,
 \frac{1}{\sqrt2}\,,\,
  -\frac{1}{\sqrt2}\,,\,0\,
\biggr)^T.
\end{equation}
The measurements of the two parties are represented by the following
real symmetric $4 \times 4$ matrices:
$$ \begin{small}
A_1 \, = \,\begin{pmatrix}
\cos(\alpha) & 0 & \sin(\alpha) & 0 \\
 0 & \cos(\alpha) & 0 & \sin(\alpha) \\
 \sin(\alpha) & 0 & -\cos(\alpha) & 0 \\
 0 & \sin(\alpha) & 0 & -\cos(\alpha)
\end{pmatrix}, \quad
A_2  \, = \,\begin{pmatrix}
\cos(\gamma) & 0 & -\sin(\gamma) & 0 \\
 0 & \cos(\gamma) & 0 & -\sin(\gamma) \\
 -\sin(\gamma) & 0 & \!\! -\cos(\gamma) & 0 \\
 0 & -\sin(\gamma) & 0 & \!\! -\cos(\gamma) \,
\end{pmatrix}, \end{small}
$$
$$ \begin{small}
B_1 \, = \,\begin{pmatrix}
\, -1 & 0 & 0 & 0 \,\\
\,   0 & 1 & 0 & 0 \,\\
\, 0 & 0 & -1 & 0 \,\\
\, 0 & 0 & 0 & 1 \,\end{pmatrix},\qquad
B_2 \, = \,\begin{pmatrix}
-\cos(\alpha+\beta) & -\sin(\alpha+\beta) & 0 & 0 \\
-\sin(\alpha+\beta) & \phantom{-}\cos(\alpha+\beta) & 0 & 0 \\
 0 & 0 & -\cos(\alpha+\beta) & -\sin(\alpha+\beta) \\
 0 & 0 & -\sin(\alpha+\beta) & \phantom{-}\cos(\alpha+\beta)
\end{pmatrix}. \end{small}
$$
These matrices satisfy the hypotheses stated in (\ref{eq:ABhypotheses}).
We now compute the four correlations $c_{ij}$ in (\ref{eq:getCfromAB}).

\begin{lem}\label{lem:paramQ}
The point $c$ with parameters $(\alpha, \beta, \gamma, \delta)$ in \reff{prop:paramExt}, even without the
inequality constraint on $\Delta$, satisfies
$\,c_{ij}\,=\,\tr(\rho A_iB_j)$.
\end{lem}

This is a formal identity, for all $1 \leq i,j \leq 2$, and can be checked using computer algebra. However, this takes time and some careful typing, and is the approach of minimal insight.
It is better to remember the physics, and to realize that this family is the ``minimal'' quantum model at the historical root of the subject. It arises naturally when Alice and Bob each see the polarization degree of freedom of a pair of photons, or the spin degree of freedom of a spin-$1/2$ particle, and their measurements are just polarization (or spin component) measurements. The only interesting quantity is then the angle between the orientations of the polarizers, and the correlation is just the cosine of this angle. In this form it was already used in John Bell's first paper \cite{Bell}, and in the work of CHSH \cite{CHSH}. The problem that needed to be solved for these authors was to adjust the angles in order to maximize the violation of the CHSH inequality. What we claim here is that the same physical setup suffices to realize the entire quantum correlation body $\Qm$.

To see how this family is ``minimal'', note that when either Alice or Bob holds a classical system, for instance if the observables $A_1$ and $A_2$ commute, no entangled states are possible. This even characterizes classical systems \cite{Raggio}. Therefore
only classical correlations can be constructed. Hence Alice's and Bob's subsystems need at least $2$-dimensional Hilbert spaces. Since $[A_i,B_j]=0$, these subsystems must be combined in a tensor product, so $\HH=\Cx^2\otimes\Cx^2=\Cx^4$. Actually, we can take the $A_i$ and $B_j$, as well as the density operator $\rho$, to be real rather than complex, so the Hilbert space is
actually $\Rl^4$. Of course, for almost all of quantum physics, e.g., the Schr\"odinger equation, this would be an untenable restriction. As shown only recently, there are even correlation experiments of just the type considered here (only not minimal) that prove that ``real quantum mechanics'' is insufficient \cite{Reality}. With the further simplification that $A_i^2=B_j^2=\idty$, these matrices must be given on each side by reflections in planar geometry, a one parameter family. Explicitly, the reflections are
\begin{equation}\label{quantumA}
  M(\tau) \,:= \, \begin{pmatrix}
              \cos(\tau) & \sin(\tau) \\
              \sin(\tau) & \! \!\! -\cos(\tau)
            \end{pmatrix}
           \,=\,\cos(\tau)\cdot \sigma_3\,+\, \sin(\tau)\cdot \sigma_1
\quad \hbox{for} \, \tau \in[0,2\pi].
\end{equation}
The matrices above are then $A_1=M(\alpha)\otimes\idty$, $A_2=M(-\gamma)\otimes\idty$, $B_1=\idty\otimes M(\pi)$, and $B_2=\idty\otimes M(\alpha+\beta+\pi)$.

The state $\rho$ must be entangled. We choose a maximally entangled one, which is by definition pure on the whole system, $\rho=\kettbra\Psi$, and maximally mixed on the subsystems, i.e.~for any matrix $A$ we have
\begin{equation}
\label{eq:tr/2}
 \braket\Psi{(A\otimes\idty)\Psi}\,=\,\braket\Psi{(\idty\otimes A)\Psi} \,=\, \tr(A)/2.
\end{equation}
This would be too special for the full marginals case \cite{allpure}. But here we can get away with it,
and even fix the vector $\Psi$ in \eqref{psi0marge} for the whole family.  This is the
unique vector (up to scaling) which is antisymmetric with respect to the exchange of the two tensor factors.
For any matrix $A$, the vector $(A\otimes A)\Psi$ is again antisymmetric.
Hence  $(A\otimes A)\Psi=\varepsilon(A)\Psi$, for some homogeneous quadratic function $\varepsilon$, which is also multiplicative, i.e., equals the determinant. Inserting for $A$ a one-parameter subgroup of SU($2$), i.e., $A=\exp(itM)$  for traceless $M$, we
find that $\varepsilon(A)=1$ is constant. Hence, in first order in $t$,
we have $(M\otimes\idty+\idty\otimes M)\Psi=0$. Applying this to the matrices from \eqref{quantumA} and combining with  \eqref{eq:tr/2} we thus find
\begin{eqnarray}\label{psiMMpsi}
  \tr \bigl(\rho\,M(u)\otimes M(v)\bigr)& = &
\braket\Psi{M(u)\otimes M(v)\Psi}=\braket\Psi{(M(u)M(v))\otimes\idty\Psi}\nonumber\\
      &=&\tr\bigl(M(u)M(v)\bigr)/2 \,=\,-\cos(u-v).
\end{eqnarray}
The last identity is seen by observing that $\tr(A B^{T})$ is the standard scalar product in matrix space,
and using the addition theorem for the cosine, or, geometrically that the product of reflection across lines at an angle $u-v$ is the rotation by that angle. Inserting the choice of angles after \eqref{quantumA} gives the result.

\subsection{Geometric aspects of quantum key distribution}
\label{sec:qkd}

Quantum key distribution (QKD) is an important task in quantum information technology \cite{BB84}.
It furnishes the main practical reason for studying the body $\Qm$.
Here we discuss geometric features that are fundamental for that task.
The goal of QKD is for two distant parties, Alice and Bob, to utilize quantum correlations for
generating a key which is guaranteed to be secret from any eavesdropper, here called Eve.
Eve is only assumed to be constrained by the laws of quantum mechanics, but otherwise enjoys every
possible freedom. In particular, she is allowed to manipulate the correlated systems on which the scheme is based, the quantum channels by which they are transmitted, and even
the measurement devices. She also gets a copy of the communications exchanged between Alice and Bob. However, she can only read these but not change them.
One must also assume that once the data collection starts, Eve cannot reach into Alice's and Bob's lab and access their measurement settings or outcomes. Indeed, if Eve could do that, she would not even need to bother with the whole quantum setup, or she could play a trivial woman-in-the middle attack, and secrecy would be obviously impossible.
So the rules of the game force her to gain at least some information from the quantum systems.  According to the laws of quantum mechanics, this introduces a disturbance detectable by Alice and Bob. When they do detect such deviations from the expected statistics the key distribution has failed. Eve can always achieve that, but this is counted as a failure for her, because she will also not learn any secrets.

We want to show here that the minimal setup  in this paper is already sufficient to support QKD.
Moreover, the main security argument is based directly on the geometry of $\Qm$. In undisturbed operation the setup
 leads to some correlations $c\in\Qm$. Alice and Bob will use a random sample of their particles
to verify this via the public classical channel, and will abort the process if they find significant deviations from $c$.
We claim that QKD is possible whenever  $c$ is a {\it non-classical extreme point} (cf.~\cite{FranzWerner})

Suppose Alice and Bob test their correlations and find them to be such a point $c\in\ext\Qm$. What
could Eve know about their measurement results?
Let $\evev$ be a random variable that summarizes her findings.
The conditional correlation $c_\evev$ is the $2 \times 2$ matrix that pertains to those cases
  where Eve found $\evev$.
We have $c_\evev\in\Qm$  since Eve is constrained by quantum mechanics.
Combining the data with the probabilities $p_\evev$ for observing $\evev$,
we get $c=\sum_\evev p_\evev\, c_\evev$. But since $c$ is extremal, all $c_\evev$ that appear
with nonzero probability must be equal to $c$. That is the same as saying that $\evev$ is
statistically independent of Alice's and Bob's information. So
while Eve knows $\evev$, she learns nothing about $c$.

Note, however, that the argument applies equally to classical extreme points.
Only, in that case the extreme points are completely deterministic.
In the case of deterministic agreement probabilities indeed factorize, say $c_{ij} = a_i b_j$.
With $c_{ij} \in \{0,1\}$ and $a_i, b_j \in [0,1]$, this factorization would be $1\cdot1=1$ or $0\cdot0=0$.
Of course, this is utterly useless for drawing a secret key.
So what Alice and Bob use to generate the key are the non-trivial correlations that may be present in {\it non-classical}
extreme points. Any non-classical correlation is fine for that purpose, since this may be further distilled into perfect agreement.

Note also that in case of non-classical extremal points, {\em any} pair (out of four) of measurement settings can be chosen to be the ``key (generating) basis" due to the very definition of cryptographic security~\cite{FranzWerner}. Weakening this to only the existence of one pair of measurement settings that is uncorrelated with Eve still supports a perfect QKD protocol. This necessitates moving to the full 8-dimensional body where a non-extremal point $P$ has been shown to exhibit this phenomenon~\cite{weakST}. Intuitively, this means while the behaviour $P$ admits a nontrivial convex decomposition $P=\sum p_\evev P_\evev$, Eve does not gain additional information in the key basis: $P_\evev(a,b|\text{key})=P(a,b|\text{key})$ for all $a,b,\evev$; she only has an advantage in other pairs of settings where the outcomes are not used for making a pair of secret key.

The full analysis of QKD takes not only error correction into account, but also the
overhead of statistically verifying that the given source is really described by $c$.
This necessarily involves errors, and the experimental implementation will have additional errors of its own.
The analysis, done carefully also for $c$ sufficiently close to $\partial_e \Qm$, results not in a blanket statement that Eve will know ``nothing'', but in a quantitative bound on how much she might know in the worst case. So, in addition to error correction (getting the keys to be really the same) one needs ``privacy amplification'', a process  that had already been studied in purely classical settings~\cite{privacy} prior to the advent of QKD. The traditional information theoretic view  focuses on rates in the asymptotic regime, i.e., for a large number of exchanged raw key bits. This systematically neglects the overhead of reliably estimating $c$. This can be considerable in real, and therefore finite, runs. A usable QKD security proof always has to include the finite key analysis, and all imperfections.
This is far beyond the current paper, and we refer to \cite{DIQKD1,DIQKD}.
To connect with the literature, we emphasize that here we have described ``device independent'' QKD,
for which the experimental entrance ticket is a ``loophole free'' Bell test, which has been achieved only
recently \cite{giustina15,lopholefree1,shalm15}. Nevertheless, recent advances on the theoretical
side \cite{DIQKD1,tan2020qkd,DIQKD} have resulted in proof-of-principle experiments such as~\cite{diqkdexp1,diqkdexp2,diqkdexp3} with varying degree of closeness to ``loophole free".
On the other hand, systems not realizing the ideal of device independence are already commercially available
(see~\cite[Section 3.2]{commercialQKD}).

Coming back to the geometric features of $\Qm$ relevant for QKD, we can first see some of the tradeoffs that enter the choice of $c\in\Qm$. Choosing $c$ of type \Qt3 might seem advantageous, because then the outcomes of Alice and Bob agree for one of the settings, making the error correcting step superfluous. However, this comes at the price of moving closer to $\Cl$, so that relatively low noise may make privacy amplification harder. The traditional working point therefore has been the $c$ from \eqref{CHSH}, maximally violating a CHSH inequality. As an exposed point this seems to simplify the tomography, because only one combination of correlations needs to be estimated. But this actually leaves unnecessary leeway in the tangent directions, which are only fixed to $\sqrt\varepsilon$, when the correlation is fixed to order $\varepsilon$. It is therefore better to use all the correlation data, rather than focusing on just one linear combination.

We further have to correct a simplification in the above argument for non-classical extreme points. It really applies only to the full statistics of Alice's and Bob's measurements including marginals, i.e., the $8$-dimensional body of which $\Qm$ is a projection. The key must be generated from the local measurement data, not just the combined outcomes entering a correlation. So it will be important to get extremality of $c\in\Qm$ also when it is extended to $8$ dimensions with zero marginals. This will be done in the following section,
in particular in items (4),(5) of \autoref{prop:selftest}.

\subsection{Uniqueness of quantum models}\label{sec:selftest}

What can we deduce about the uniqueness of quantum realization of points in $\Qm$?
Prima facie  there is no reason for any uniqueness.
In \reff{sec:desquribe} we made special choices to realize the extreme points {\it somehow}.
Why $m=4$ in \reff{lem:paramQ}?
Why not $m=3$? What about $m=\infty$?
That there is limited choice if we try to find realizations in minimal dimension says very little about the non-uniqueness if we allow more spacious Hilbert spaces.
We begin by noting that there are some trivial ways in which uniqueness fails:
\begin{itemize}
\item{\it Unitary transformation}\\
This refers to a change of basis in the Hilbert space $\mathcal{H}$.
It makes the quantum model looks different but does not change the correlations $c_{ij}$.
They are invariants of the action of the unitary group $U(m)$ on
 quintuples of matrices $(\rho,A_1,A_2,B_1,B_2)$ as in \reff{sec:introPhys},
with  $\mathcal{H} = \Cx^m$.
Two unitarily equivalent models, i.e., two quintuples
in the same $U(m)$-orbit, are considered to be ``the same''.
\item{\it Expansion}\\
This means enlarging the Hilbert space $\HH$ of the model by an additional summand $\HH_0$
where the states act trivially. If we use the space $\HH\oplus\HH_0$, the state $\rho\oplus0$ and the
observables $A_i\oplus A_i^0$ and $ B_j\oplus B_j^0$,
 with $A_i^0$ and $B_j^0$ arbitrary, provided \eqref{eq:ABhypotheses} still holds,
 then the correlations do not change.
\item{\it Adding multiplicity}\\
This means enlarging $\HH$ by an additional tensor factor where the measurement acts trivially.
If we use  $\HH\otimes\HH_\nu$ with  $A_i\otimes\idty$, $B_j\otimes\idty$,
and a state $\widetilde\rho$ whose partial trace over the second factor is $\rho$, then the correlation do not change.
Writing the tensor product as a direct sum with respect to a basis of $\HH_\nu$ makes this a direct sum of possibly correlated copies of the given model.
\end{itemize}

The best we can hope for is uniqueness of the quantum model for  $c$ up to these three operations. This
is Condition (2) in \reff{prop:selftest} below.
Condition (1) is the cryptographic security  discussed in \reff{sec:qkd},
i.e., anything an eavesdropper might know is statistically independent of $c$.
Remarkably, the two conditions are equivalent.
Moreover, cryptographic security is extended in (4) to all measurements made with observables in the algebra
$\mathcal{A}$ generated by $A_1,A_2,B_1,B_2$. The central condition is (3), the uniqueness of a certain kind of model, defined by removing the redundancy of the operations of expansion and adding multiplicity.  These {\it cyclic models} are defined by the property that $\rho=\kettbra\Psi$ is pure and $\Psi$ is cyclic for the operators $A_i,B_j$, i.e.,  we get a dense subspace
of $\HH$ by acting with these operators repeatedly on $\Psi$.

Uniqueness results like~\reff{prop:selftest} appear in the literature under the
keyword {\it self-testing} \cite{MayersYao,reviewST}. This
indicates that the correlations  can be used cryptographically without
first verifying that the devices act as they should, or that the prepared state is as planned:
Security is based directly on the observed correlations.
Current definitions of self-testing \cite{Goh,MayersYao,reviewST,singletST} implicitly
use some of the equivalences below. Unlike these definitions and statements therein, our formulation below does not make purifying assumptions on the unknown models, i.e. ``going to the church of the larger Hilbert space'', nor require additional ancillas into which the ideal system is swapped.

\begin{thm}\label{prop:selftest}\proofin{sec:qproofs}
Fix a point $c\in\Qm$. Let $(\rho,A_i,B_j)$ be any quantum model
in a Hilbert space $\HH$ for the correlations $c$, and write
$\Alg$ for the
 norm closed operator
algebra generated by $A_i,B_j$. When applicable, denote by
$(\hat\rho,\hat A_i,\hat B_j)$ the specific model for $c$ given in \autoref{sec:quantummodels}. The following are equivalent:
\begin{itemize}
\item[(1)] $c$ is nonclassical and extremal in $Q$.
\item[(2)] There is a unique cyclic model for $c$. Explicitly, when $\rho=\kettbra\Psi$ is pure and cyclic for $\Alg$ (i.e., $\Alg\ket{\Psi}$ is dense in $\HH$), then it is unitarily equivalent to the model specified for $c$ in  \autoref{sec:quantummodels}.
\item[(3)] The quantum model is unique up to unitary transformation, expansion and adding multiplicity. That is, by a unitary transformation any model can be brought into canonical form $\HH=(\hat\HH\otimes\HH_\nu)\oplus\HH_0$, $\rho=\hat\rho\otimes\rho_\nu\oplus0$, $A_i=\hat A_i\otimes\idty_\nu\oplus A_i^0$, and ditto for $B_j$.
\item[(4)] $c$ has a unique extension to a quantum behavior $p$ (necessarily extreme) in the 8-dimensional body including both correlations and marginals.
\item[(5)] $c$ is cryptographically secure, i.e.~its unique extension $p$ is cryptographically secure (cf.~\cite{FranzWerner}).
\item[(6)] $c$ is algebraically secure, i.e.~$\tr(\rho EX) =\tr(\rho E)\,\tr(\rho X)$
      for $X\in\Alg$ and any $E$ commuting with~$\Alg$.
      \end{itemize}
\end{thm}

We stress that these equivalences are claimed only for the minimal scenario
assumed in this paper. For larger parameters NMK, the condition (1) may well be weaker than the others.
This was noted by Tsirelson who treated the 2M2\marge0 case and
showed \cite[Thm.\,3.3]{Tsi85} that for ``odd rank'' there are
two inequivalent cyclic representations (but connected by transposition, a non-unitary operation). Recently,~\cite{stMUB} gave another example thanks to the existence of inequivalent MUBs (this time even up to transposition). Parallel to Tsirelson's work,
 uniqueness was shown in \cite{SW1,SW2}.  At the time,
the authors were not aware of cryptographic applications,
and  R.F.W.~(the second author of \cite{SW1}) thought of
this result as a fortuitous algebraic property
of the CHSH expression, and unlikely to generalize.
After an earlier attempt in \cite{FranzWerner}, he only realized with the current work that the methods of \cite{SW1} provide
uniqueness for the largest scope possible within the 222 scenario.

In cryptography, one generally has only an approximation to $c$. Luckily, the
uniqueness result holds robustly. As shown first in \cite{SW2}, one can find
 a nearby set of operators and a state realizing $c$ precisely.

\subsection{The role of the elliptope}
Historically, the elliptope has played an important role for quantum correlations.  John Bell's
first inequality \cite{Bell} concerned three measurement settings on each side, a 232 setting.
The underlying quantum model was the singlet state $\rho$ above, and the settings on the two sides pairwise equal, so that the outcomes where pairwise perfectly anticorrelated, hence opposite with probability one. Even without accepting Bell's conclusion that they are therefore ``predetermined'', or going into a discussion what that means, this simplifies the analysis to a discussion of just three possibilities. Although Bell does not include the picture, the quantum body is exactly the elliptope shown in \reff{fig:elliptope}, and the classical body is the embedded tetrahedron.
Bell's original scenario and the elliptope continue to serve as a simplified example for explaining quantum correlations, from Mermin's classic article \cite{mermin} to today~\cite{janas2019}.

However, perfect anticorrelation, while easily verified in the model, is much harder to get in the quantum experiments.
It was the decisive advance due to Clauser, Horne, Shimony and Holt \cite{CHSH} to eliminate this experimentally doubtful assumption, and to identify the 222 case as the minimal setting.

Our analysis of the convex body $\Qm$ shows that the elliptope still occurs in the 222 case, and that it suffices to assume (or experimentally verify) full correlation between only two of the observables. The geometric picture of Bell's analysis (he does not draw it, though) then holds. This is of interest in cryptography, since drawing key from the perfectly correlated settings eliminates the error correction step, albeit at the expense of the distance from the classical polytope, and hence harder noise requirements.

We close with the remark that the elliptope is iconic in
many branches of the mathematical spectrum,
notably in convex optimization, as the feasible region of
a semidefinite program, and in
Gaussian statistics, as the
set of $3 \times 3$ correlation matrices.
See the pointers surrounding \cite[Figure 1.1]{Michalek} and \cite[Figure 1]{Uhler}.

\subsection{Fixed Hilbert space dimension} \label{sec:dimH}

The models above use a $4$-dimensional Hilbert space. This is an option, that happens to be sufficient
for getting the extreme points of $\Qm$, but no assumption like this is made in the definition of $\Qm$. As seen in~\reff{sec:qkd},
this is vital for QKD. However, one can ask how such an assumption would change $\Qm$. For non-minimal settings, the question what inference about dimension can be drawn from an observed correlation has been studied extensively (see, e.g., \cite{dimwit,dimwit2016,dimwit2017}). So what can be said in the minimal setting?
Here we just collect some basic observations, that can be skipped without loss for other sections.

Let us denote by $\Qm_m$  the (warning: nonconvex!) set of quantum correlations
that are obtainable by models as in \reff{sec:introPhys} under the additional assumption $\dim\HH=m$. We also write
$\Cl_m\subset\Qm_m$ for the corresponding set of classical correlations realized in a sample space of $m$ points. Since we do not require $A_i^2=B_j^2=\idty$, or tensor product separation,
the set $\Qm_m$ is increasing with $m$, and so is $\Cl_m$.
Since building a model for a convex combination of correlations requires the direct sum of models, and this increases $m$, these sets are not in general convex.
However, we know that extreme points of the full bodies can be realized in fixed dimensions, so $\ext\Cl\subset\Cl_1$ and $\ext\Qm\subset\Qm_4$. Thus Carath\'eodory's Theorem allows us to put a bound on the required $m$: Since in an $n$-dimensional
convex body every point is the convex combination of at most $n+1$ extreme points, and $n=4$, we conclude that $\Cl\subset\Cl_5$ and $\Qm\subset\Qm_{20}$.

The Carath\'eodory bound
is typically tight only for polytopes, where most points require
$n+1$ of the vertices for a convex representations. When there is a continuum of extreme points, that {\em Carath\'eodory number}
is usually smaller.   One extreme example for this are Euclidean balls of any dimension, in which every point is the convex combination of two antipodal boundary points. Another familiar example is the quantum state space with $\dim\HH=m$, which has dimension $m^2-1$, although every density operator has a spectral resolution into $m$ pure states.  Our body $\Qm$ will be an intermediate case in this respect. The following result is also part of~\cite{Tsi85}.

\begin{prop}\label{prop:dimH}\proofin{sec:dimHproof}
Let $\Qm_m$ be the set of quantum correlations obtainable from models with $m$-dimensional Hilbert space.
Then $\Qm_m\subset\Cl\neq\Qm$  for  $m\leq3$, and $\Qm_m=\Qm$ for $m\geq4$.
\end{prop}

The analogous set $\Cl_m$ for the cross polytope $\Cl$ is also interesting
and worth further study. Obviously, $\Cl_1$ is the subset of rank $1$ matrices $c\in\Rl^{2\times2}$, a condition which is invariant under the natural symmetry group of correlation bodies but {\it not} under the extended symmetry group discussed in \reff{sec:symm}.

In both cases an interesting variant of the problem is to look at the subset of models with $A_i^2=B_j^2=\idty$ and/or a pure state $\rho$. Also it may be of interest to fix the observables, and consider the set of correlations resulting from varying $\rho$.

\subsection{Fixed State}\label{sec:fixedState}
\begin{figure}\centering
	\includegraphics[width=7cm]{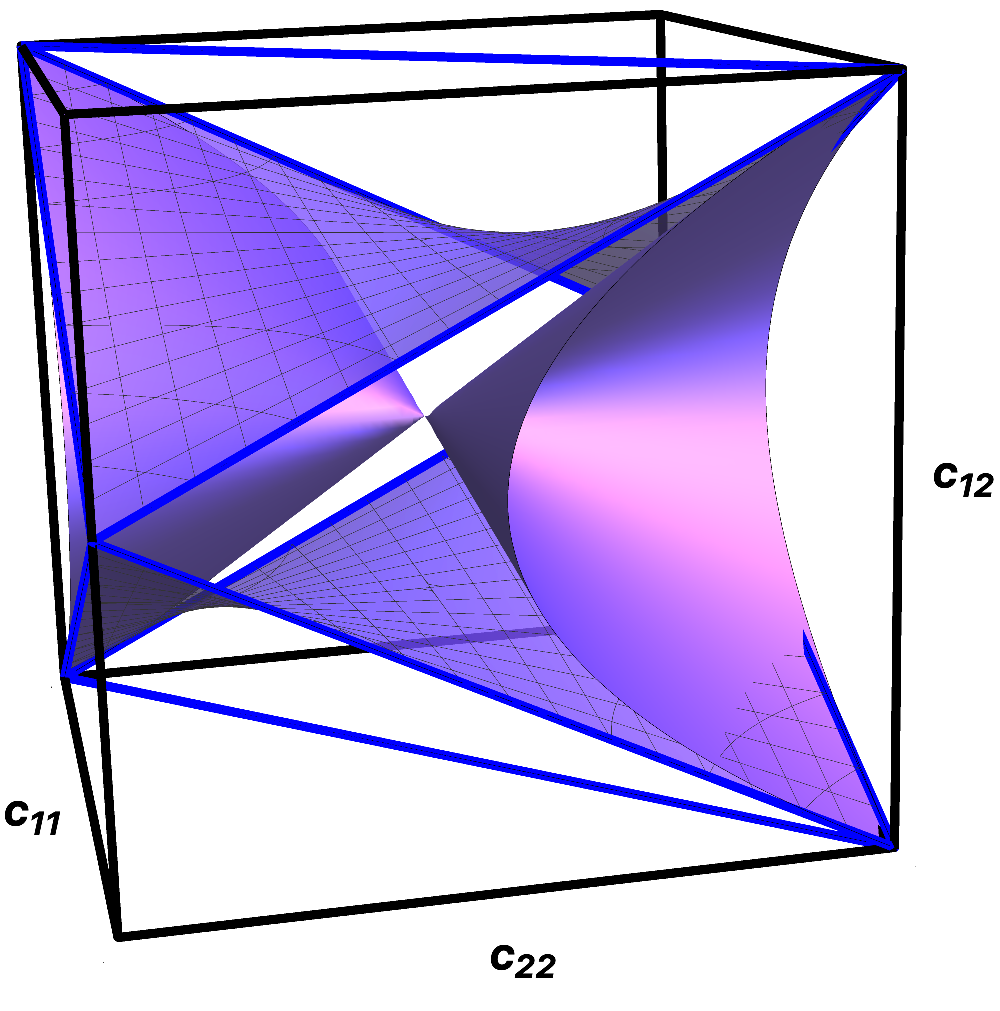} 
	\caption{The set $\Cl_1\equiv Q(\rho_A\otimes\rho_B)=\Ns\cap\{c:\det(c)=0\}$ in the Alice-Bob symmetric plane. The black bounding cube is $\Ns$, and the edges of $\Cl$ are given as blue lines. The section of $\Cl_1$ with this plane is the (purple) double cone $c_{11}c_{22}=c_{12}^2$ emanating from the origin. The projection contains additional points bounded between the double cone and the (top and bottom) surfaces $2\abs{c_{12}}-1=c_{11}c_{22}$.}
\label{fig:C1}
\end{figure}

For many applications it is crucial that no assumption about the quantum state is made: the linear bounds given by $f\in\Qm\polar$ hold for {\it any} quantum state. In some applications, however, the state is well-known and the question is what correlations might be realized on that basis. The question really only makes sense, when the Alice-Bob split is also known, so we ask that $A_i\in\Alg$, $B_j\in\BB$, where $\Alg$ and $\BB$ are commuting operator algebras, typically even a tensor product split $\Alg=\BB(\HH_A)\otimes\idty$ and $\BB=\idty\otimes\BB(\HH_B)$. We thus consider
$$
\Qm(\rho) := \{c_{ij}\in\Rl^4: c_{ij}=\tr\rho A_iB_j, -\idty\leq A_i,B_j\leq\idty, A_i\in\Alg, B_j\in\BB\}\,.
$$
This is, in general, not a convex set. For example, when $\rho$ is a product state, $c_{ij}=a_ib_j$ factorizes, and $\Qm(\rho)=\Qm_1=\Cl_1$ in the sense of the previous section (see also \autoref{fig:C1}). For entangled states close to that, the non-convexity will persist, even though there are some proper quantum correlations. A closed form or complete algebraic description of $\Qm(\rho)$ can hardly be expected, so we just make a few remarks.

First, if we fix a a linear functional $f$, finding the maximal quantum value is a bilinear optimization problem. We can fix all observables except one, e.g. $A_i$, and isolate the linear functional by which it enters, i.e. find $X_i$ such that $\sum_{j}f_{ij}\tr \rho (A \otimes B_j)=\tr X_i A$ or equivalently $X_i=\sum_jf_{ij}\tr_B(\rho \idty\otimes B_j)$. Maximizing over $A_i$, is now the problem of finding the $A$ with $-\idty\leq A\leq\idty$ for which $\tr(A X_i)$ is maximized. Clearly, that is $A_i=\sign(X_i)$ in the functional calculus of the hermitian operator $X_i$. The maximum is then the trace norm of $X_i$, which would seem to eliminate the $A_i$. However, it is better to just fix the optimal $A_i$, and optimize again with respect to the $B_j$ in the next step, in a see-saw style. This works nicely in practice, and improves the target functional in every step. However, this iteration has no convergence guarantee.

The second remark is that we can give a ``closed solution'' in the case
of a $2\otimes2$-dimensional Hilbert space, the minimal dimension where
nonclassical correlations could manifest by~\reff{prop:dimH}. This
generalizes a well-known formula by the Horodeckis
\cite{Horodecki2qubits} for the maximal CHSH-violation of a given
two-qubit state to an arbitrary linear inequality. As in the case of the
Horodecki formula, the result depends only on the two largest singular
values of  the matrix $R$ described in the proposition.

\begin{figure}[t]\centering
	\includegraphics[width=6cm]{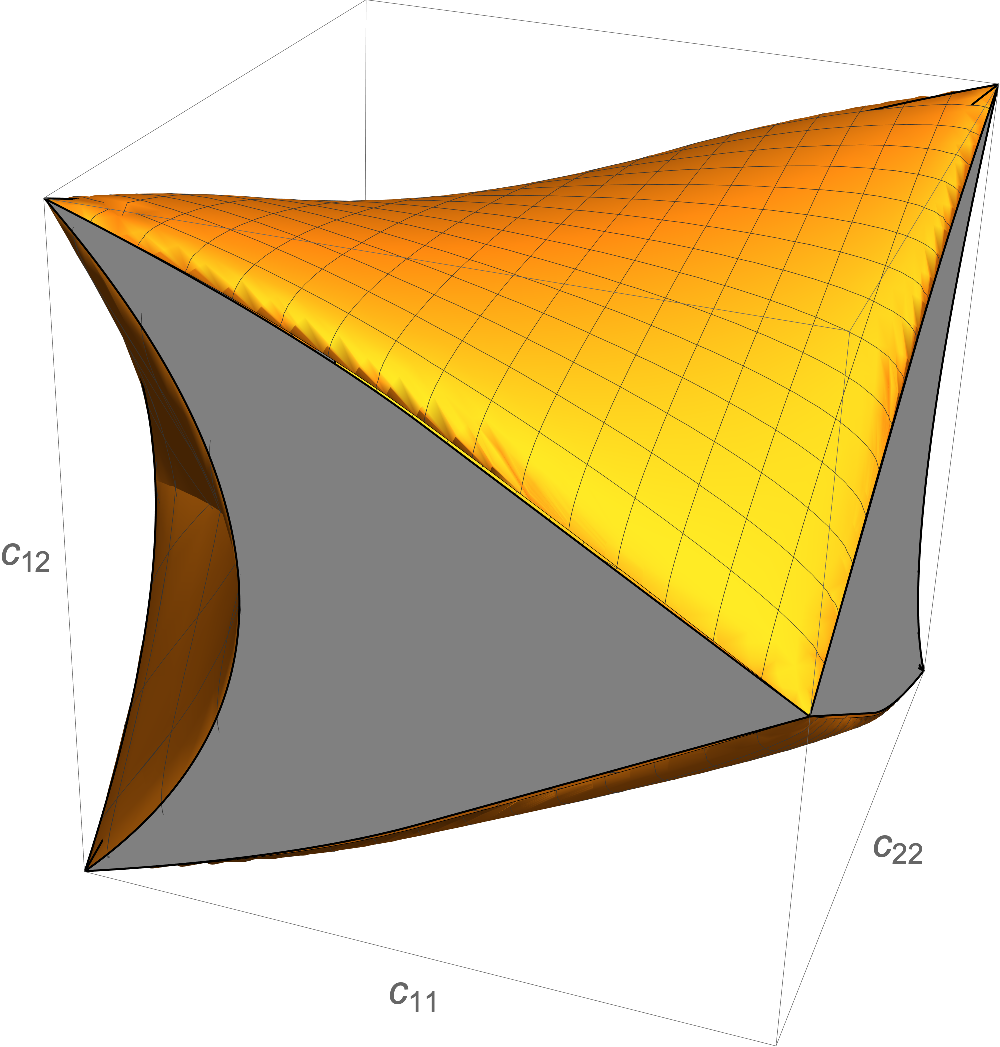} 
	\caption{The projection of the set $\Qmst(1,0.5)$ of \reff{prop:fixrho} to the AB-symmetric plane. This is a 3-dimensional connected nonconvex semialgebraic set, the midpoint of a continuous family connecting
     $\Qmst(1,0)=\Cl_1$ from \reff{fig:C1} with $\Qmst(1,1)=\Qm$ from \reff{fig:symAB}. The two-parameter family is homogeneous: $\Qmst(t\lambda_1,t\lambda_2)=t \Qmst(\lambda_1,\lambda_2)$, which for $t\leq1/\sqrt2\approx.7$ is included in $\Cl$.}
\label{fig:fixedstate}
\end{figure}

\begin{prop}\label{prop:fixrho}\proofin{sec:fixedStateproof}
Let $\rho$ be a density operator on $\HH=\Cx^2\otimes\Cx^2$, and let $\lambda_1\geq\lambda_2\geq0$ be the two largest singular values of the $3\times3$-matrix with entries
$R_{k\ell}=\tr[\rho\,(\sigma_k\otimes\sigma_\ell)]$ for Pauli matrices $\sigma_1,\sigma_2,\sigma_3$. Then the convex hull $\conv\Qm(\rho)=\conv\bigl(\Qmst(\lambda_1,\lambda_2)\cup\Cl\bigr)$, where $\Qmst(\lambda_1,\lambda_2)$ is the set of correlations of the form
\begin{equation}\label{cfixrho}
   c_{ij}=\lambda_1 \cos(\alpha_i)\cos(\beta_j)+\lambda_2 \sin(\alpha_i)\sin(\beta_j).
\end{equation}
Moreover, we have $\Qm(\rho)\subset\Cl$ if and only if $\lambda_1^2+\lambda_2^2\leq1$.
\end{prop}

\begin{figure}\centering
	\includegraphics[width=6cm]{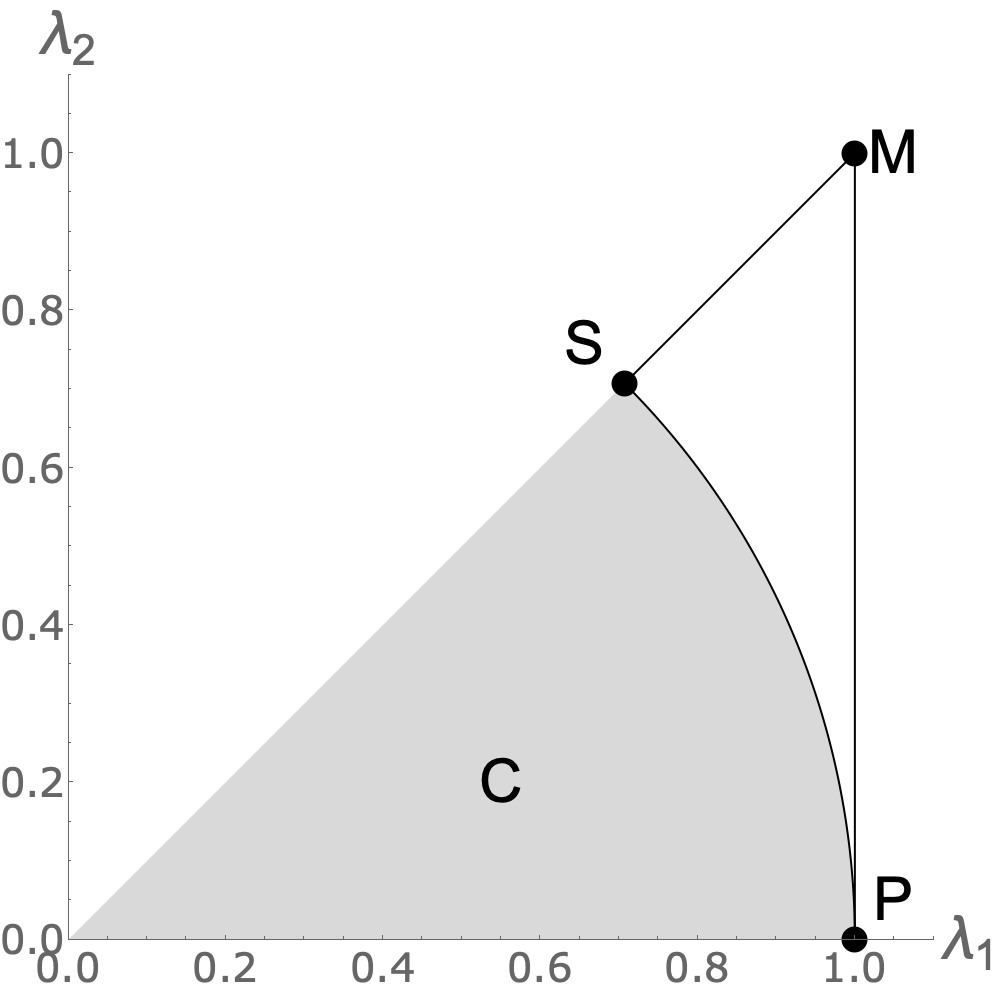} 
	\caption{The map of singular value pairs $(\lambda_1,\lambda_2)$ in \reff{prop:fixrho}. In the shaded region $C$, $\Qmst(\lambda_1,\lambda_2)\subset\Cl$. The labelled points are: M: the maximally entangled case, P: a pure product state, all pure two-qubit states, parametrized for this purpose by their Schmidt spectra, lie on the segment PM. The diagonal gives scaled versions of $\Qm$. At S the set just touches $\Cl$ from within. }
\label{fig:singval}
\end{figure}

The map of possibilities is shown in \reff{fig:singval}. The singular values for a vector with Schmidt spectrum $s_1\geq s_2\geq0$, e.g., $\psi=s_1\ket{++}+s_2\ket{--}$ where $\ket{\pm}:=(\ket{0}\pm\ket{1})/\sqrt{2}$ are $\lambda_1=s_1^2+s_2^2=1$ and $\lambda_2=2s_1s_2$, which parametrizes the line segment PM. The fact that all these states (other than P) give correlations outside $\Cl$, and hence violate the CHSH inequality, is known as Gisin's Theorem \cite{gisin1991}.

\subsection{Fixed Observables}\label{sec:fixedObs}
Dually we can fix the operators $A_i,B_j$, so that the resulting compact convex semialgebraic set
\begin{equation}
\Qm(A_i,B_j) := \{c_{ij}\in\Rl^4\,:\, c_{ij}=\tr\rho A_iB_j, \rho\geq0, \tr\rho = 1\}
\end{equation}
is just an affine image of the quantum state space, a spectrahedron. Still an explicit description may be difficult. As in the case of a fixed state, the boundary in any single direction, i.e., the maximization of any affine functional $f$ reduces to an eigenvalue problem, namely the largest eigenvalue of the operator $T(f)=\sum_{ij}f_{ij}\,A_iB_j$. Giving an algebraic solution might be harder because this corresponds to computing an algebraic formula for the largest root of a polynomial of high degree, depending on the dimension of the underlying Hilbert space.

The general case $-\idty\leq A_i,B_j\leq\idty$ can, in principle, be dealt with by a minimal dilation $U:\HH\to\KK$ so that $\Qm(A_i,B_j)$ is the affine image of a specific isometric copy of the state space of $\HH$ in $\KK$. This is harder to work out, so we give a characterization only for the special case $A_i^2=B_j^2=\idty$.
\begin{prop}\label{prop:fixedObs}\proofin{sec:fixedObsproof}
Let $A_i,B_j$ be hermitian operators on a finite dimensional Hilbert space with $[A_i,B_j]=0$ and $A_i^2=B_j^2=\idty$. Denote the spectrum of $(A_1A_2+A_2A_1)/2$ by $\Sigma_A$, and define $\Sigma_B$ analogously.
Then
\begin{equation}\label{transU}
\Qm(A_i,B_j)=\conv\{\Qmobs(u,v):=S(u)\QmObs S(v)^\top|\ u\in\Sigma_A,\ v\in\Sigma_B\},  \qquad \text{where}\ S(u)=\begin{pmatrix} 1 & 0 \\ u & \sqrt{1-u^2}\end{pmatrix},
\end{equation}
and $\QmObs$ is the set of $2\times2$ orthogonal matrices, i.e. the circles of rotations and reflections about some axes.
\end{prop}

There is a subtlety here when $A_i, B_j$ are ``genuinely infinite dimensional''. Then the \reff{prop:fixedObs} holds, if we interpret the convex hull as the closed convex hull and $\Qm(A_i,B_j)$ as the closure of the set of correlations attainable with arbitrary density operators. But for points in the continuous spectrum (not eigenvalues), certain boundary points will be unattainable by density operators in the given space. This is a straightforward consequence of the uniqueness theorem \reff{prop:selftest}. When some self-testing correlation inequality $f$ is attained, the maximum must be an eigenvector of the corresponding $T_f$. Such points can only be attained by singular states, i.e. positive normalized linear functionals not induced by density operators, because the continuous spectrum can be approximated by eigenvectors.

\begin{figure}[h]\centering
	\includegraphics[width=7cm]{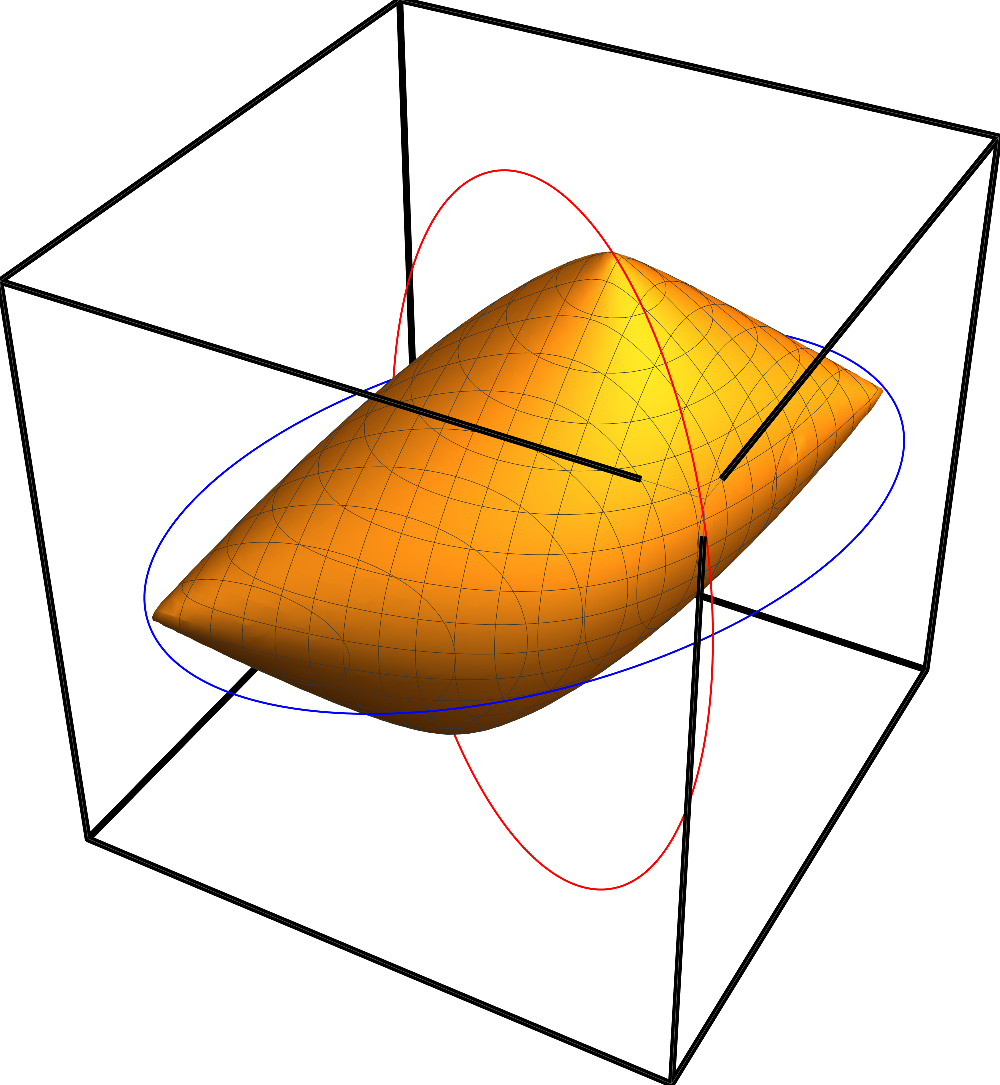} 
	\caption{The fixed observable set $\Qmobs(0.7,-0.9)$ of \reff{prop:fixedObs} in the Alice-Bob symmetric plane. The section with that plane is the orange body, while the projection is larger and is generated by the two circles shown. The red circle
     corresponds to maximal (positive) determinant, the blue one to minimal determinant.}
\label{fig:tilt}
\end{figure}

Since $\QmObs$ is the basic model for the fixed-observable correlation sets, we consider it in more detail.\\
\begin{prop}\label{prop:O2cando}\proofin{sec:fixedObsproof}
\begin{itemize}
\item[(1)] The set  $\Qmobs(0,0)=\conv\QmObs$ is the convex hull of two circles. The circles both have the origin as their center, and lie in orthogonal planes of $\Rl^4$.
\item[(2)] Both circles consist of extreme points, and any line connecting a pair of points from different circles is an exposed edge of $\Qmobs(0,0)$. The union of these edges is the entire boundary.
\item[(3)] The necessary and sufficient semialgebraic condition for $c\in\Qmobs(0,0)$ is that
\begin{equation}\label{Q00semalg}
   \ell(c):=1-(c_{11}^2+c_{12}^2+c_{21}^2+c_{22}^2)+(\det c)^2\geq0 \qquad\text{and}\qquad -1\leq \det c\leq1.
\end{equation}
\end{itemize}
\end{prop}

\begin{figure}[h!]\centering
	\includegraphics[scale=0.25]{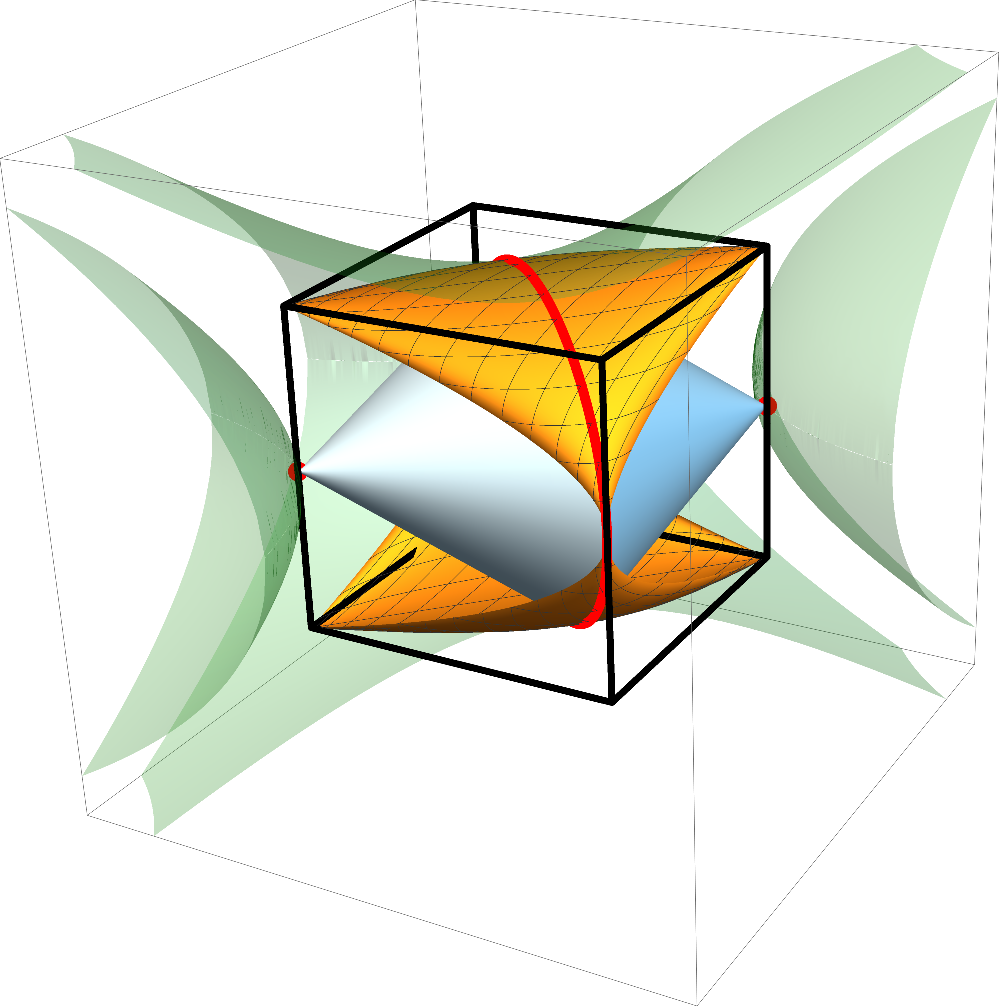} 
	\caption{The determinant function in an Alice-Bob symmetric plane.
Green: The two surfaces $\det c=-1$ (one-sheeted hyperboloid, cut open
for visibility) and $\det c=+1$ (two-sheeted hyperboloid touching $\Ns$
only in two points). Orange: the boundary of $\Qm$. The double cone
inside $\Qm$ is both a section and a projection of $\Qmobs(0,0)$. The
intersection of all three sets, i.e., the case of equality in
\reff{prop:det}  is the red line, together with the two cone tips.}
\label{fig:det}
\end{figure}

Since the observables used by Alice and Bob for $\Qmobs(0,0)$ are the same, the Alice-Bob swap symmetry leaves this set invariant, and so its intersection with the plane $\{c_{12}=c_{21}\}$ is at the same time the projection (see the double cone in \reff{fig:det}). On the other hand, for general values of $u$ and $v$ this is not true, as shown in \reff{fig:tilt}.

The above Proposition suggests that quantum points with $\abs{\det c}=1$ are self-testing.
\begin{prop}\label{prop:det}\proofin{sec:fixedObsproof}
For $c\in\Qm$, $\abs{\det c}\leq1$ with equality if and only if $c$ is an orthogonal matrix. In that case, in the unique cyclic quantum model for such correlations the identities  $A_1A_2=-A_2A_1$, and $B_1B_2=-B_2B_1$ hold, i.e., up to unitary isomorphism all measurements operators $A_i,B_j$ are Pauli matrices.
\end{prop}

\section{Proofs}\label{sec:proofs}

In this section we prove all theorems and propositions seen  so far. Since many results in our paper have appeared previously in the literature, we could give many proofs by citation. However, we aim to make our text self-contained. Where the pedestrian argument, tailored to the case at hand, can be understood as an example of a more general theory, we provide this background as well. As a help for monitoring the logical flow we included some summaries of what has been shown up to some point. These are formatted like the definitions and propositions throughout the paper, and numbered consecutively with these.

There are, of course, many ways to organize the proofs. We now briefly describe our overall strategy and the structure of
\reff{sec:proofs}.
We begin in \reff{sec:pround} with the equivalences of \reff{thm:main}, centered around the matrix completion problem. Self-duality (\reff{thm:selfdual}) follows in \reff{sec:self-dual}.
On this path we already need some information about the boundary (e.g., in \reff{sec:cos_param}), which is extended to the detailed classification of boundary points in \reff{sec:dualMcomplete}. The study of boundary points is perhaps a bit more detailed than necessary, since it uses a local criterion for excluding certain points from the extreme boundary of the convex hull of a variety. This technique might be helpful more generally. \reff{sec:supproof} contains the computation of the support function (\reff{prop:hom}), which proved to be more subtle than expected, even with a full understanding of the boundary. Finally the quantum properties, self-testing and all that, are established in \reff{sec:qproofs}.

\vspace{-0.08in}

\subsection{Proof of Theorem 1}\label{sec:pround}
For the sake of this proof let us denote by $\Qset x\subset\Rl^4$ the set characterized by item $(x)$ in \reff{thm:main}. We have to show that these are all equal. The backbone of the proof is the chain
\begin{equation}\label{thmChain}
  \Qset a \subset \Qset e = \Qset d = \Qset b \subset \Qset a
\end{equation}
and separate arguments for $\Qset e = \Qset f$ and $\Qset b = \Qset c$. The main work will be getting the solution set $\Qset e$ of the matrix completion problem in great detail. The boundary information coming out of that, in particular for $\rank C=2$, will then be used to get the further equivalences and finally go back to $\Qset a$.

\subsubsection{Making matrix completion real}\label{sec:Peasy}
The inclusion $\Qset a\subset\Qset e$ is an important step, because the definition of  $\Qset a$ admits infinite Hilbert space dimension, while $\Qset e$ only allows finite dimension. This reduction step works for more parties, settings, and outcomes, as well, which is the whole point of the semidefinite hierarchies \cite{doherty2008, navascues2008}. But outside the minimal scenario the inclusion is strict.

We saw the inclusion $\Qset a\subset\Qset e$ at the beginning of \reff{sec:specshadow}. From the quantum model we naturally get a positive definite matrix with some unknown complex entries, and diagonal not equal to $1$. These assumptions are part of the description of $\Qset e$, however. In order to prove $\Qset a\subset\Qset e$, we thus have to make sure that the additional assumptions do not make $\Qset e$ smaller. This is the content of \reff{prop:Cpos}.
\begin{proof}[of \reff{prop:Cpos}]
\indent Consider a matrix $C\geq0$ of the form \eqref{bigC} with diagonal entries $d_i\leq1$, and let $C'=\re(C)=(C+\overline{C})/2$ be its entrywise real part. Then $C'\geq0$ since the complex conjugate preserves positive semidefiniteness. Add to $C'$ the matrix with diagonal entries $1-d_i$ to get $C''$. Then $C''\geq C'\geq0$ is a matrix with the same off-diagonal $c_{ij}$, but in the standard form with $u,v$ real and $d_i=1$.
\end{proof}

\subsubsection{Solving the real  completion problem}\label{sec:compl_probl}
In this and the following two sections, we proceed to actually solve the matrix completion problem defining $\Qset e$, i.e. decide which $c$   permit real values $u,v$ that make the following matrix positive semidefinite:
\begin{equation}\label{eq:specshadow}
C= \left(\begin{matrix}
    1 & u & c_{11} & c_{12} \\
    {u} & 1 & c_{21} & c_{22} \\
    c_{11} & c_{21} & 1 & v \\
    c_{12} & c_{22} & {v} & 1\end{matrix}\right)\,.
\end{equation}
Equivalently, this is the decision problem for membership in the set $\{C|C\geq0\}$, which is clearly the intersection of the cone of positive semidefinite with the hyperplane $\{M|\diag(M)=(1,1,1,1)\}$. The solution uses Sylvester's criterion. We use the standard notation $C_I$ for the principal submatrix selecting the rows and the columns specified by the indices in $I$, and $m_I=\det C_I$ for the corresponding principal minor. Sylvester's necessary and sufficient criterion for $C\geq0$ is that $m_I\geq0$ for all $2^n-1$ index sets $I$, while for  $C>0$ is that $m_1,m_{12},m_{123},m_{1234}>0$.
The positivity of principal $2 \times 2$-minors gives $c_{ij}^2\leq 1$ or equivalently $\Qmc\subset\Ns$ as a necessary condition. Consider next the principal $3\times3$-minors
\begin{equation}\label{C3x3}
  C_{123}\,=\,\begin{pmatrix}
            1&u&c_{11}\\
            u&1&c_{21}\\
            c_{11}&c_{21}&1\end{pmatrix}\,\geq \,0 \qquad\mbox{and}\qquad
 C_{124}\,=\,\begin{pmatrix}
            1&u&c_{12}\\
            u&1&c_{22}\\
            c_{12}&c_{22}&1\end{pmatrix}\,\geq\,0.
\end{equation}
These do not involve $v$, and they give a condition for $u$. We show that \eqref{C3x3} is all we need to consider:

\begin{lem}\label{lem:Qu=Qe}
Suppose that, for some $u\in\Rl$, the matrices in \eqref{C3x3} are positive semidefinite.
Then one can find $v$ such that in \eqref{eq:specshadow} we have $C\geq0$, i.e., $c\in\Qmc$.
\end{lem}
\begin{proof}\\
\indent We consider first the case of strict positivity, i.e., we are given $u\in\Rl$ satisfying~\eqref{C3x3} and additionally $C_{123}>0$ and $C_{124}>0$. This necessarily implies $\abs u<1$, by Sylvester's criterion. So far we have $m_1,m_{12},m_{123}>0$ so by Sylvester's criterion, we only need to find $v$ such that $m_{1234}=\det C>0$ in order to guarantee $C>0$ for this $v$ and the given $u$. Note that $\det C$ is a quadratic polynomial in $v$ with a negative leading coefficient; therefore, if the desired $v$ exists, the $v$ maximizing this polynomial will do just as well. This is
$v = (c_{11}c_{12}+c_{21} c_{22}-(c_{11} c_{22}+c_{12} c_{21})u)/(1-u^2)$, and at that point we can compute the determinant and get
\begin{equation}\label{minorId}
  \max_{v\in\Rl}m_{1234}=(1-u^2)^{-1} m_{123}m_{124}>0.
\end{equation}
This choice of $v$ gives a completion $C>0$ which is stronger than $C\geq0$.

In general, we are only given $u\in\Rl$ satisfying~\eqref{C3x3}. To reduce to the previous case, we consider a family $u(\lambda)\in\Rl$ satisfying $C_{123}(\lambda)>0$ and $C_{124}(\lambda)>0$, that depends on a parameter $0<\lambda<1$, constructed as follows: $C_{123}(\lambda) = \lambda C_{123} + (1-\lambda)\idty$, ditto for $C_{124}(\lambda)$, and with $u(\lambda)=\lambda u$. Applying our previous result, there exists $v(\lambda)$ making a completion $C(\lambda)>0$. Observe that the $v(\lambda)$ that we took is then a continuous function of $\lambda$. Taking the limit $\lambda\to1$, we have $u(\lambda)\to u$ by construction, $v(\lambda)\to v$ for some $v\in\Rl$ by continuity, and $C(\lambda)\to C\geq0$. The principal submatrices 123 and 124 of $C$ coincides with the given $C_{123},C_{124}$ we started with. Thus we found $v\in\Rl$ giving a completion $C\geq0$ and the proof is complete.
\end{proof}

\begin{figure}[h]\centering
	\includegraphics[width=1.8cm]{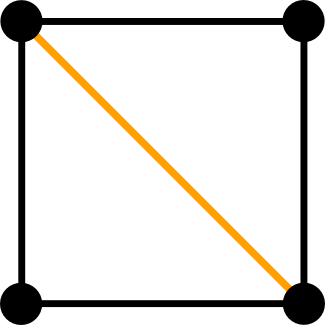}
	\caption{The black graph is the $4$-cycle associated to $\Qmc$ with vertices $1$, $3$, $2$, $4$ starting from the upper-left corner.
	It is not chordal.
	On the other hand, adding the orange edge, which corresponds to assuming the existence of $u$ satisfying \eqref{C3x3}, makes it chordal.}
\label{chordal_graph}
\end{figure}

Alternatively, we could invoke a  general result on semidefinite matrix completion \cite[Theorem~7]{wolk}:
Consider the graph $G$ whose edges represent the given entries of a partial matrix whose completion we seek.
The obvious necessary conditions for completability are that for those subsets of vertices, where all matrix
entries are specified (cliques of $G$),
the corresponding submatrices are positive semidefinite. Then, if the graph is {\it chordal} (meaning any cycle of length $\geq4$ allows a shortcut),
this condition is also sufficient. Moreover, this holds for both strict positive definiteness and semidefiniteness.
In the case at hand, the graph given for $\Qmc$ is a $4$-cycle shown in \autoref{chordal_graph}. Assuming the existence of $u$ with $C_{123}, C_{124}\geq0$ adds a diagonal, making the graph chordal. Hence no further condition needs to be considered.

\reff{lem:Qu=Qe} eliminates $v$  and leaves us with
the nonnegativity of the following three principal minors:
\begin{equation}\label{threequad}
  \begin{array}{rcl}
    m_{12}  \,\, = &1-u^2                                            &\qquad(A)\\
    m_{123} \,\, = &1 - c_{11}^2 - c_{21}^2 + 2 c_{11} c_{21} u - u^2 &\qquad(B)\\
    m_{124} \,\, = &1 - c_{12}^2 - c_{22}^2 + 2 c_{12} c_{22} u - u^2 &\qquad(C)
  \end{array}
\end{equation}
We have $c \in\Qmc$ if and only if these are simultaneously satisfied for the same $u$. By Helly's Theorem in $\Rl^1$~\cite{Barvinok}, the three positivity intervals have a common point
if and only if they intersect pairwise. Hence we can consider pairwise intersections.
The maximum of $(B)$ is
\begin{equation}\label{max123}
  \max_u m_{123}= (1-c_{11}^2)(1-c_{21}^2)=m_{13}m_{23}\geq0,
\end{equation}
attained at $u=c_{11} c_{21}\in[-1,1]$, so its positivity interval
intersects that of $(A)$, and similarly for the pair $(A,C)$.
Hence only the pair $(B,C)$ needs to be considered. We conclude:

\begin{sumry}\label{lem:BCintersect}
If $c\in\Ns$, then $c\in\Qmc$ if and only if  $(B)$ and $(C)$
 are both non-negative for some  $u\in\Rl$.
\end{sumry}

\subsubsection{Joint positivity of two parabolas}\label{sec:parab}

We next examine the criterion in \autoref{lem:BCintersect} independently of the specific context.
We do the quantifier elimination carefully, because the structure of the solution explains the disjunction in \reff{thm:main}~(d).
Thus at the end of this section we will achieve the equality $\Qset e=\Qset d$.

\begin{figure}[ht]\centering
	\includegraphics[width=8cm]{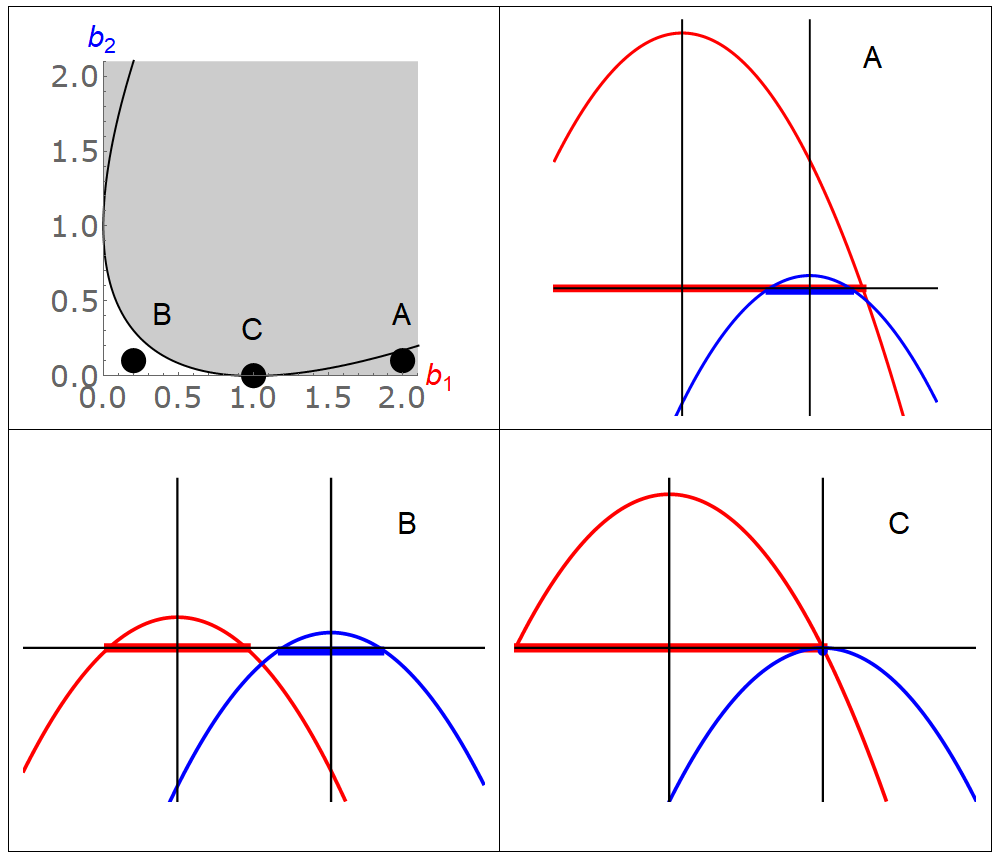}
	\caption{The configurations of two parabolas $f_i(x)=b_i-(x-a_i)^2$ with $a_1=0$ and $a_2=1$. Top Left: Parameter plane for $(b_1,b_2)$. Shaded: region, where intersection of positivity ranges is nonempty.
       Points {\sf A,B,C}: parameters for the other panels. Black Parabola: line at which the intersection of the parabolas lies on the $x$-axis.}
\label{fig:2para}
\end{figure}

The configurations of two parabolas  in the following lemma are shown in \reff{fig:2para}. Both are given by quadratic polynomials $f(x)=b-(x-a)^2$ with the same negative quadratic term.
The parameters are chosen so that $(x,y)=(a,b)$ is the location of the maximum. This function is positive in the interval $[a-\sqrt b,a+\sqrt b]$. The question is when for two such parabolas the positivity intervals overlap. It is clear that the problem is invariant under shifts (adding a constant to both $a_1$ and $a_2$), and $(a_1-a_2)^2$ just sets a scale for the $b$s. Hence we could choose $a_1=0$ and $a_2=1$. \reff{fig:2para} is drawn with this choice.

\begin{lem}\label{lem:2parab}
Given two quadratic polynomials $f_i(x)=b_i-(x-a_i)^2$, $\,i=1,2$, the following are equivalent:
\begin{itemize}
\item[(1)] There exists a point $u \in \Rl$ such that $f_1(u)\geq0$ and $f_2(u)\geq0$.
\vspace{-0.08in}
\item[(2)] $b_1\geq0
\,\land\, b_2\geq0 \,\land \,
\left[\,b_1+b_2-(a_1-a_2)^2\geq0\,\lor \,
4 b_1 b_2 - (b_1+b_2-(a_1-a_2)^2)^2 \geq0\,\right]$.
\end{itemize}
If one demands strict inequality then the following are also equivalent:
\begin{itemize}
\item[(1')] There exists a point $u\in \Rl$ such that $f_1(u)>0$ and $f_2(u)>0$.
\vspace{-0.08in}
\item[(2')] $b_1>0\,\land \,b_2>0\,\land\,
\left[\,b_1+b_2-(a_1-a_2)^2\geq0\,\lor\,4 b_1 b_2 - (b_1+b_2-(a_1-a_2)^2)^2 >0\,\right]$.
\end{itemize}
\end{lem}

There is a very simple algebraic proof of this result which starts by solving $f_i(u)\geq0$ for the domain of $u$ that ensures positivity. Here we instead present a geometric proof based on \reff{fig:2para}. This proof contains additional insights that allow verifying that $\Qm$ is semialgebraic but not basic semialgebraic, and the semialgebraic description is not unique.

\begin{proof}\\
\indent Clearly, both positivity ranges must be non-empty. Therefore, it is necessary that $b_1,b_2\geq0$. We can also trivially take care of the cases with $a_1=a_2$, because then one positivity interval is contained in the other.
The maximum of each parabola is clearly in the positivity interval. So if $b_1\geq(a_1-a_2)^2$ the maximum of the second parabola is in the positivity range of the first, and so there is a non-empty intersection. Symmetrically, this holds for $b_2\geq(a_1-a_2)^2$. This gives two closed rectangles in the $(b_1,b_2)$-plane with non-empty intersection.

Next consider the intersection of the two parabolas. Their unique intersection point is
\begin{equation}\label{xsprab}
 \bigl(x_s,f_i(x_s)\bigr)\,=\,\left(\frac{a_1+a_2}2-\frac{b_1-b_2}{2(a_1-a_2)},\ \frac{4 b_1 b_2-\bigr(b_1+b_2-(a_1-a_2)^2\bigl)^2}{4(a_1-a_2)^2}\right).
\end{equation}
Now suppose $f_i(x_s)\geq0$. Then $x_s$ is in the intersection of the positivity ranges. The corresponding region, defined by the positivity of the numerator of $f_i(x_s)$ in \eqref{xsprab} is the closed parabola in the top left panel of \reff{fig:2para}. Hence we have  non-zero intersection for the union of the two rectangles and the parabola in \reff{fig:2para}. This can be simplified as the union of just two regions, namely the parabola and the region $\{b_1+b_2\geq(a_1-a_2)^2\}$, which contains the two rectangles plus a triangle, which is contained in the parabola.

To complete the proof of the first part, we need to show that, for any point outside this region, the positivity ranges have empty intersection. This complement is defined by the conditions
$0\leq b_i<(a_1-a_2)^2$ and $f_i(x_s)<0$. Now the first condition implies  $\abs{b_1-b_2}<(a_1-a_2)^2$. Hence  the second term for $x_s$ in \eqref{xsprab} is bounded by $\abs{a_1-a_2}/2$,
so $x_s$ lies between $a_1$ and $a_2$. Therefore one of the maxima lies to the right of $x_s$ and the other one lies on its left (as depicted in panel B of~\reff{fig:2para}), and therefore the same holds for the positivity ranges of the parabolas. Since $f_i(x_s)<0$ the ranges do not intersect. This completes the proof of the first part.

The primed statements characterize the interior of the parameter set, i.e., the region just described without the boundary points, for which $b_1=0$, or $b_2=0$ or $f_i(x_s)=0$.
\end{proof}

We now apply \autoref{lem:2parab}  to the
quadratic polynomials (B,C) in \eqref{threequad}. The
 parameters are
\begin{equation}\label{22ab}
 \begin{array}{rlrl}
   a_1 &= \,\,c_{11}c_{21},& a_2 &=\,\,c_{12}c_{22}, \\
   b_1 &=\,\,(1-c_{11}^2)(1-c_{21}^2), \qquad   &b_2&=\,\,(1-c_{12}^2)(1-c_{22}^2).
 \end{array}
\end{equation}
With this, the two polynomials in \autoref{lem:2parab}~(2) are $g$ in \eqref{polyg}
and $h$ in  \eqref{polyh}.  The combination of \autoref{lem:BCintersect} and \autoref{lem:2parab} now
establishes the equivalence of $(d)$ and $(e)$ in
\autoref{thm:main}. Furthermore, if $\pi:\Rl^6\to\Rl^4$ denotes the projection $\pi(c,u,v)=c$, where we identify $\Rl^6$ with the real hermitian $4\times4$-matrices with unit diagonal, then $\mathrm{int}(\Qset e) = \mathrm{int}(\pi(\{C\geq0\})) = \pi(\mathrm{int}(\{C\geq0\})) = \pi(\{C>0\})$. Thus, we have that the strict inequality result in~\autoref{lem:2parab} gives the characterization of the interior. We record this as follows:

\begin{sumry}\label{lem:Qmu} $\Qset e=\Qset d$: For  $c \in \Rl^4$, we have
$c\in\Qmc$ if and only if $c\in\Ns$ and $\bigl(g(c)\geq0\ \text{or}\ h(c)\geq0\bigr)$.
Furthermore, we have $c\in\mathrm{int}(\Qmc)$ if and only if  $c\in\mathrm{int}(\Ns)$ and $\bigl(g(c)\geq 0\ \text{or}\ h(c)>0\bigr)$.
\end{sumry}

\subsubsection{Properties of the boundary and extreme points}\label{sec:cos_param}

\autoref{lem:Qmu} characterizes membership in the spectrahedral shadow $\Qmc$ and in its interior. We now come to its boundary which is the complement $\Qmc\setminus\mathrm{int}(\Qmc)$, beginning with an extended version of the first statement in
\autoref{prop:Cposb}.

\begin{lem}\label{lem:Qmc}
Given any point $c \in \Qmc$, the following three conditions are equivalent: \vspace{-0.07in}
\begin{itemize}
\item[(1)] $c\in\partial\Qmc$. \vspace{-0.08in}
\item[(2)] The matrix completion problem for $c$ has a unique solution $(u,v)$. \vspace{-0.07in}
\item[(3)] Either ($\,c\in\partial\Ns$ and $g(c)\geq 0 \,$)
\ or \ ($\,g(c)<0$ and $\rank(C) \leq  2$ for every completion $C$ of $c$).  \vspace{-0.1in}
\end{itemize}
\end{lem}

\begin{proof}\\
\indent As before, let $\pi:\Rl^6\to\Rl^4$ denote the projection $\pi(c,u,v)=c$, where we identify $\Rl^6$ with the real hermitian $4\times4$-matrices with unit diagonal.
\vskip7pt\noindent{\it not(1)$\Rightarrow$not(2)}\\
Suppose that $c$ is in the interior of $\Qmc$. Then we can write $c=\lambda 0+(1-\lambda)c''$ for some other $c''\in\Qmc$ and $0<\lambda\leq1$. Then if $\pi(C'')=c''$, we get $C=\lambda \idty+(1-\lambda)C''$  with $\pi(C)=c$. This $C$ has rank 4 since the eigenvalues is $\lambda+(1-\lambda)\mathrm{eig}(C'')$, and a small ball $\BB$  around $C$ is still contained in the positive cone. Therefore the intersection $\pi^{-1}(c)\cap\BB$ is a two-dimensional set, in the parameters $(u,v)$, of matrix completions for $c$.

\vskip7pt\noindent{\it not(1)$\Rightarrow$not(3)}\\
The first condition in (3) fails because ${\rm int}(\Qmc) \subset{\rm int}(\Ns)$, and the second one fails because the preimage $C$ just constructed has rank $4$.

\vskip7pt\noindent{\it((1) and $c \in \partial \Ns$)$\Rightarrow$((2) and (3))}\\
When $c$  is in the boundary of the cube we must have $c_{ij}=\pm 1$ for some $(i,j)$. Up to symmetry we may assume $c_{11}=-1$.
Then from the form \eqref{threequad} of the principal minors we get that $m_{123}(u)=-(u+c_{21})^2$ and $m_{134}(v)=-(v+c_{12})^2$. These must be non-negative for any matrix completion, fixing  $u=-c_{21}$ and $v=-c_{12}$.
So (2) holds. Moreover, from \eqref{polyh1} we see that $h(c) = -g(c)^2\leq0$. With $h(c)<0$, \autoref{lem:Qmu} implies $g(c)\geq0$, and with $h(c)=0$ this is anyhow true. So the first alternative in (3) holds.

\vskip7pt\noindent{\it((1) and $c \notin \partial \Ns$)$\Rightarrow$((2) and (3))}\\
By negation, the second sentence in \autoref{lem:Qmu} provides a characterization of $c\in\partial\Qmc$.
Together with $c \notin \partial \Ns$  this implies ($g(c)<0$ and $h(c)\leq0$).
On the other hand,  $g(c)\geq0$ or $ h(c)\geq 0$ from $c\in\Qmc$. Together we have $h(c) = 0$. Hence  the parabolas from \autoref{lem:2parab} have a unique point of intersection at level $0$.
This point is given either by
$u=a_1-\sqrt{b_1}=a_2+\sqrt{b_2}$, or $u=a_2-\sqrt{b_2}=a_1+\sqrt{b_1}$.
Analogously there is also a unique $v$, and (2) holds. Moreover, these unique choices $u,v$ make all third order principal minors $m_{123},m_{124},m_{234},m_{134}$ vanish.
This means  ${\rm rank}(C) \leq 2$, and we obtain the second sentence of (3).
\end{proof}

The next lemma shows how the trigonometric functions arise in the parametrization of the boundaries.

\begin{lem}\label{lem:cosfromrank} \,
\begin{itemize}
\item[(1)] If $c$ is extremal in $\Qmc$ then it must have a unique matrix completion with $\rank C\leq2$.
\item[(2)] It must then be of the form $c_{ij}=a_i\cdot b_j$ for unit vectors $a_1,a_2,b_1,b_2$ in the Euclidean space $\Rl^2$.
\item[(3)] This means that it can be written as $c_{ij}=\cos(\alpha_i-\beta_j)$ for some $\alpha_1,\alpha_2,\beta_1,\beta_2\in\Rl$.
\end{itemize}
\end{lem}

\begin{proof}

(1) Of course, extreme points are always part of the boundary, because any line segment through an extreme point necessarily contains some points outside the body. Hence the previous Lemma applies. So by part (3) of the previous Lemma they either have $\rank C\leq2$ anyway, or else belong to a face of the cube $\Ns$. Up to symmetry this means $c_{22}=1$. Then $C_{124}\geq0$ forces $u=c_{12}$, so the extendability is just the elliptope condition $\det C_{123}=1-c_{11}^2-c_{12}^2-c_{21}^2+2c_{11}c_{12}c_{21}\geq0$. Since $c$ is furthermore extremal, it remains extremal in this face (direct verification). This implies $c$ must belong to the boundary of the elliptope, namely this inequality must be tight, and so $\rank C_{123}\leq2$. Since all $3\times3$ minors of $C$ are zero, $\rank C\leq2$.

(2) By the spectral theorem, every positive semidefinite $d\times d$ matrix can be written as a Gram matrix, i.e, $C_{\alpha\beta}=w_\alpha\cdot w_\beta$ for vectors $w_1,\ldots,w_d$  in some Euclidean space $\Rl^r$. Here $r=\rank C$ is the number of non-zero eigenvalues. In our case the diagonal matrix entries are $1$, so these vectors are unit vectors. Moreover, the $c_{ij}$ are themselves
matrix elements of $C$, so we just need to rename the vectors according to whether the dimension belongs to Alice or to Bob, i.e.,  $a_1=w_1$, $a_2=w_2$, $b_1=w_3$, and $b_2=w_4$.

(3)  Unit vectors in $\Rl^2$ lie on the unit circle, parameterized by angles. Scalar products between such vectors are the cosines of the enclosed angle.
Setting $w_i=(\cos\alpha_i,\sin\alpha_i)$ for some $\alpha_i\in\Rl$, we thus have

\begin{equation}\label{eq:uniquecompletion}
C\,\,= \,\,\begin{small} \begin{pmatrix}
1 & \cos(\alpha_1-\alpha_2) & \cos(\alpha_1-\alpha_3) & \cos(\alpha_1-\alpha_4) \\
\cos(\alpha_1-\alpha_2) & 1 & \cos(\alpha_2-\alpha_3) & \cos(\alpha_2-\alpha_4) \\
 \cos(\alpha_1-\alpha_3) & \cos(\alpha_2-\alpha_3) & 1 & \cos(\alpha_3-\alpha_4) \\
\cos(\alpha_1-\alpha_4) & \cos(\alpha_2-\alpha_4) & \cos(\alpha_3-\alpha_4) & 1
\end{pmatrix}. \end{small}
\end{equation}
Renaming according to the Alice/Bob distinction, i.e. $\alpha_3=\beta_1$ and $\alpha_4=\beta_2$,
we obtain the claim.
\end{proof}

Part (2) of \reff{lem:cosfromrank}
 says that $\ext\Qmc\subset\Qset f$, and  part (3) that $\ext\Qmc\subset\Qset b$. Since $\Qset f$ is convex by a direct sum construction, and $\Qset b$ is anyhow defined to be convex, this also means that $\Qmc\subset\Qset f$ and $\Qmc\subset\Qset b$. The first of these inclusions can be inverted trivially,  because we can just set $u=a_1\cdot a_2$ and $v=b_1\cdot b_2$ to get a matrix completion from the unit vectors $a_i,b_j$. To revert the second inclusion, we have to combine \reff{lem:Qmu}, the inclusion $\Qset a\subset\Qset e$ from \reff{sec:Peasy}, and the concrete quantum models from \reff{sec:quantummodels}, which realize all cosine parametrized $c$, and thus show $\Qset b\subset\Qset a$. Altogether this gives

\begin{sumry}\label{sum:minus1}
$\Qmc=\Qset f$, $\Qset e= \Qset d$, and $\Qset a\subset \Qset e\subset \Qset b\subset \Qset a$, i.e., all these sets are equal.
\end{sumry}

This leaves only one part of \reff{thm:main}, namely the equality with $\Qset c$. This will be shown in  \reff{sec:pushout}.
Also, we have established one part of \reff{prop:paramExt}, the cosine parametrization, in a slightly different parametrization. The parametrization used in
\reff{prop:paramExt} and \reff{prop:paramExtbd} follows from the form \eqref{eq:uniquecompletion} by setting
\begin{equation}\label{angless}
  \alpha=\alpha_1-\alpha_3,\quad \beta=\alpha_4-\alpha_1,\quad  \gamma=\alpha_3-\alpha_2,\quad  \delta=\alpha_2-\alpha_4.
\end{equation}
Note that the signs were chosen (not changing the respective cosine) such that $\alpha+\beta+\gamma+\delta=0$. This constraint reflects the fact that only differences $\alpha_i-\alpha_j$ enter. What we have not shown yet, however, is the criterion $\Delta<0$ for a tuple of angles to give an extreme point.
This will be done in \reff{sec:fnondeg}.

\subsubsection{The pushout characterization}\label{sec:pushout}

The continuous map $t\mapsto\sin(\pi t/2)$ takes the interval $[-1,1]$
to itself, and it has a continuous inverse. By applying this map
coordinate-wise, we conclude that the pushout map  in \eqref{defpush}
is a homeomorphism from the cube $\Ns = [-1,1]^4$ to itself.
Here we shall establish \autoref{prop:push}, which states that $\Qm=\Sin(\Cl)$. Note that in the literature of matrix completion, this is often presented as $\Qm=\Cos(\pi\mathrm{MET}(K_{2,2}))$ which is obvious if one realizes that the metric polytope $\mathrm{MET}(K_{2,2})$ of the complete bipartite graph $K_{2,2}$ is isomorphic to the classical polytope $\Cl$. See~\cite{L97} and references therein.

Our strategy is to show this
for the boundaries, i.e., $\Sin(\partial\Cl)=\partial\Qm$.
Suppose that we know this, then notice that $\Sin$ maps connected components of $\Ns \setminus \partial \Cl$ to connected components of $\Ns \setminus \partial \Qm$. Since $0$ is in both the classical polytope and the quantum set, and $\Sin(0)=0$ we get that $\Cl$ is precisely mapped to $\Qm$.
Hence equality of the boundaries is sufficient.

We examine the boundary of the demicube $\Cl$ facet by facet.
This task is greatly reduced by symmetry, given that
$\Sin$ commutes with all symmetry operations as explained in
\autoref{sec:symm}.
Only one CHSH face and one $\Ns$-face need to be considered.
Consider first the CHSH facet
$\,\{(x,y,z,w)\in[-1,1]^4:x+y+z-w=2\}$.
The images of the points on this facet under the pushout map $\Sin$ are
\begin{equation}
 \Sin(x,y,z,w)\,\,=\,\,
\left( \sin{\left(\frac{\pi}{2}x\right)}, \sin{\left(\frac{\pi}{2}y\right)},
\sin{\left(\frac{\pi}{2}z\right)}, \sin{\left(\frac{\pi}{2}w\right)} \right)
     \,\,=\,\,(\cos\alpha,\cos\beta,\cos\gamma,\cos\delta),
\end{equation}
where  $\alpha=\pi(1{-}x)/2$, $\beta=\pi(1{-}y)/2$, $\gamma=\pi(1{-}z)/2$,
$\delta=\pi(w{-}1)/2$. This gives $\alpha+\beta+\gamma+\delta=\pi -(x+y+z-w)\pi/2=0$.
Moreover,  $x,y,z,w\in[-1,1]$ implies $\alpha,\beta,\gamma\in[0,\pi]$ and $\delta\in[-\pi,0]$.
These inequalities imply $\Delta=\sin\alpha\,\sin\beta\,\sin\gamma\,\sin\delta\leq0$.
Hence we get exactly those parametrized patches \Qt4 from \autoref{prop:paramExt},
for which $\alpha,\beta,\gamma > 0 > \delta$.
The other seven patches \Qt4 are obtained by symmetry.

Next consider an $\Ns$-face of $\Cl$, say that defined
by $w=1$. The pushout map preserves this equation.
We can now go through the same considerations as above, but in
one dimension lower. The geometric statement
is that the pushout of the tetrahedron equals the elliptope
(see \autoref{fig:elliptope}). This also follows from the fact
that $\Sin$ identifies their boundaries. We know this
from the above discussion of the CHSH facets.

Together with \reff{sum:minus1} this completes the proof of \autoref{thm:main}.

\subsection{Self-duality}\label{sec:self-dual}

Self-duality is a special feature of the pair $\Qm,\Qm^\circ$. It is {\it not} the kind of property that tends to hold for any sufficiently simple
case. Indeed, in one dimension smaller, the elliptope $\mathcal{E}^3$ (see \reff{fig:elliptope}) is not self-dual: It has extreme points with a pointed normal cone, which translate to flat faces in $(\mathcal{E}^3)\polar$. But $\mathcal{E}^3$ has no such faces, and hence is not self-dual. For an illustration of this phenomenon see
\cite[Figure 5.2]{Blekherman}.

With the boundary information obtained so far one could try to derive self-duality extreme point by extreme point. We take a more global approach based on the duality of the semidefinite matrices $C$ and $F$, and their representations as Gram matrices. We begin with the characterization of the $F$ matrices.
\begin{proof}[of~\autoref{prop:dualMatComp}]
\indent The primal/dual characterization of spectrahedral shadows described briefly before the statement of
\reff{prop:dualMatComp} is a standard result, so we omit its proof and are left with the additional property $p_1+p_2=p_3+p_4$ (necessarily $=1$). We can take $p_1+p_2$ and $p_3+p_4$ both non-zero, since otherwise we have the trivial case $f=0$, $F=\frac12\idty$. Indeed, if $p_1+p_2=0$ then since $p_1,p_2\geq0$ it must be that $p_1=p_2=0$. Now analyze the positivity of the $2\times2$ minor containing $f_{11}$ to get $f_{11}=0$ and likewise for all $f_{ij}$. (The reader might be tempted to conclude $F=0_{2\times2}\oplus\idty_{2\times2}$, but this will not achieve what we want.) Forgetting what we have just argued, given $f=0$ there is clearly a completion $F=\frac12\idty$ that has the property $p_1+p_2=p_3+p_4=1$.

Now observe that $F'=\Lambda F\Lambda$, where $\Lambda$ is the diagonal matrix with diagonal $(\lambda,\lambda,\lambda^{-1},\lambda^{-1})$ is also positive, and has the same off-diagonal block $f_{ij}$.
The diagonal is changed to $(p'_1,p'_2,p'_3,p'_4)=(\lambda^2 p_1,\lambda^2 p_2,\lambda^{-2}p_3,\lambda^{-2}p_4)$. In order to satisfy the sum constraint we need $\lambda^2=(p_3+p_4)/(p_1+p_2)$. Then $(F+F')/2$ is the desired extension with $p_1+p_2=p_3+p_4$.
This is different from $F$, if $\lambda^2\neq1$.

Finally, using a similar argument, we show that if the point $f$ has a completion $F$ for which the condition $p_1+p_2=p_3+p_4=1$ does not hold, then $f$ is not extreme. Using the diagonal matrix with $(\lambda_A,\lambda_A,\lambda_B,\lambda_B)$, and assuming, without loss that $x=p_1+p_2-1>0$, the normalization condition becomes $\lambda_A^2(1+x)+\lambda_B^2(1-x)=2$. Then choosing both terms equal to $1$ maximizes $\lambda_A\lambda_B=(1-x^2)^{-1/2}>1$. We conclude that $f'_{ij}=\lambda_A\lambda_Bf_{ij}$ is again in $\Qm\polar$, so $f$ is not a boundary point.
\end{proof}

We now consider matrices $C$ and $F$ as Gram matrices of suitable vectors in a real Euclidean space. That is, we choose $a_1,a_2,b_1,b_2$ from the characterization (f) in \reff{thm:main}, and similarly
four vectors $x_1,x_2,y_1,y_2$ whose scalar products give $F$, and $-f_{ij}=x_i\cdot y_j$. The conditions for these vectors for $C$ to be of the form \eqref{eq:specshadow} and $F$ to be of the form \eqref{Fmatrix} with $p_1+p_2=1$ are
\begin{equation}\label{Gramcond}
  \norm{a_1}=\norm{a_2}=1 \qquad\mbox{and}\qquad x_1\perp x_2,\quad \norm{x_1}^2+\norm{x_2}^2=1,
\end{equation}
and analogous conditions for $b_j$ and $y_j$.
The sets of vectors can be mapped to each other by
\begin{equation}\label{rhombi}
   a_1=x_1+x_2,\quad a_2=x_1-x_2 \quad \mbox{and}\qquad x_1=(a_1+a_2)/2,\quad  x_2=(a_1-a_2)/2.
\end{equation}
With the analogous relations for the $b_j$ and $y_j$ we get a bijective correspondence between the allowed vectors. Expressing the relation by the $2\times2$-Hadamard matrix
$H_2=2^{-1/2}\left(\begin{smallmatrix}  1&1\\1&-1\end{smallmatrix}\right)$ and inserting the relations into $c_{ij}=a_i\cdot b_j$ this correspondence relates the $4$-vector $c=(c_{11},c_{12},c_{21},c_{22})$ to the corresponding $f\in\Rl^4$ by $H=H_2\otimes H_2$, i.e., \eqref{Hadamard}. This proves the duality theorems \reff{prop:duality} and \reff{thm:selfdual}.

\subsection{Further boundary properties}\label{sec:dualMcomplete}

The pushout characterization makes the disjoint union structure of \autoref{prop:boundary} obvious, but we need to verify various geometric properties of strata mentioned in that proposition.

\subsubsection{Classification of boundary points by rank}\label{sec:facerank}
We now show the rank statements in \reff{prop:Cposb}, and the parametrization of boundary points in \reff{prop:paramExtbd}. Rather than treating these separately, we prove them together, following the classification of boundary points
in \autoref{prop:boundary}. We organize these items firstly by the options given in \autoref{lem:Qmc}, and secondly by $\rank C$, where $C$ is a  matrix completion for $c$. Again, let us recall the projection $\pi:\Rl^6\to\Rl^4$ used in the proof of  \autoref{lem:Qmc}.

\vskip7pt\noindent{\bf \Qt6} $\,\Leftrightarrow\,$\ $\rank C$ may be $4$. \\
If $C$ has rank $4$, then so have the points in some open ball $\BB\ni C$. Since the projection is an open map,  $\pi(\BB)$ is open.  Hence $c=\pi(C)\in\pi(\BB)\subset\Qm$ is in the interior.
For the converse note that, for any pair of interior points $c,c'$, we can write $c$ as a convex combination in which $c'$ has positive weight. Hence if $c'$ has a full rank completion, so does $c$, i.e., every interior point.
(An explicit version of this, with $c'=0$, $C'=\idty$, was used in the beginning of the proof of \autoref{lem:Qmc}.)

\vskip7pt\noindent From here to the end of this proof, $c$ will be in the boundary, and so $c$ has a unique completion $C$ by \reff{lem:Qmc}. In the twofold case distinction of \reff{lem:Qmc}~(3) we now begin with the first case, $c\in\partial\Ns$, so that for at least one pair $(i,j)$ we have $\abs{c_{ij}}=1$. Up to symmetry we can take $c_{11}=1$. Then the vector $(1,0,-1,0)$ is in the kernel of $C$, and, consequently rows/columns $1$ and $3$ of $C$ are proportional, and $\rank C\leq3$.

\vskip7pt\noindent{\bf \Qt5} $\,\Leftrightarrow\,$\ $c\in\partial\Qm$ and $\rank C=3$.\\
Consider the submatrix $C_{124}$ of $C$ obtained by deleting the redundant $3^{\rm rd}$ row/column, that is the second matrix in \eqref{C3x3} with $u=c_{21}$. Its semidefiniteness means that $(c_{12},c_{21},c_{22})$ is in the closed elliptope.
It has full rank, i.e.,  $\rank C=\rank C_{124}=3$, precisely if $c$ is in the interior of this elliptope. Note that the second case of \reff{lem:Qmc}~(3) contributes no further cases with rank $3$, so that the implication ``$\Leftarrow$'' holds in all cases.

\vskip7pt\noindent From now on we have $\rank C\leq2$, and so by \reff{lem:cosfromrank}~(3), the matrix $C$ lies in the cosine-parametrized family written out in \eqref{eq:uniquecompletion}. We consider instead differences of angles as variables $(\alpha,\beta,\gamma,\delta)$, as in  \reff{prop:paramExt} and \reff{prop:paramExtbd}. Since $(c_{11},c_{12},c_{21},c_{22})=(\cos\alpha,\cos\beta,\cos\gamma,\cos\delta)$, the condition $c_{ij}=\pm1$ defining a facet of $\Ns$ just correponds to the corresponding angle being a multiple of $\pi$.

\vskip7pt\noindent{\bf \Qt3} $\,\Leftrightarrow\,$\ $c\in\partial\Qm$, $\rank C=2$, and for exactly one pair $(i,j)$ we have $\abs{c_{ij}}=1$. \\
                  {\bf \Qt2} $\,\Leftrightarrow\,$\ $c\in\partial\Qm$, $\rank C=2$, and for exactly two pairs $(i,j)$ we have $\abs{c_{ij}}=1$. \\
We are still in the closed elliptope (\reff{fig:elliptope}). The boundary is singled out either by the reduced rank, or equivalently the cosine parametrization with one angle a multiple of $\pi$. The generic boundary points are exposed in $3$ dimensions, but (as we will see in \reff{sec:normals}) not in $4$ dimensions. (This is a general feature of exposedness: it depends on the ambient convex set and space.) If $c$ in \reff{fig:elliptope} lies in a facet of the $3$-cube, i.e., one more angle becomes a multiple of $\pi$, we get a point on an edge of $\Qm$. This proves the two  equivalences above.

\vskip7pt\noindent{\bf \Qt1} $\,\Leftrightarrow\,$\ $c\in\partial\Qm$ and $\rank C=1$\\
Of course, we may set even more $\abs{c_{ij}}=1$. When three angles are multiples of $\pi$, so must be the fourth, since they add up to zero. This is also geometrically obvious from \reff{fig:elliptope}: The only points of $\Qm$ that lie on an edge of the $3$-cube are endpoints. So in this case we have a classical extreme point. It is easy to see that this implies rank $1$: Up to symmetry this is the case $c_{ij}=1$ for all $i,j=1,2$, which clearly has the rank $1$ matrix $C_{ij}\equiv1$ as a completion. For the converse, note that $\rank C=1$ and $C\geq0$ imply $C_{ij}=x_i\,x_j$ for some vector $x$, and $c_{ij}=x_i\,x_{j+2}$. If we now insist on the special form of $C$ from \reff{prop:Cpos}, i.e., unit diagonal, we must have an extreme point just from the rank condition. However, with $\abs{x_i}<1$, we get even many interior rank $1$ points. Therefore, in the above equivalence  $c\in\partial\Qm$ needs to be imposed as well.

\vskip7pt\noindent{\bf \Qt4} $\,\Leftrightarrow\,$\ $c\in\partial\Qm\cap({\mathrm{int}}\,\Ns)\Leftrightarrow$\, $\Delta<0$ in \reff{prop:paramExt}\\
Finally, we come to the other branch of~\autoref{lem:Qmc}~(3). From the semialgebraic description of $\Qm$ in \reff{prop:semialg} this type consists of the points $c\in\Ns$ with $h(c)=0$ and $g(c)<0$. But for the cosine parametrized correlations with $\delta=-(\alpha+\beta+\gamma)$ we get the identity
\begin{equation}\label{gcosines}
  g(\cos\alpha,\cos\beta,\cos\gamma,\cos\delta)= 2\Delta.
\end{equation}

This completes the proofs of  \reff{prop:Cposb} and \reff{prop:paramExtbd}.

\subsubsection{Exposing functionals: first order analysis}\label{sec:fnondeg}
\begin{proof}[of \reff{prop:paramExtf}]
Since $c$ is a boundary point, we have  $c\cdot f=\max\{c\cdot f'|f'\in\Qm\polar\}=1$ for
 some $f\in\Qm\polar$. Indeed, if the maximum over the compact  set $\Qm\polar$ were $M<1$, we would have $c/M\in\Qm\bipolar=\Qm$, and $c$ would be in the interior. We fix some such $f$.
Thus $c'\cdot f$ has a global maximum at $c'=c$, and therefore a local maximum on the surface $c_{ij}=\cos\theta_{ij}$ with $\sum_{ij}\theta_{ij}=0$ where the $\theta_{ij}$ are the given angles.

 We analyze this extremum problem by introducing a Lagrange multiplier $\lambda$. The critical equations~are
\begin{equation}\label{lagmult}
  0\,=\,d\Bigl(\sum_{ij}f_{ij}\cos(\theta_{ij})+\lambda\theta_{ij} \Bigr)\,=\,\sum_{ij}\bigl(-f_{ij}\sin(\theta_{ij})+\lambda\bigr)\,d\theta_{ij}.
\end{equation}
Because $\Delta<0$, all $\sin(\theta_{ij})\neq0$, and $f_{ij}=\lambda/\sin(\theta_{ij})$. The multiplier is determined from $c\cdot f=1$ to be $\lambda^{-1}=\sum_{ij}\cot\theta_{ij}\equiv K$. Hence a maximizing $f$ is unique, and given by the formula.
\end{proof}

This proof utilizes our knowledge from other sources that such $c$ are extremal. When computing the convex hull of a parametrized surface (as in \cite{ciripoi2018computing}) this is the tricky part. We can compute the tangent hyperplane at every point of such a surface, but is the local extremum a global maximum? It is then natural to look first at the local criterion and to rule out saddle points. That is, by looking at the parametrization in second order Taylor approximation, we can determine whether the surface is locally on one side of the tangent plane. If not, the point can be omitted from the list of potential extreme points. We carried this out for the case at hand, and found that  {\it all} points with $\Delta>0$ are saddles.

When a variety is not given by a parametrization, but by implicit equations, the first order analysis is just the definition of the dual variety. From the self-duality of the convex body we thus conclude that $\{h=0\}$ is a self-dual variety. For the study of $\Qm$ this is not only modified by intersecting with the cube, but also by eliminating the branch of the variety in the interior of $\Qm$. Under dualization these operations are connected, i.e., the duals of the normalized tangents of interior points end up outside the dual body.

\subsubsection{Unique CHSH violation}\label{sec:uniqueChsh}
It is clear that each curved tetrahedron, being the pushout of a CHSH-face, violates exactly that CHSH inequality. That this is true for all non-classical correlations is an elementary fact that we have not used otherwise. Its self-dual version has likewise nothing to do with $\Qm$, but only with the enclosing polytopes $\Cl$ and $\Ns$. It states that for any non-trivial Bell inequality, i.e., for any affine inequality valid for all classical correlations ($f\in\Cl\polar$), which is not true by virtue of positivity constraints alone ($f\notin\Ns\polar$) there is a unique non-local box ($c\in\ext\Ns$) making this evident ($c\cdot f>1$).

\begin{proof}[of \reff{prop:1chsh}]
\indent The $8$ CHSH inequalities are $\pm c_{11}\pm c_{12}\pm c_{21}\pm c_{22}\leq2$, where the product of the signs is $-1$. Given two distinct inequalities of this sort, the signs cannot be equal in all four places (then the inequalities would be the same), and also not in three places (because this would imply equality on the fourth place). Adding two violated inequalities ($\cdots>2$) thus gives
$\pm2c_{ij}\pm2c_{kl}>4$, which contradicts the inequality $\abs{c_{ij}}\leq1$ following from non-signalling. Hence at most one such inequality is violated for $c\in\Ns$. On the other hand, for $c\notin\Cl$ at least one is violated.
\end{proof}

\subsubsection{Face orthogonality}\label{sec:normals}

We now consider the normal cycle and the orthogonality relations of faces. Some claims about exposedness are already made in \reff{prop:boundary}. These and the entries in  \reff{table:normalcycle} will now be treated in detail. As in~\reff{sec:dualMcomplete},  statements are grouped by  boundary types. For points $\{c\}$ of each type we must identify the functionals $f\in\Qm\polar$ attaining the maximal value $1$ at $c$, i.e., the face $\{c\}^\dufa$. Note that the orthogonal face operation
\begin{equation}\label{dualfaceK}
  S^\dufa=\{f\in\Qm\polar|\, \forall c\in S \subset \Qm, \ c\cdot f=1\}
\end{equation}
also depends on the whole convex body in which this is taken (here: $\Qm$). We will write $S\dufak\Cl$ and $S\dufak\Ns$ when we take orthogonal complements with regards to $\Cl$ and $\Ns$ respectively.  Obviously, the operation is monotone in the sense that for $S\subset \Cl$ we have
$S\dufak\Ns\subset S^\dufa\subset S\dufak\Cl$.

\vskip7pt\noindent{\bf\Qt1$^\dufa=\: $\Qtc5\:=\:\Qt 5$\cup$\Qt 3$\cup$\Qt 2$\cup$\Qt 1:} \\
Consider a classical extreme point in \Qt1; without loss of generality $c=(1,1,1,1)$. Then $\{c\}^\dufa\subset\{c\}\dufak\Cl$, i.e., $\{c\}^\dufa$ is a face contained in a facet of the cube $\Cl\polar$. This was our definition of an $\Ns$-face. Since these characterizations are taken over by self-duality, let us look at $2Hf$ for $f\in\{c\}^\dufa$. The first component $(2Hf)_{11}$ is just $c\cdot f=1$; the others (with maybe a sign added) are the scalar products of $f$ with all the other classical points, which need to be $\leq1$ for $f\in\Qm\polar$. Hence $2Hf$ is indeed in the standard cube, with $(2Hf)_{1}=1$.
The condition $2Hf\in\Qm$ singles out an elliptope satisfying the third order inequality, which in terms of $f$ is the elementary symmetric polynomial appearing in \eqref{cubicbdary}, albeit with the opposite inequality. Note that $\{c\}^\dufa$ is thus the closed elliptope \Qtc5, which also contains boundary points of type \Qt1,\Qt2,\Qt3. Computing their complements, as we will do presently, will give faces including $c$.

\vskip7pt\noindent{\bf\Qt5$^\dufa=\:$\Qt1:}\\
Starting from an interior elliptope point $c$, self-duality gives us essentially the dual of the previous paragraph. More explicitly, set $c=(1,x,y,z)$ with $1-x^2-y^2-z^2+2xyz\geq0$, and $x^2,y^2,z^2<1$, to avoid edges and classical extreme points. For an interior point, the equality
$c\cdot f=1$ clearly extends from $c$ to the face generated by this. Hence $f$ must be $(1,0,0,0)$, and $2Hf=(1,1,1,1)$ is the classical extreme point considered in the previous paragraph.

\vskip7pt\noindent{\bf\Qt3$^\dufa=\:$\Qt1:}\\
More care is needed for elliptope boundary points, because, in principle, this could allow more freedom for $f$. We will use the local extremality in the sense of \reff{sec:fnondeg}, with angles $\theta_{11}=0$, and $\sin(\theta_{ij})\neq0$ for the other $(i,j)\neq(1,1)$. Then the equation $f_{ij}\sin(\theta_{ij})=\lambda$ implies $\lambda=0\cdot f_{11}=0$; since the other sines are $\neq0$, then once again the only possibility is $f=(1,0,0,0)$.
This shows that the \Qt3-points are indeed non-exposed extreme points as claimed in \reff{prop:boundary}.

\vskip7pt\noindent{\bf\Qt2$^\dufa=\:$\Qtc2\:=\:\Qt 2$\cup$\Qt 1:}\\
Now consider a point $c$ on an edge, but not an endpoint. This forces exactly two angles, say $\theta_{11}=\theta_{12}$, to be zero (cf.\ \reff{prop:paramExtbd}) and by the same reasoning used in the previous paragraph, $\lambda=0$, and $f_{21}=f_{22}=0$. However, $f_{11}$ and $f_{12}$ remain unconstrained and merely have to add up to $1$. Remarkable here is that we get a drop of expected dimension from $\Cl$, where the complement of and edge is a $2$-dimensional face:
\begin{equation}\label{dualCedge}
  \{(1,1,0,0)\}\dufak\Cl=\Bigl\{\bigl( (1+x)/2,(1-x)/2,y/2,-y/2\bigr)\bigm|\ \abs x+\abs y\leq1\Bigr\}.
\end{equation}
Geometrically, $3$-dimensional faces of $\Cl$ containing this edge meet at an angle, whereas the corresponding elliptopes are tangent. This difference is explicitly seen comparing the rows of \reff{fig:projective}.

\vskip7pt\noindent{\bf\Qt4$^\dufa=\:$\Qt4:}\\
Here points go to points, as discussed in detail in \reff{sec:fnondeg}.

\vskip10pt
 \reff{table:normalcycle} also gives the manifold dimensions of each boundary type. We only have one continuous family of type \Qt4, of dimension $3$. All other types occur only in discrete instances, i.e., with dimension zero.

\subsection{Support function}\label{sec:supproof}

As described in \reff{sec:suppfct}, we seek the criteria for a ray to hit the boundary of $\Qm\polar$ either in an exposed extreme point or in an $\Ns$-face. The relevant geometry is shown in \reff{fig:projective}. A point in these diagrams represents a ray, and the separating surfaces precisely mark the distinction we need to study here. The problem becomes almost trivial, however, if we already know by which face the ray leaves the surrounding cube: Then we just have to check whether the boundary point is inside or outside the elliptope, for which we have a convenient third order criterion. The following proof is based on the case distinction by cube faces. But since these are all connected by symmetries, it boils down to just considering one case.

\begin{proof}[of~\autoref{prop:hom}]
\indent $(1){\Leftrightarrow}(2)$:\
We use the normal cycle: The set of pairs $(c,f)\in\ncyc(\Qm)$ such that both $c$ and $f$ are uniquely determined when the other is fixed is just the stratum \Qb44. Hence, if $f$ is of type \Qt4, so is $c$.

{\it Reduction by symmetry:}\\
Applying a symmetry to $f$, i.e., a permutation of the components or an even number of sign changes, clearly does not change the validity of (1), (2), (3), or (4), while (5) respects the symmetry by requiring the necessary transformation to be made first. Hence it suffices prove for $f$ in a standard form achievable by symmetry transformation. Since all extreme points of $\Cl$ are connected by symmetry we can assume, as required in (5) that a maximizer for $c\cdot f$ in $\ext\Cl$ is $c=(1,1,1,1)$. Now suppose that two or more coordinates $f_{ij}<0$. Then, by applying an even sign change to $c$ we could increase $c\cdot f$. So our assumption actually rules out more than one negative sign. By the same argument, if there is a negative sign, this must be on an $f_{ij}$ which minimizes $\abs{f_{ij}}$. By a permutation we may assume that this element is $f_{22}$.

We can quickly handle the case that all $f_{ij}\geq0$. In that case, $\max\{c\cdot f| c\in\Cl\}=\sum_{ij}f_{ij}=\sum_{ij}|f_{ij}|=\max\{c\cdot f| c\in\Ns\}$. So the maximum over $\Qm$ is attained at a classical point, $p(f)\geq0$ and the product in \eqref{cubicbdary} cannot be $<0$, so all conditions evaluate to false.
So we may assume that there is just one negative sign and sort the remaining $f_{ij}$. That is, from now on we take
\begin{equation}\label{assumeSymm}
  f_{11}\geq f_{12}\geq f_{21}\geq\abs{f_{22}}>0> f_{22}.
\end{equation}

$(1){\Leftrightarrow}(5)$:\
We use the same classification of boundary points for $\Qm$ and $\Qm\polar$ via duality transform. So (1) means that $c=2Hf$ is a multiple of a \Qt4 point. We know that the component with the largest absolute value is the first, $c_{11}=\sum_{ij}f_{ij}$. Thus the point where the ray $\Rl c$ intersects the boundary of $\Ns$ is $(1,c_{12}/c_{11},c_{21}/c_{11},c_{22}/c_{11})=(1,x,y,z)$. This is in an $\Ns$-facet of $\Qm$
if and only if $1-x^2-y^2-z^2+2xyz\geq0$. Otherwise, this intersection is already outside of $\Qm$, and hence the ray intersects $\partial\Qm$ at a \Qt4 point. So the necessary and sufficient condition for (1) is that the cubic is $<0$. Multiplying by $c_{11}^3$, which we know to be positive, and rearranging the resulting homogeneous polynomial in the $f_{ij}$, gives condition~(5).

$(5){\Leftrightarrow}(3)$:\
Under the above symmetry reduction we have
\begin{equation}\label{mfinpf}
  m(f)\,\,=\,\,(-f_{22})\Bigl(\frac1{f_{11}}+\frac1{f_{12}}+\frac1{f_{21}}-\frac1{f_{22}}\Bigr)\,\,=
  \,\,1-\frac{f_{22}}{f_{11}}-\frac{f_{22}}{f_{12}}-\frac{f_{22}}{f_{21}}.
\end{equation}
Subtracting $1$ from both sides of the inequality $m(f)>2$, and multiplying by $f_{11}f_{12}f_{21}$, gives~(5).

$(4){\Leftrightarrow}(5)$:\
The first factor in $\tilde m(f)$ is the polynomial in \eqref{cubicbdary}, divided by $p(f)$. So while $p(f)<0$, the positivity of this factor is equivalent
to (4). We complete the proof by showing that under the symmetry reduction \eqref{assumeSymm} the other three factors are always, respectively, positive, negative and positive. Indeed, in the second factor the $f_{11}$ and $f_{22}$-terms together are positive, and so are the $f_{12}$ and $f_{21}$-terms. Similarly, in the third factor we group the $f_{11},f_{12}$ and the $f_{21},f_{22}$-terms (both negative), and use the same grouping in the forth factor, giving two positive terms.
\end{proof}

It is interesting to see what a purely algebraic approach as in \cite[Section 5.3]{Blekherman}
 would say about the situation. First of all, the support function asks us to compute a maximum of a linear functional over a variety. Evaluating just the first order conditions, we get an expression for the maximum for each patch of $\partial\Qm$. For the curved \Qt4 patches this will be just $\widetilde\phi$. So we get little progress over the simple story told in \reff{sec:suppfct}. In many well-known problems of duality, we can simply take the maximum of the
 different branches of the algebraic function.
  For example when computing the Legendre transform $({\mathcal L}F)(p):=\sup_x\{p\cdot x-F(x)\}$ of a non-convex function like $F(x)=x^4-x^2$ one gets a multi-valued function defined by eliminating $x$ from the cubic $p=dF(x)$. This algebraic version of the Legendre transform
  is easily corrected by taking the supremum from the convex definition just given as a maximum over the branches. One might therefore guess (and we tried that out in an early stage of this work) that our support function
  is simply the maximum of the two branches $\widetilde\phi$ and $\phi_\Cl$. Suffice it to say that this fails, and one easily finds points where $\widetilde\phi(f)>\phi_\Cl(f)=\phi(f)$. To get this to work, one must ensure that the associated maximizers are contained in the quantum set.

\subsection{Quantum representations of extreme points}\label{sec:qproofs}

We now prove \reff{prop:selftest}. A central role will be played by the uniqueness of a cyclic model.
Recall that {\it every} state of a C*-algebra has a unique cyclic model known as the GNS (Gelfand-Naimark-Segal) representation.
What makes uniqueness of a cyclic model non-trivial for correlations $c\in\Qm$ is that the algebra itself is not known from expectation values: We can usually not infer the multiplication rules for the $A_i,B_j$ from just these expectations, nor the expectation of, say, $A_1B_2^2$. If the cyclic model is unique up to unitary isomorphism, as condition (2) of \reff{prop:selftest} asserts, then all algebraic relations are fixed.

According to \reff{prop:selftest} uniqueness fails for the {\it classical} extreme points, so these provide a key example. The correlations are $c_{ij}=a_ib_j$ with $a_i,b_j=\pm1$. These numbers constitute a model with one dimensional Hilbert space, which is obviously cyclic. The models related by $a_i'=-a_i$ and $b_j'=-b_j$ clearly give the same $c$, but the unitary operator connecting the model Hilbert spaces cannot take $a_1$ to $a_1'$, so these are not unitarily equivalent. More generally, any classical correlation $c\in\Cl$ allows a model with $a_1=+\idty$, and another with $a_1=-\idty$. So no classical correlation can have property (2).

Any $c\in\Qm$ has {\it some} cyclic model: By definition of $\Qm$, it has a model. The state induced on the algebra $\Alg$ generated by all $A_i,B_j$ in that model then has a GNS-representation, which is cyclic.
Moreover, it has a {\it finite dimensional} cyclic representation. Indeed, by Carath\'eodory's Theorem we can represent $c$ as a finite convex combination of extreme points. Since we have provided finite dimensional models for each of these, we can combine them into a single cyclic model by the direct sum construction
\begin{equation}\label{cyclicsum}
  \HH=\bigoplus_k\HH_k \text{ with } A_i=\bigoplus_k A_i^{(k)},\ B_j=\bigoplus_k B_j^{(k)},\ \text{ and } \Psi=\bigoplus_k\sqrt{p_k}\Psi^{(k)},
\end{equation}
where $p_k$ is the convex weight of the $k^{\rm th}$ contributing extreme point. This could fail to be cyclic, for example, if two of the extreme points used were actually (needlessly) the same. In that case one can take the cyclic subspace generated from $\Psi$ instead. In any case, from Carath\'eodory's bound we can get by with $5$ terms, so the overall dimension can be chosen to be $\leq20$.

The direct sum construction can be reversed: If $P_k$ are orthogonal projections in the Hilbert space of some cyclic model commuting with all $A_i,B_j$, then setting $\HH_k=P_k\HH$, $p_k=\braket\Psi{P_k\Psi}$, $\Psi^{(k)}=\sqrt{p_k}P_k\Psi$, and $A_i^{(k)},B_j^{(k)}$ the restrictions of $A_i,B_j$ to $\HH_k$ we have just written the given model in the form \eqref{cyclicsum}.

\begin{proof}[of \reff{prop:selftest}]
\noindent$(1){\Rightarrow}(2)$:\\
Let $c$ be a non-classical extreme point, and let $\Psi$ be the cyclic vector of some quantum model. By \reff{prop:Cposb}, there
is a unique matrix completion  $C$ with $-1<u,v<1$ and rank $2$. The real part of the Gram matrix of the four vectors $\chi_1,\ldots,\chi_4=A_1\Psi,A_2\Psi,B_1\Psi,B_2\Psi$ is such a completion, so it must equal  $C$. Thus $\re \braket{A_1\Psi}{A_2\Psi}=u$, but it is not immediately obvious that this scalar product has to be~real.

Since $\rank C=2$ we have two linearly independent {\it real} vectors $\xi$ in the kernel of $C$. Any such vector
satisfies $\sum_{ij}\xi_i\braket{\chi_i}{\chi_j}\xi_j+\sum_{ij}\xi_i\overline{\braket{\chi_i}{\chi_j}}\xi_j=0$. Since both terms are positive, they must vanish separately, and $\sum_j\xi_j\chi_j=0$. Thus the linear relations between the $\chi_i$ are given by real coefficients.  Since the $1,2$ and $3,4$ submatrices of $C$ are both nonsingular (as a consequence of $-1<u,v<1$), either $(\chi_1,\chi_2)$  or $(\chi_3,\chi_4)$ can serve as a basis for the kernel of $C$, and the two are related by a non-singular real $2\times2$-matrix $\gamma$, so
\begin{equation}\label{Bpsi}
  B_j{\Psi}=\sum_i\gamma_{ji}A_i{\Psi}\,.
\end{equation}
In particular, $c_{ij}=\braket{A_i\Psi}{B_j\Psi}=\sum_k\braket{A_i\Psi}{A_k\Psi}\gamma_{jk}$. Since the matrices $c$ and $\gamma$ are real,
so is the scalar $\braket{A_1\Psi}{A_2\Psi}=u$. Thus, there was no need after all to take the real part in the first paragraph of this~proof.

It follows from \eqref{Bpsi} that non-commutative polynomials in $A_1,A_2$ acting on $\Psi$ already span the whole space. By cyclicity of $\Psi$ this is true for polynomials involving also the $B_j$.
Now we can successively get rid of the factors $B_j$ in any polynomial acting on $\Psi$: In any monomial consider the rightmost factor $B_j$, so we have an expression of the form
\begin{equation}\label{commthrough}
  M B_j M_A\Psi=M\,M_A B_j\Psi=\sum_i\gamma_{ji}\ M\,M_A A_i{\Psi} ,
\end{equation}
where $M$ is a monomial involving factors $A_i,B_k$ and $M_A$ is a monomial containing only factors $A_i$. In the evaluation we used that $B_j$ commutes with all $A_i$, hence with $M_A$, and \eqref{Bpsi}.
By downwards induction on the number of $B_j$-factors we get that the vectors $M_A\Psi$ span the space.  Symmetrically the same is true for polynomials in $B$. Next suppose that $f(A_1,A_2,\idty)\Psi=0$ for some non-commutative polynomial.
Then even the operator equation $f(A_1,A_2,\idty)=0$ holds, because we can multiply with any polynomial in $B$, commute through and use the cyclicity just established.

The first application is to $f=\idty-A_1^2$, which is positive, and has zero expectation $\braket\Psi{f\Psi}$ because $C_{11}=1$. This implies $f\Psi=0$, and hence $f=0$.
We conclude that $A_1^2=A_2^2=B_1^2=B_2^2=\idty$.
Similarly,
\begin{eqnarray}\label{anticom}
  \Psi&=&B_j^2\Psi=B_j\sum_i\gamma_{ji}A_i{\Psi}=\sum_i\gamma_{ji}A_i B_j{\Psi}=\Bigl(\sum_i\gamma_{ji}A_i\Bigr)^2\Psi\nonumber\\
      &=&(\gamma_{j1}^2+\gamma_{j2}^2)\Psi +\gamma_{j1}\gamma_{j2}(A_1A_2+A_2A_1)\Psi.
\end{eqnarray}
If $\gamma_{j1}\gamma_{j2}\neq0$ then $(A_1A_2+A_2A_1)\Psi$ is a multiple of $\Psi$, and so $(A_1A_2+A_2A_1)$ is a multiple of the identity, and from $C_{12}$ this multiple is $2u$.
This conclusion does not depend on $j$, so it is valid whenever $\gamma_{j1}\gamma_{j2}\neq0$ holds either for $j=1$ or for $j=2$.
Now if  $\gamma_{j1}\gamma_{j2}=0$, one of the coefficients is zero, and by \eqref{Bpsi} $B_j\Psi$ is proportional to some $A_i\Psi$. By taking the norm, the factor (the non-zero $\gamma_{ji}$) is $\pm1$ and $c_{ij}=\pm1$. That is, the correlation is on an $\Ns$-facet. But this cannot hold for the other $B_{j'}$ at the same time except for a classical $c$. Hence in either case
\begin{equation}\label{JordanA1A2}
  (A_1A_2+A_2A_1)=2u\idty.
\end{equation}
Using this identity and $A_i^2=\idty$, every word in $A_1$ and $A_2$ simplifies to a linear combination of $\idty,A_1,A_2$, and $A_1A_2$. Hence the *-algebra $\mathcal{P}_2$ generated by $A_1$ and $A_2$ is at most four dimensional. It is non-commutative, because $A_1A_2=A_2A_1$ would imply $4u^2\idty=(A_1A_2+A_2A_1)^2=4\idty$, hence $\abs u=1$. Therefore $\mathcal{P}_2$ is isomorphic to $\text{Mat}_2(\Cx)$, the smallest noncommutative finite dimensional C*-algebra of $2\times2$-matrices with complex entries.

Next, we claim that the state induced by the vector $\Psi$ on our algebra $\mathcal{P}_2$ has the form
\begin{equation}\label{tracial}
M \mapsto \braket\Psi{M\Psi}=:\tau(M)={\textstyle\frac12}\tr M
\end{equation}
when $M$ is a polynomial in the $A_i$. Note that the normalized trace is uniquely  characterized among states on the $2\times2$-matrices by the algebraic ``tracial property'' $\tau(AB)=\tau(BA)$. Indeed, setting $A=\ketbra{e_i}{e_1}$ and $B=\ketbra{e_1}{e_j}$ in the equation gives $\tau(\ketbra{e_i}{e_j})=\delta_{ij}/2$. Therefore, we only have to establish the tracial property for the given state. Now, using \eqref{Bpsi},
\begin{equation}\label{cyclictrace}
  0=\braket\Psi{MB_j\Psi}-\braket{B_j\Psi}{M\Psi}=\sum_i\gamma_{ji}\braket{\Psi}{(MA_i-A_iM)\Psi}.
\end{equation}
Since $\gamma$ is non-singular, we conclude $\braket{\Psi}{(A_iM-MA_i)\Psi}=0$. By induction on the degree of an arbitrary other polynomial $M'$ in the $A_i$, we have the tracial property $\braket{\Psi}{(M'M-MM')\Psi}=0$, proving \eqref{tracial}. We note that $M'M-MM'=0$ is false in general for any two polynomials (e.g.~\eqref{JordanA1A2}), but is true under expectation value as we have shown.

Since $A_j^2=\idty$ and $A_j^*=A_j$, we must either have $A_j=\pm1$, which would entail classical correlation, or else $\tr A_j=0$. The trace of \eqref{JordanA1A2} gives $\tr A_1A_2=u$ and so we have evaluated $\braket\Psi{M\Psi}$ on all polynomials in $A_i$. By the reduction process used above we can also compute the expectations involving polynomials in both $A_i$ and $B_j$. Hence by the GNS construction~\cite{dixmier} the cyclic model is unique. Moreover, it is equivalent to the model in \reff{sec:quantummodels} via a unique unitary isomorphism that maps the linearly independent (since $\abs{u}<1$) vectors $\Psi,A_1\Psi,A_2\Psi,A_1A_2\Psi$ to the corresponding vectors of the standard model (then also cyclic). This finishes the proof. We remark that the family of standard models was defined for points other than those in (1) may very well be cyclic but not unique for these points.

\vskip7pt\noindent$(2){\Rightarrow}(1)$:\\
Assume $c$ has a unique cyclic model. Any property that holds in some cyclic model must be true for this unique one. For example, the unique model must be finite dimensional.
Similarly, dilation theory tells us that $A_1^2=\idty$.  To this end, we first decompose the algebra into irreducible summands. In each summand the algebras generated by $A_1,A_2$ and $B_1,B_2$ must themselves be irreducible, so full matrix algebras. Hence the algebras are combined in a tensor product, and are represented on a Hilbert space tensor product $\HH_A\otimes\HH_B$. We can write $A_1=V^*\widehat A_1V\otimes\idty_B$, where $V:\HH_A\to\widehat\HH_A$ is isometric, and $\widehat A_1^2=\idty$. This construction preserves the commutativity of $A_i$ and $B_j$, and mapping the cyclic vector by $V\otimes\idty_B$ to the larger space we get a model with $A_1^2=\idty$. Restricting to the cyclic subspace preserves this property. Hence the unique cyclic model must also satisfy it, and similarly for the other operators.

Now suppose there is a factor $\lambda>1$ such that $\lambda c\in\Qm$. Then using a cyclic model for $\lambda c$ and scaling down the $A_i$ by a factor $1/\lambda$ we get a cyclic model with $A_1^2\neq\idty$. Hence $c$ does not have a unique model. It follows that $c$ must be on the boundary of $\Qm$.

Consider now the boundary classification of \reff{prop:boundary}. The classical types \Qt1 and \Qt2 anyhow fail to have unique models, as mentioned in the second paragraph of~\autoref{sec:qproofs}). So in order to show that $c$ is of type \Qt3 or \Qt4, a non-classical extreme point, we only need to exclude points in an elliptope interior (type \Qt5). Let us assume without loss that $c_{11}=1$. Consider the one parameter family of correlations $(1,\lambda c_{12},c_{21},\lambda c_{22})$ with $\lambda$ increasing from $1$. This will intersect the elliptope boundary for some $\lambda>1$. Starting from a model at that point, and scaling $B_2\mapsto B_2/\lambda$ we can obtain a model with $\norm{B_2}<1$. As argued before, this contradicts the uniqueness of the cyclic model.

\vskip7pt\noindent$(3){\Rightarrow}(2)$:\\
Consider a cyclic model of the type described in (2). In $\rho=\kettbra\Psi$ the state vector must be of the form  $\Psi=\Psi\otimes\Psi_\nu\oplus0$. The cyclic subspace contains only vectors of the form
$(X\Psi)\otimes\Psi_\nu\oplus0$, so $\HH_0=\{0\}$ and $\HH_\nu=\Cx\Psi_\nu\cong\Cx$, and this tensor factor can be omitted.

\vskip7pt\noindent$(2){\Rightarrow}(3)$:\\
Given a model $\HH$ for $c$, let us denote by  $\Alg$ the norm closed algebra generated by the $A_i,B_j$, by $\Alg'$ its commutant, i.e., the algebra of operators commuting with every element of $\Alg$.  By $\KK_\rho$, we denote the vector space of $\phi\in\HH$ such that $\kettbra\phi\leq\lambda\rho$ for some factor $\lambda$. The closure of $\KK_\rho$ is called the support of $\rho$. (In finite dimension, $\KK_\rho$ is closed and also equal to the range of $\rho$, but in infinite dimension $\overline{\KK_\rho}$ may be strictly larger than $\KK_\rho$.) Then we define $\HH_0$ as the subspace orthogonal to all vectors of the form $XY\phi$ with $X\in\Alg$, $Y\in\Alg'$, and $\phi\in\KK_\rho$. Since this $\HH_0$ is an invariant subspace for all $X\in\Alg$, every $X\in\Alg$ splits into a direct sum of a component on $\HH_0$, about which we can say nothing at all, and a rest, which we will need to characterize. So we may assume $\HH_0=\{0\}$ in the sequel.

For any unit vector of the form $\Psi_1=Y\phi$ with $Y\in\Alg'$ and $\phi\in\KK_\rho$ consider the cyclic representation subrepresentation of $\Alg$ on $\Alg\Psi_1$. Then for positive $X\in\Alg$ we get
\begin{eqnarray}
  \braket{\Psi_1}{X\Psi_1} &=& \braket{\phi}{Y^*\sqrt X^2Y\phi}=\braket{\phi}{\sqrt X Y^*Y\sqrt X\phi} \nonumber\\
   &\leq& \norm Y^2 \braket{\phi}{\sqrt X \,\sqrt X\phi}=\norm Y^2\, \tr(\kettbra\phi X)\leq \lambda\norm Y^2\, \tr \rho X.  \nonumber
\end{eqnarray}
Thus as a functional on $\Alg$ the state defined by $\Psi_1$ is dominated by a multiple of $\rho$, hence is a convex component of $\rho$. Since we have already established from (2) that $c$ is extremal,
$\Psi_1$ defines again a cyclic model for $c$, and is hence unitarily isomorphic to the standard model for $c$. In particular, all algebraic identities of that model hold on $\Alg Y\phi$, and since such vectors span dense subspace, they hold on all of $\HH$.

There is now a polynomial $G$ in the generators $A_i,B_j$ that in the standard model is equal to the one-dimensional projection onto the state vector.
When $\hat X\in\BB(\hat\HH)$ and $X\in\Alg$ are given by the same polynomial in the generators, we therefore have $GXG=\braket{\hat\Psi}{\hat X\hat\Psi}\, G$.
Moreover, all vectors $Y\phi$ from the previous paragraph are in $G\HH$. We can therefore define $\HH_\nu=G\HH$, and get a unitary operator $U:\hat\HH\otimes\HH_\nu\to\HH$ defined by $U(\hat X\hat\Psi\otimes\Phi_\nu)=X\Phi_{\nu}$. Since $\rho$ has support in $G\HH$ we have $U^*\rho U=\kettbra{\hat\Psi}\otimes \rho_\nu$ for some state $\rho_\nu\in\BB(\HH_\nu)$, as claimed in the standard form (3).

\vskip7pt\noindent$(3){\Rightarrow}(6)$:\\
From the explicit form of the model it is clear that any operator $E$ commuting with $\Alg$ is of the form $E=\idty\otimes E_1\oplus E_0$. Hence
$\tr(\rho EX)=\tr(\rho_\nu E_1)\braket{\hat\Psi}{X\hat\Psi}=\tr(\rho E)\,\tr(\rho X)$.

\vskip7pt\noindent$(6){\Rightarrow}(5)$:\\
This is trivial, because the factorization is no longer demanded for all $X\in\Alg$, but only for those $X$ actually measured in the experiment (including the marginals!).

\vskip7pt\noindent$(5){\Rightarrow}(4)$:\\
Since the marginals are now included, their expectations, which define $p$ must also be those of the unique model, i.e., zero.

\vskip7pt\noindent$(4){\Rightarrow}(1)$:\\
First we eliminate the classical extreme points as they do not have unique extensions $p$, as explained in the beginning of this Section. If $c$ is not extremal, i.e., on an edge, in an elliptope interior or the interior of $\Qm$, then $c=\lambda c'+(1-\lambda)c''$ for some $\lambda>0$ and $c'$ classical. So we get distinct extensions by fixing the extension of $c''$ but changing that of $c'$.
\end{proof}

\subsection{Dependence on Hilbert space dimension}\label{sec:dimHproof}
We now return to  \reff{sec:dimH}, and we prove the remaining statement on the Hilbert space dimension. We remark that $\Qm_m=\Qm$ for $m\geq4$ is already proved in Tsirelson's original paper~\cite{Tsi85}. Here we provide a different geometric proof.

\begin{proof}[of \reff{prop:dimH}]
\indent (1) Let $c\in\Qm_m$ be a quantum correlation in an $m$-dimensional Hilbert space. Then the algebra $\Alg$ generated by $A_i,B_j$ can be decomposed into irreducible components, represented on orthogonal subspaces $\HH_\alpha$. $c$ is then a convex combination of correlations realized in $\HH_\alpha$ with weights equal to the expectation of the projection onto this subspace. When $m\leq3$, we also have $\dim\HH_\alpha\leq3$, so it suffices to consider the case of a single summand, i.e., $\Alg$ acting irreducibly.

Consider now the subalgebra $\Alg_A$ generated only by the $A_i$. An element in the center of $\Alg_A$, i.e., one that commutes with $A_1$ and $A_2$, also commutes with the $B_j$, and is therefore in the center of $\Alg$. By irreducibility it has to be a multiple of $\idty$. Therefore, $\Alg_A$ is a finite dimensional C*-algebra with trivial center, and has to be isomorphic to a full matrix algebra, say the $n\times n$ matrices. Similarly, we get a matrix dimension $n'$ for the algebra generated by the $B_j$. Since the two sets of operators commute, and even generate commutants of each other, $\dim\HH_\alpha=nn'$. Hence, by assumption, $nn'\leq m\leq3$, so either $n=1$ or $n'=1$.

Taking the first case without loss, $A_i=a_i\idty$ and the contribution from this summand is $c_{ij}=\tr_\alpha(\rho A_iB_j)=a_i\tr_\alpha(\rho B_j)$, which is a product, hence classical.

We note that this argument also applies in any quantum model with $[A_1,A_2]=0$ (\cite{Raggio}), and hence if $A_i=a_i\idty$ for some $i$: Then every irreducible component gives a product correlation, and the resulting convex combination is still classical.

(2) By monotonicity, we only need to show that $\Qm\subset\Qm_4$. We already know that many boundary points are in $\Qm_4$, since we built explicit $4$-dimensional models in \autoref{sec:quantummodels}. In particular, $\hbox{\Qtc4}\subset\Qm_4$. Here \Qtc4 is the closure of the exposed extreme points \Qt4, and by the discussion after \reff{prop:boundary} covers all boundary elements except the elliptope interiors.

Some points in the interior of $\Qm$ can now be obtained by simple scaling: When $A_i,B_j,\rho$ is a model for $c$, and $0\leq\lambda\leq1$, then $A'_i=\lambda A_i$, $B_j'=B_j$, $\rho'=\rho$ is a model for $\lambda c$ with the same dimension. Turning this around, we can consider, for any $c\in\Qm$ with $c\neq0$, the unique point $\overline c=tc$, which is the intersection of the ray $\{t c|t\geq0\}$ with $\partial\Qm$ (possible because $\Qm$ is a compact convex set with origin in the interior). Then if $\overline c\in\hbox{\Qtc4}$, we have $c\in\Qm_4$.

This leaves points, for which $\overline c$ is in an elliptope interior. By the argument just used, we only need to show that elliptope interiors themselves are in $\Qm_4$, and by symmetry we may choose the elliptope in the $\Ns$-facet $c_{11}=1$. 
To this end, we consider a modified rescaling, namely $A_1'=A_1$, $B_1'=B_1$, $A_2'=\lambda\,A_2$, and $B_2'=\lambda\,B_2$, and $\rho'=\rho$. This also leaves the dimension unchanged, and realizes the correlation $c'(\lambda)=(1,\lambda\,c_{12},\lambda\,c_{21},\lambda^2\,c_{22})$.
Clearly, with $\lambda=0$ we get the center of the elliptope. For all other points $c$ we find a suitable boundary point by inverse scaling: For $\lambda>1$ the curve $c'(\lambda)$ is unbounded, and so eventually crosses the elliptope boundary. So we can use the model for that boundary point and scale with $1/\lambda$. 
\end{proof}

\subsection{Fixed state}\label{sec:fixedStateproof}
\begin{proof}[of \reff{prop:fixrho}]
\indent Suppose one of the $A_i,B_j$ has a proper convex decomposition into extreme points of the allowed set, where ``proper'' means that several distinct points each appear with positive weight. This would make any correlation $c$ obtained with these operators a convex combination of others. Suppose this $c$ is an extreme point of $\conv\Qm(\rho)$. Then the convex combination of correlations cannot be proper, i.e., all the correlations appearing with a positive weight must be equal. So either all $A_i,B_j$ are extremal in the first place, or the extreme points in their convex decomposition also represent c. In either case, the extreme points of $\conv\Qm(\rho)$ can be realized by operators such that each $A_i$ and $B_j$ is extremal. This means that $A_i^2=B_j^2=\idty$, because if $A$ with $-\idty\leq A\leq\idty$ has any eigenvalues {\em other than} $\pm1$, it can be convexly decomposed in this set into two operators with the same eigenvectors, but with eigenvalues $\pm1$.

In two dimensions there are just two eigenvalues $\pm1$ which are either the same or distinct. This gives the union structure in the statement of the proposition to be proved. When they are the same, we have $A_i=\pm\idty$. Then by part (1) of the proof in~\autoref{sec:dimHproof} the resulting correlation is classical. Moreover, all extremal classical correlations can be generated with $A_i,B_j=\pm\idty$. When none of $A_i,B_j$ are $\pm\idty$, we obtain correlations that belong to a set denoted $\Qmst$. By definition, $\Qmst$ is the set of all correlations obtained with {\em extremal} $A_i,B_j$, none of which is $\pm\idty$. In total, $\conv\Qm(\rho)$ is the convex hull of $\Cl$ joined with $\Qmst$. It is worth noting, however, that extreme points of $\conv\Qmst$ might not be extremal in $\conv\Qm(\rho)$, because for some $\rho$, e.g., $\rho$ close to the maximally mixed state, this set may be small, and even $\Qmst\subset\Cl$. In this case $\Qm(\rho)\subset\Cl$, and the given state can only ever produce classical correlations (see below).

Our next step is to determine $\Qmst$. Hermitian $2\times2$-matrices $A$ with eigenvalues $\pm1$ can be expanded in Pauli matrices $\sigma_1,\sigma_2,\sigma_3$ as $A=\sum_{k=1}^{3}a_k\sigma_k=:\vec a\cdot\vec\sigma$, where $\vec a\in\Rl^3$ is a Euclidean unit vector. That is, we have now to choose $4$ unit vectors $\vec a_1,\vec a_2,\vec b_1,\vec b_2$. This  parametrization gives
\begin{equation}\label{2qbc}
  c_{ij}=\tr\rho A_i\otimes B_j=\vec a_i\cdot R\vec b_j,
\end{equation}
where $R$ is the matrix of Pauli expectations given in the proposition. Suppose now that $R$ is replaced by $URV$ with orthogonal matrices $U,V\in\mathrm{Mat}_3(\Rl)$. This corresponds to a change of the state (by a local unitaries), but since we can use $U$ and $V$ to change the vectors $\vec a_i,\vec b_j$,  the {\it set} of correlations remains the same. Now the singular value decomposition allows us to choose $URV$ diagonal, where, by definition the diagonal elements of this matrix are the singular values $\lambda_1,\lambda_2,\lambda_3$ of the original $R$. The set of the resulting correlations is now $\Qmst=\Qmst(\lambda_1,\lambda_2,\lambda_3)=\{c\in\Rl^4|c_{ij}=\vec a_i\cdot\mathrm{Diag}(\lambda_1,\lambda_2,\lambda_3)\vec b_j,\, \text{unit vectors } \vec a_i,\,\vec b_j\in\Rl^3\}$ with $D:=\mathrm{Diag}(\lambda_1,\lambda_2,\lambda_3)$ the diagonal matrix with these elements along the diagonal.

Moreover, the choice $\vec a_i,\vec b_j\in\Rl^3$ can be decomposed into first choosing two-dimensional subspaces $W_A,W_B$ and then choosing $\vec a_i\in W_A, \vec b_j\in W_B$, resulting in $\Qmst$ the union of sets of correlations with fixed $W_A,W_B$. What does each of these set look like? It clearly has the same form as above but with $D$ projected to two-dimensions via orthogonal projections $P_A$ onto $W_A$ and $P_B$ onto $W_B$, that is
\begin{align}
\Qmst(W_A,W_B)=\{c\in\Rl^4|c_{ij}=\vec a_i\cdot P_ADP_B\vec b_j,\, \text{unit vectors } \vec a_i,\,\vec b_j\in\Rl^2\}\,.
\end{align}
Then everything said in the previous paragraph applies with the only change that the $3\times3$-matrix $R$ is replaced by a $2\times2$-matrix $P_ADP_B$, so there are only two singular values denoted $\lambda_1,\lambda_2$ (abuse of notation, these are generally distinct from the original singular values of $R$ in three-dimension). Denote the resulting set of correlations by $\Qmst(\lambda_1,\lambda_2)$. Since unit vectors in two dimensions are of the form $\vec a_i=\bigl(\cos(\alpha_i),\sin(\alpha_i)\bigr)$, and $R$ is now the diagonal matrix with entries $(\lambda_1,\lambda_2)$,  this set  is exactly the one described in the proposition.

It remains to compute the convex hull. Observe that there is a monotonicity relation when taking convex hull: if $\lambda_1\leq\mu_1$ and $\lambda_2\leq\mu_2$ then $\conv\Qmst(\lambda_1,\lambda_2)\subset\conv\Qmst(\mu_1,\mu_2)$. Indeed, it suffices to consider the case $\lambda_1\leq\mu_1$. Starting from any model $c_{ij}=\mu_1\cos(\alpha_i)\cos(\beta_j)+\mu_2\sin(\alpha_i)\sin(\beta_j)$ with the larger $\mu_1$, we take the convex combination
\begin{equation*}
p\left(\mu_1\cos(\alpha_i)\cos(\beta_j)+\mu_2\sin(\alpha_i)\sin(\beta_j)\right)+(1-p)\left(\mu_1\cos(\pi-\alpha_i)\cos(\beta_j)+\mu_2\sin(\pi-\alpha_i)\sin(\beta_j)\right)
\end{equation*}
to get $c_{ij}'=\mu_1 (p-(1-p))\cos(\alpha_i)\cos(\beta_j)+\mu_2 \sin(\alpha_i)\sin(\beta_j)$. That is, precisely to the corresponding correlation with $\mu_1'=(2p-1)\mu_1$, which can thus be made equal to any $\lambda_1\leq\mu_1$. 

Therefore, in the union over all $W_A,W_B$ we can ignore all subspace pairs for which the singular values $\lambda_1',\lambda_2'$ of the projected $P_ADP_B$-matrix are dominated by those of another one. We claim there is one pair of subspaces with largest pair of projected singular values: $W_A=W_B$ is a/the eigenspace (nonunique if $D$ is degenerate) of the top two eigenvalues $\lambda_1,\lambda_2$ of $D$ or equivalently of $R$. This follows by the observation that the squares of the non-zero singular values of any matrix $M$ are the non-zero eigenvalues of $M^*M$. When $M=P_ADP_B$, the squares of its two singular values are the eigenvalues of $P_BD^*P_ADP_B$. Now $D^*P_AD$ is positive semidefinite rank~$2$ matrix, so we can apply the Rayleigh-Ritz variational principle \cite[Theorem XIII.3]{ReedSimon4}: The eigenvalues of $P_BD^*P_ADP_B$ are largest, when $P_B$ is just the projection onto the eigenspace associated with the two largest eigenvalues of $D^*P_AD$, or equivalently of $P_ADD^*P_A$ (since the eigenvalues of $X^*X$ and $XX^*$ are the same). The same Rayleigh-Ritz argument can be applied again to maximize the two eigenvalue by picking $P_A$ as the projection to the eigenspace of two largest eigenvalues of $DD^*$. To summarize, when we take the $\lambda_1\geq\lambda_2\geq\lambda_3$ to be the singular values of $R$, we have 
\begin{equation}\label{singval3to2}
  \conv\Qmst(\lambda_1,\lambda_2,\lambda_3)=\conv\Qmst(\lambda_1,\lambda_2). 
\end{equation}
This concludes the proof of the main statement of \reff{prop:fixrho}. 

Finally, we have to check when $\Qm(\rho)$ becomes classical. This happens iff all CHSH inequalities are satisfied, also in $\Qmst(\lambda_1,\lambda_2)$. By symmetry and from \eqref{CHSH} we have
\begin{equation}\label{chsh2Qbit}
\text{CHSH}(c)=\frac12\Bigl(\vec a_1\cdot R(\vec b_1+\vec b_2)+\vec a_2\cdot R(\vec b_1-\vec b_2)\Bigr),
\end{equation}
which is clearly maximized with respect to the unit vectors $\vec a_i$ as $\frac12\bigl(\norm{R(\vec b_1+\vec b_2)}+\norm{R(\vec b_1-\vec b_2)}\bigr)$. Since the $\vec b_1\pm\vec b_2$ are orthogonal vectors, we can write
$\vec b_1\pm\vec b_2=\lambda_\pm \vec e_\pm$, where $\vec e_+$ and $\vec e_-$ are orthogonal unit vectors and $\lambda_+^2+\lambda_-^2=\norm{\vec b_1+\vec b_2}^2+\norm{\vec b_1-\vec b_2}^2=4$. Thus by Cauchy-Schwarz inequality
\begin{equation}\label{maxchsh2Qbit}
  \max \text{CHSH}(c)=\frac12\bigl(\lambda_+ \norm{R\vec e_+}+\lambda_- \norm{R\vec e_-}\bigr)\leq \Bigl(\norm{R\vec e_+}^2+\norm{R\vec e_-}^2\Bigr)^{\frac12}.
\end{equation}
The choice of orthogonal unit vectors $\vec e_\pm$ maximizing this expression is the eigenvectors of $R^*R$ for its to largest eigenvectors, which are $\lambda_1^2$ and $\lambda_2^2$.
\end{proof}

\subsection{Fixed observables}\label{sec:fixedObsproof}
\begin{proof}[of \reff{prop:fixedObs}]
\indent By Jordan's Lemma~\cite{Masanes05} or C*-algebra of two projections~\cite{twoproj}, any choice of operators $A_1,A_2$ and $B_1,B_2$ is a representation this algebra and so it decomposes into irreducibles. We first show that $\Qm(A_i,B_j)=\conv\{\Qmobs(u,v)|\,u\in\Sigma_A,v\in\Sigma_B\}$ for some correlation sets $\Qmobs(u,v)$ to be defined later. Indeed, let $\Pi_i$ denote the central projection onto the $i^{\mathrm{th}}$ irreducible summand, then restricted to this summand the operator $A_1A_2+A_2A_1=2u\idty$ for some $u\in\Sigma_A$ because it commutes with all generators. This can be used to drastically simplify polynomials in $A_1,A_2$, so that they generate the algebra of $2\times2$-matrices, except at the endpoints $u=\pm1$, where they form a commutative subalgebra. That is, up to isomorphism we can take $A_1=\sigma_1$, and $A_2=\cos\alpha\,\sigma_1+\sin\alpha\,\sigma_3$, so that $u=\cos\alpha$. The same holds for Bob with central parameter $v$, and the operators are combined as $A_i\otimes B_j$. Whenever $\lambda_i=\tr(\rho \Pi_i)\neq0$, we can define the state $\rho_i=\lambda_i^{-1} \Pi_i\rho \Pi_i$, and the corresponding correlation $c^i\in\Qmobs(u,v)$, the affine image of the state space of $\Cx^2\otimes\Cx^2$ under $\rho\mapsto\tr(\rho A_i(u)\otimes B_j(v))$. Thus, $c\in\Qm(A_i,B_j)$ implies that it is a convex hull of points in $\Qmobs(u,v)$. The converse is also true.

The sets $\Qmobs(u,v)$ are affinely isomorphic to each other except when $u,v=\pm1$. To see this, observe that $A_i(u)=\sum_{i'}S(u)_{ii'}A_{i'}(0)$, likewise for $B_j(v)$, and consequently $c(u,v)=S(u)c(0,0)S(v)^\top$ where both $c(u,v)\in\Rl^4$ and $c(0,0)\in\Rl^4$ are formed with the same density operator $\rho$, namely $c_{ij}(u,v)=\tr\rho A_i(u)\otimes B_j(v)$. This gives $\Qmobs(u,v) = S(u)\Qmobs(0,0)S(v)^\top$, which is sufficient for the purpose of this Proposition. Note, in addition, that $\Qmobs(u,v)$ are all non-isometrically isomorphic to $\Qmobs(0,0)$ since $S(u)$ are invertible whenever $-1 < u < 1$.

It remains to show that $\partial_e\Qmobs(0,0)=\QmObs$. Since all $A_i(0)\otimes B_j(0)$ commute with $U=\sigma_2\otimes\sigma_2$, we can assume without loss that $\rho$ commutes with $U$. Moreover, extreme points must come from pure states $\rho=\kettbra\psi$, which by the argument in \reff{sec:Peasy} we can choose as given by a real vector $\psi$. Together these imply $\psi_+=(\cos t,\sin t,\sin t,-\cos t)/\sqrt2$ or $\psi_-=(\cos t,\sin t,-\sin t,\cos t)/\sqrt2$, which gives two families of extreme points
\begin{equation}\label{O2corrs}
  \begin{pmatrix} -\cos2t & \sin2t \\ \sin2t  & \cos2t \end{pmatrix} \qquad\text{and}\qquad \begin{pmatrix} \cos2t & -\sin2t \\ \sin2t  & \cos2t \end{pmatrix},
\end{equation}
belonging to two orthogonal subspaces of $\Rl^4$. These are exactly all the $2\times2$ orthogonal matrices: the reflections and rotations.
\end{proof}

\begin{proof}[of \reff{prop:O2cando}]
\indent The general element of $\Qmobs(0,0)$ can be written as
\begin{equation}\label{cQ00}
   c=\lambda\,\begin{pmatrix} \cos\alpha & -\sin\alpha \\ \sin\alpha  & \cos\alpha \end{pmatrix}
      +\mu\,\begin{pmatrix} \cos\beta & \sin\beta \\ \sin\beta  & -\cos\beta \end{pmatrix},
\end{equation}
where $\lambda,\mu\geq0$, and $\lambda+\mu<1$. The two extremal circles are, by definition, the points with $\{\lambda,\mu\}=\{0,1\}$. This covers convex combinations form different circles, when $\lambda+\mu=1$. If some contributions to a convex combination come from the same circle, this just reduces the corresponding coefficient. From this description it is clear that all points with $\lambda+\mu<1$ belong to the interior of the body. The respective planes are given by $c_{11}-c_{22}=c_{12}+c_{21}=0$, and $c_{11}+c_{22}=c_{12}-c_{21}=0$, respectively. They are clearly Euclidean orthogonal. The body is symmetric under rotations in any of these planes, so all connecting edges are symmetry equivalent, and one can also check directly that they are all exposed. Note that the image of a cylinder that three-dimensional geometric intuition maybe suggests as the convex hull of two circles is entirely misleading here: The interiors of the circles are not faces but in the interior of the body.

To get the semialgebraic description, note that, with $s= \sum_{ij}c_{ij}^2$,
\begin{eqnarray}\label{Q00radii}
  4\lambda^2&=&(c_{11}+c_{22})^2+(c_{12}-c_{21})^2=s +2\det c \nonumber\\
  4\mu^2    &=&(c_{11}-c_{22})^2+(c_{12}+c_{21})^2=s -2\det c.
\end{eqnarray}
So in order to get a polynomial condition for the $c_{ij}$ we need to express the necessary and sufficient condition $\lambda+\mu\leq1$ in terms of $\lambda^2$ and $\mu^2$. By successive squaring we get  first $2\lambda\mu\leq(1-\lambda^2-\mu^2)$, and then $0\leq(1-2\lambda^2-2\mu^2+(\lambda^2-\mu^2)^2)$.
Using $s=2(\lambda^2+\mu^2)$ and $\det c=(\lambda^2-\mu^2)$ gives the  inequality for $\ell$ given in the proposition. On the boundary $\lambda+\mu=1$ these inequalities are equalities. So this polynomial is clearly the algebraic description of the boundary. However, the second squaring step is not reversible, so we get some unwanted elements in the Zariski closure. As for $\Qm$, where we had to augment the boundary polynomial $h$ by an additional (non-unique) polynomial inequality $g\geq0$, there are different ways of achieving this. \reff{fig:Q00-ears} is the map for this purpose. We see that we can either supplement the polynomial inequality $\lambda^2+\mu^2\leq1$, meaning $s\leq2$, or by the determinant inequality given in the proposition.

\begin{figure}\centering
	\includegraphics[width=5cm]{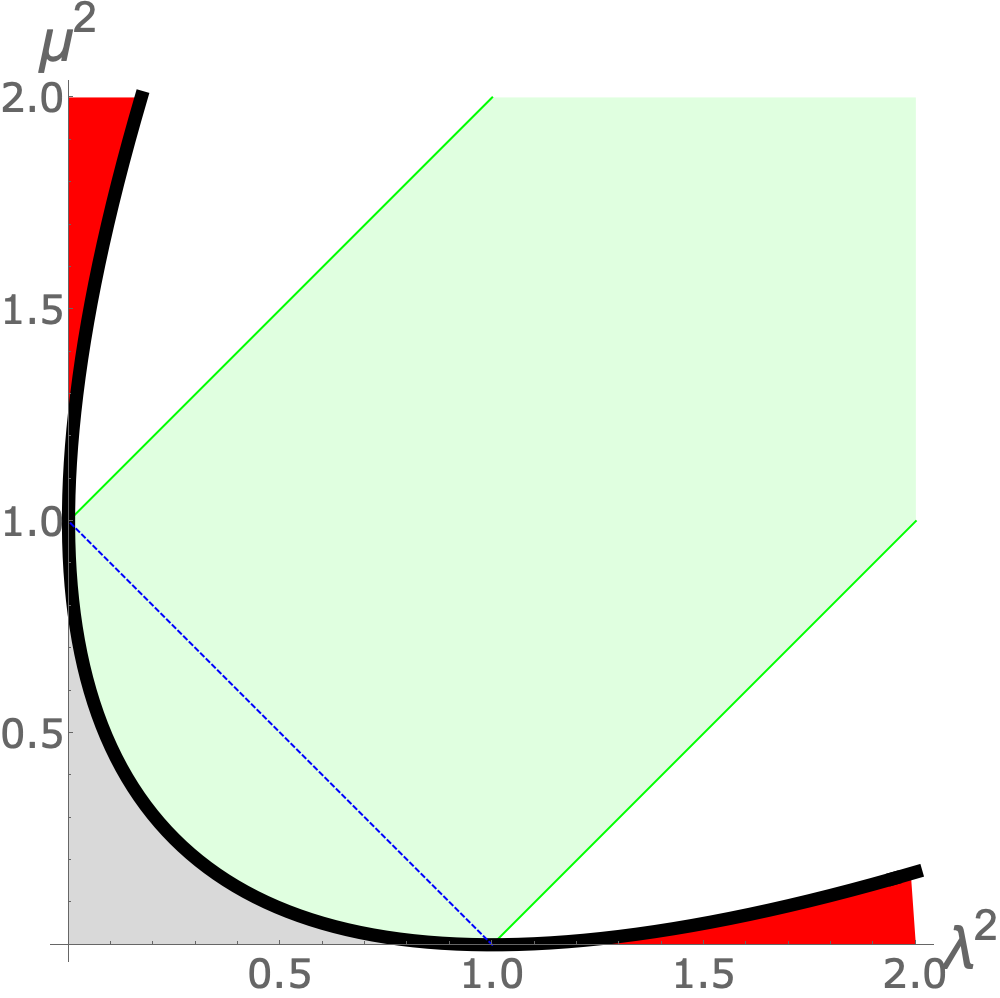} 
	\caption{The condition $\lambda+\mu\leq1$ is expressed in terms of $\lambda^2$ and $\mu^2$ as the gray shaded region near the origin. The bounding parabola extends further, so the inequality which it expresses needs to be supplemented by further conditions to exclude the red shaded areas. We choose the intersection with the green strip, which amounts to a condition on $\det c$. But other choices, e.g., the triangle bounded by the dashed blue line would also work. The latter would express that $\Qmobs(0,0)$ is contained in a ball of radius $\sqrt2$.}
\label{fig:Q00-ears}
\end{figure}
\end{proof}

\begin{proof}[of \reff{prop:det}]
\indent We have shown in \reff{prop:det} that on $\Qmobs(0,0)$ the determinant inequality holds, and by \reff{prop:fixedObs} and the observation that $\det S(u)\leq1$, the same holds for all components $\Qm(A_i,B_j)$ with different singleton spectra $\Sigma_A,\Sigma_B$. This would finish the argument if $\det$ were a convex function. However, it clearly is not.

We therefore take a more global approach using the detailed knowledge compiled in the paper. First of all, it suffices to show the determinant inequality on the boundary $\partial\Qm$, because the determinant is homogenous and every point $c\in\Qm$ can be written as $c=\lambda c'$ with $\lambda\leq1$ and $c'\in\partial\Qm$. This shows that also the cases of equality can only lie on the boundary.
For the curved tetrahedra we can use the cosine parametrization \reff{prop:paramExt}. The determinant is easily evaluated and simplifies to
\begin{equation}\label{Q00detc}
  \det c= -\sin(\alpha+\beta)\,\sin(\alpha+\gamma) \in[-1,1].
\end{equation}
On an $\Ns$-facet, say $\{c_{11}=1\}$, the determinant is an affine function of $c_{22}$. Hence its maximum will be attained on the bounding elliptope, where the previous argument applies.

For the case of equality it suffices to find points in the parametrized surface with $\det c=-1$, since the sign can be reversed by a symmetry transformation. From \eqref{Q00detc} we find, up to an overall sign for all angles and modulo $2\pi$ that $\alpha+\beta=\pi/2=\alpha+\gamma$. So we get $\beta=\gamma$, making $c$ Alice-Bob symmetric. Moreover, the fourth angle is
$\delta=-(\alpha+\beta+\gamma)=-\alpha-2(\pi/2-\alpha)=\alpha-\pi$, so that $c_{22}=-c_{11}$. This singles out the circle of reflection matrices, shown in red in \reff{fig:det}.
\end{proof}

\section{Outlook}
\subsection{What more of $\Qm$?}
There are some geometric aspects of $\Qm$ we did not discuss, although they might be natural and interesting.

\begin{itemize}
\item {\it Robustness of self-testing}\\
In the practice of quantum key distribution, the correlation $c$ is determined by statistical evaluation. The assumption that $c$ is an extremal point is never exactly verifiable. Therefore, one needs explicit bounds of the sort: When $c$ is known up to accuracy $\varepsilon$ (confidence intervals), then the eavesdropper cannot know more than $\delta$ about the bits used for key generation. Various definitions of ``accuracy'' and ``know more'' can be given. In any case, the concrete bound $\delta$ as a function of $\varepsilon$ will depend also on the local geometry of $\Qm$, particularly the curvature. Giving more details here would have turned this into a paper on key distribution. Generally speaking, self-testing is a robust phenomenon: Near-extremal correlations allow the conclusion that the observables involved can be deformed (norm-)slightly so that they turn into the minimal two-qubit example \cite{SW2}.

\item {\it Integral curvatures}\\
As mentioned in \reff{sec:convNC}, the Steiner volume polynomial contains information about curvature integrals of the boundary. In \reff{sec:Vol} we merely determined the volume, but no further coefficients.

\item{\it Weakly self-dual geometry}\\
The Hadamard matrix defines an indefinite pseudo-Euclidean metric on $\Rl^4$, for which $\Qm$ is self-dual. Some natural questions come with this structure, but it is unclear whether it sheds any light on  $\Qm$.

\item{\it Constrained Hilbert space dimension}\\
The basics were discussed in \reff{sec:dimH}, but a full characterization is still lacking.
\end{itemize}

\subsection{How does this generalize?}
Many of the techniques described above were originally drafted to address more general situations.
Let us briefly indicate their natural levels of generality.

\begin{itemize}
\item \emph{The full $222$ case, and the C*-algebra generated by two projections}\\
  In the minimal $222$ case, without the zero marginals condition, the quantum body $\Qm$ is a convex body in $\Rl^8$. The semidefinite matrix completion point of view is not directly effective for breaking this down to a finite dimensional problem. As has been noted also by Masanes \cite{Masanes05} (actually more generally for the N22 case) this can be achieved by the representation theory of the universal C*-algebra generated by two projections \cite{twoproj}. This provides a description of the extreme points parametrized by the product of spheres $\mathbb{S}^1\times\mathbb{S}^1\times\mathbb{S}^3$, analogous to \reff{prop:paramExt}.   A {\tt Macaulay2} computation reveals that this variety has degree $40$ and its prime ideal is generated by $28$ polynomials whose degrees
  are $5,6,7$ and $8$. This may be a starting point for the analysis of the $222$ case.
\item {\it Higher universal correlation bodies}\\
  It is clear that the analog of $\Qm$ can be defined for larger $N$, $M$, or $K$, or any specification of parties settings and outcomes.
  However, the construction is notoriously non-constructive; see \cite{ART}.
\item {\it Semidefinite hierarchies}\\
   We used this extensively, and got a complete characterization of the minimal $\Qm$ out of it, just using the bottom level (as explained in \autoref{sec:specshadow}) of the hierarchy. In fact, this characterizes the minimal case within the full 222 scenario with marginals: If the bottom-level semidefinite relaxation of the full 222 case is tight, i.e., $c\in$\Qt4, self-testing ~\autoref{prop:selftest} (2) implies zero marginals, so we must be in the $222|0$ case. For larger $N$, $M$, or $K$, a gap can be expected to exist and become rapidly larger. Hierarchies are still the key tool for getting upper bounds, but tightness is too much to ask.
   It should also be noted that in spite of the proven convergence of the hierarchy it is computationally unfeasible to really push this to high levels.
   It is unclear (to us) for which other cases a tight bound at some higher but finite hierarchy level holds.

   \item{\it Parametrized extreme points}\\
  Although this is perhaps best understood via the two-projections theory, another parametrization of the extreme points (not explicitly using this theory) in the N22 case was given in~\cite{WW01a}.
\item{\it Duality}\\
  The duality in \reff{thm:selfdual} clearly depends on minimality: The duality of $\Ns$ and $\Cl$ requires at least the dimensions of these sets to coincide, which fails for $MK>4$.
\item{\it Correlation matrices and Clifford algebras}\\
  All the above extensions drop the 0-marginal condition or increase $N$. We
  stress that Tsirelson's technique of correlation matrices is an extension in another direction, i.e., to 2M2\marge0, with general~$M$.
\item{\it Algebra}\\
  Algebraic methods are expected to apply once a reduction to finite dimension has been achieved by other means. We found them directly useful in the full 222 case (work in progress), but there are clearly also applications to N22 and 2M2 cases. However, the complexity of algebraic characterizations can be expected to increase very rapidly. Even worse are the case distinctions and inequalities of a {\it semi}-algebraic description. We can understand this from the classical case, the characterization of $\Cl$. Here the algebra is linear, but the inequality part, i.e., the determination of all Bell inequalities, is a family of convex hull problems, which is known to grow badly \cite{Pitowsky}.  So no general results can be expected, but as in the classical case \cite[Problem~1]{problemSite} one can look for trails into the wilderness, e.g., infinite families of models defined by special properties or symmetries.
\item \emph{Algebraic statistics} \\
 Consider the $\pm 1$-valued random variables $A_1,A_2,B_1,B_2$ satisfying
\eqref{eq:ABhypotheses}. Their statistical model is the graphical model whose graph $G$ is the $4$-cycle with edges $A_1-B_1$, $B_1-A_2$, $A_2-B_2$, $B_2-A_1$. This is \cite[Example $4$]{GMS} up to relabelling. The sufficient statistics of this model are obtained by applying the linear map $A(G)$ in \cite[Example $4$]{GMS}. The image of this map is our classical polytope $\Cl$. In particular this map is a bijection between the model and $\Cl$ and inverting it is called \emph{maximum likelihood estimation} (MLE). We can recover also $\Ns$ from this construction, and more in general such polytopes can be defined for any toric model. For an undirected graphical model $G$ we have that $\Cl = \Ns$ if and only if the graph is decomposable; the four-cycle model is the smallest non-decomposable model so we can deduce also from this argument that the inclusion is strict.
It would be particularly interesting to examine the general 222 case through the lens of
algebraic statistics.
\end{itemize}

\newpage
\section*{Acknowledgements}
We would like to thank M. Navascu{\'e}s, N. Gisin, A. Ac{\'\i}n, J. Kaniewski, L. Masanes, and S. Pironio for helpful comments regarding the literature. We are grateful to the anonymous referees who gave detailed feedback that helped us improve our manuscript. T.P.L gratefully acknowledges support from the Alexander von Humboldt Foundation and the Austrian Academy of Sciences (project number M 2812-N).

\bibliographystyle{quantum}
\bibliography{ExCorLit}

\begin{thebibliography}{10}

\bibitem{Aspect}
Alain Aspect, Philippe Grangier, and G\'erard Roger.
\newblock ``Experimental realization of {Einstein-Podolsky-Rosen-Bohm
  Gedankenexperiment}: A new violation of {B}ell's inequalities''.
\newblock \href{https://dx.doi.org/10.1103/PhysRevLett.49.91}{Phys. Rev. Lett.
  {\bf 49}, 91--94}~(1982).

\bibitem{lopholefree1}
B.~Hensen, R.~Hanson, et~al.
\newblock ``Loophole-free {B}ell inequality violation using electron spins
  separated by 1.3 kilometres''.
\newblock
  \href{https://dx.doi.org/http://dx.doi.org/10.1038/nature15759}{Nature {\bf
  526}, 682 EP --}~(2015).
\newblock  \href{http://arxiv.org/abs/1508.05949}{arXiv:1508.05949}.

\bibitem{loopholefree2}
N.~Sangouard, J.-D. Bancal, N.~Gisin, W.~Rosenfeld, P.~Sekatski, M.~Weber, and
  H.~Weinfurter.
\newblock ``Loophole-free {B}ell test with one atom and less than one photon on
  average''.
\newblock \href{https://dx.doi.org/10.1103/PhysRevA.84.052122}{Phys. Rev. A
  {\bf 84}, 052122}~(2011).
\newblock  \href{http://arxiv.org/abs/1108.1027}{arXiv:1108.1027}.

\bibitem{Bell}
J.~S. Bell.
\newblock ``On the {E}instein {P}odolsky {R}osen paradox''.
\newblock \href{https://dx.doi.org/10.1103/PhysicsPhysiqueFizika.1.195}{Physics
  {\bf 1}, 195--200}~(1964).

\bibitem{CHSH}
John~F. Clauser, Michael~A. Horne, Abner Shimony, and Richard~A. Holt.
\newblock ``Proposed experiment to test local hidden-variable theories''.
\newblock \href{https://dx.doi.org/10.1103/PhysRevLett.23.880}{Phys. Rev. Lett.
  {\bf 23}, 880--884}~(1969).

\bibitem{problemSite}
R.~F. Werner et~al.
\newblock ``Open quantum problems''.
\newblock  url:~\url{https://oqp.iqoqi.oeaw.ac.at/}.

\bibitem{Tsi85}
Boris~S. Tsirelson.
\newblock ``Quantum analogues of the {B}ell inequalities. the case of two
  spatially separated domains''.
\newblock \href{https://dx.doi.org/10.1007/BF01663472}{J. Soviet Math. {\bf
  36}, 557--570}~(1987).

\bibitem{WW01a}
R.~F. Werner and M.~M. Wolf.
\newblock ``All multipartite {B}ell-correlation inequalities for two dichotomic
  observables per site''.
\newblock \href{https://dx.doi.org/10.1103/PhysRevA.64.032112}{Phys. Rev. A
  {\bf 64}, 032112}~(2001).
\newblock
  \href{http://arxiv.org/abs/quant-ph/0102024}{arXiv:quant-ph/0102024}.

\bibitem{Slofstra}
William Slofstra.
\newblock ``The set of quantum correlations is not closed''.
\newblock \href{https://dx.doi.org/10.1017/fmp.2018.3}{Forum of Mathematics, Pi
  {\bf 7}, e1}~(2019).
\newblock  \href{http://arxiv.org/abs/1703.08618}{arXiv:1703.08618}.

\bibitem{Tsir_Scholz}
Volkher~B. Scholz and R.~F. Werner.
\newblock ``Tsirelson's problem''~(2008).
\newblock  \href{http://arxiv.org/abs/0812.4305}{arXiv:0812.4305}.

\bibitem{TS93}
Boris~S Tsirelson.
\newblock ``Some results and problems on quantum {B}ell-type inequalities''.
\newblock Hadronic Journal Supplement {\bf 8}, 329--345~(1993).
\newblock  url:~\url{https://www.tau.ac.il/~tsirel/download/hadron.html}.

\bibitem{navascues2008}
Miguel Navascues, Stefano Pironio, and Antonio Ac{\'\i}n.
\newblock ``A convergent hierarchy of semidefinite programs characterizing the
  set of quantum correlations''.
\newblock \href{https://dx.doi.org/doi:10.1088/1367-2630/10/7/073013}{New J.
  Phys. {\bf 10}, 073013}~(2008).
\newblock  \href{http://arxiv.org/abs/0803.4290}{arXiv:0803.4290}.

\bibitem{Tproblem_Werner}
M.~Junge, M.~Navascues, C.~Palazuelos, D.~Perez-Garcia, V.~B. Scholz, and R.~F.
  Werner.
\newblock ``Connes' embedding problem and {T}sirelson's problem''.
\newblock \href{https://dx.doi.org/10.1063/1.3514538}{J. Math. Phys. {\bf 52},
  012102}~(2011).
\newblock  \href{http://arxiv.org/abs/1008.1142}{arXiv:1008.1142}.

\bibitem{Fritz}
Tobias Fritz.
\newblock ``Tsirelson's problem and {K}irchberg's conjecture''.
\newblock \href{https://dx.doi.org/10.1142/S0129055X12500122}{Rev. Math. Phys.
  {\bf 24}, 1250012}~(2012).
\newblock  \href{http://arxiv.org/abs/1008.1168}{arXiv:1008.1168}.

\bibitem{Tsirelson_solved}
Zhengfeng Ji, Anand Natarajan, Thomas Vidick, John Wright, and Henry Yuen.
\newblock ``{MIP}\textasteriskcentered={RE}''~(2020).
\newblock  \href{http://arxiv.org/abs/2001.04383}{arXiv:2001.04383}.

\bibitem{Ziegler}
G\"unther~M. Ziegler.
\newblock ``Lectures on polytopes''.
\newblock \href{https://dx.doi.org/10.1007/978-1-4613-8431-1}{Springer}.
  Berlin~(1995).

\bibitem{Michalek}
Mateusz Micha{\l}ek and Bernd Sturmfels.
\newblock ``Invitation to nonlinear algebra''.
\newblock Volume 211 of Graduate Studies in Mathematics.
\newblock AMS. ~(2021).

\bibitem{Blekherman}
Grigoriy Blekherman, Pablo Parrilo, and Rekha Thomas.
\newblock ``Semidefinite optimization and convex algebraic geometry''.
\newblock \href{https://dx.doi.org/10.1137/1.9781611972290}{MOS-SIAM Series on
  Optimization 13}. SIAM. Philadelphia~(2012).

\bibitem{Uhler}
Bernd Sturmfels and Caroline Uhler.
\newblock ``Multivariate {G}aussians, semidefinite matrix completion, and
  convex algebraic geometry''.
\newblock \href{https://dx.doi.org/10.1007/s10463-010-0295-4}{Ann. Inst.
  Statist. Math. {\bf 62}, 603--638}~(2010).
\newblock  \href{http://arxiv.org/abs/0906.3529}{arXiv:0906.3529}.

\bibitem{Scheiderer}
Claus Scheiderer.
\newblock ``Spectrahedral shadows''.
\newblock \href{https://dx.doi.org/10.1137/17M1118981}{SIAM J.~Appl.~Algebra
  Geometry {\bf 2}, 26--44}~(2018).
\newblock  \href{http://arxiv.org/abs/1612.07048}{arXiv:1612.07048}.

\bibitem{Tsi80}
B.~S. Cirel'son.
\newblock ``Quantum generalizations of {B}ell's inequality''.
\newblock \href{https://dx.doi.org/10.1007/BF00417500}{Lett. Math. Phys. {\bf
  4}, 93--100}~(1980).

\bibitem{Kiukas}
Jukka Kiukas and Reinhard~F. Werner.
\newblock ``Maximal violation of {B}ell inequalities by position
  measurements''.
\newblock \href{https://dx.doi.org/10.1063/1.3447736}{J. Math. Phys. {\bf 51},
  072105}~(2010).
\newblock  \href{http://arxiv.org/abs/0912.3740}{arXiv:0912.3740}.

\bibitem{Lan88}
Lawrence~J. Landau.
\newblock ``Empirical two-point correlation functions''.
\newblock \href{https://dx.doi.org/10.1007/BF00732549}{Found. Phys. {\bf 18},
  449--460}~(1988).

\bibitem{Masanes03}
L~Masanes.
\newblock ``Necessary and sufficient condition for quantum-generated
  correlations''~(2003)
  \href{http://arxiv.org/abs/quant-ph/0309137}{arXiv:quant-ph/0309137}.

\bibitem{singletST}
Yukun Wang, Xingyao Wu, and Valerio Scarani.
\newblock ``All the self-testings of the singlet for two binary measurements''.
\newblock \href{https://dx.doi.org/10.1088/1367-2630/18/2/025021}{New J. Phys.
  {\bf 18}, 025021}~(2016).
\newblock  \href{http://arxiv.org/abs/1511.04886}{arXiv:1511.04886}.

\bibitem{doherty2008}
Andrew~C Doherty, Yeong-Cherng Liang, Ben Toner, and Stephanie Wehner.
\newblock ``The quantum moment problem and bounds on entangled multi-prover
  games''.
\newblock In 23rd Annual IEEE Conference on Computational Complexity.
\newblock \href{https://dx.doi.org/10.1109/CCC.2008.26}{Pages 199--210}.
\newblock IEEE~(2008).
\newblock  \href{http://arxiv.org/abs/0803.4373}{arXiv:0803.4373}.

\bibitem{N22duality}
Tobias Fritz.
\newblock ``Polyhedral duality in {B}ell scenarios with two binary
  observables''.
\newblock \href{https://dx.doi.org/10.1063/1.4734586}{J. Math. Phys. {\bf 53},
  072202}~(2012).
\newblock  \href{http://arxiv.org/abs/1202.0141}{arXiv:1202.0141}.

\bibitem{MayersYao}
Dominic Mayers and Andrew Yao.
\newblock ``Self testing quantum apparatus''.
\newblock \href{https://dx.doi.org/10.26421/QIC4.4-3}{Quantum Info. Comput.
  {\bf 4}, 273–286}~(2004).
\newblock
  \href{http://arxiv.org/abs/quant-ph/0307205}{arXiv:quant-ph/0307205}.

\bibitem{SW1}
Stephen~J. Summers and Reinhard~F. Werner.
\newblock ``Maximal violation of {B}ell's inequalities is generic in quantum
  field theory''.
\newblock \href{https://dx.doi.org/10.1007/BF01207366}{Commun. Math. Phys. {\bf
  110}, 247--259}~(1987).

\bibitem{Masanes05}
L~Masanes.
\newblock ``Extremal quantum correlations for n parties with two dichotomic
  observables per site''~(2005)
  \href{http://arxiv.org/abs/quant-ph/0512100}{arXiv:quant-ph/0512100}.

\bibitem{Le18}
Le~Phuc Thinh, Antonios Varvitsiotis, and Yu~Cai.
\newblock ``Geometric structure of quantum correlators via semidefinite
  programming''.
\newblock \href{https://dx.doi.org/10.1103/PhysRevA.99.052108}{Phys. Rev. A
  {\bf 99}, 052108}~(2019).
\newblock  \href{http://arxiv.org/abs/1809.10886}{arXiv:1809.10886}.

\bibitem{bellnonlocality}
Nicolas Brunner, Daniel Cavalcanti, Stefano Pironio, Valerio Scarani, and
  Stephanie Wehner.
\newblock ``{B}ell nonlocality''.
\newblock \href{https://dx.doi.org/10.1103/RevModPhys.86.419}{Rev. Mod. Phys.
  {\bf 86}, 419--478}~(2014).
\newblock  \href{http://arxiv.org/abs/1303.2849}{arXiv:1303.2849}.

\bibitem{Goh}
Koon~Tong Goh, J\k{e}drzej Kaniewski, Elie Wolfe, Tam\'as V\'ertesi, Xingyao
  Wu, Yu~Cai, Yeong-Cherng Liang, and Valerio Scarani.
\newblock ``Geometry of the set of quantum correlations''.
\newblock \href{https://dx.doi.org/10.1103/PhysRevA.97.022104}{Phys. Rev. A
  {\bf 97}, 022104}~(2018).
\newblock  \href{http://arxiv.org/abs/1710.05892}{arXiv:1710.05892}.

\bibitem{reviewST}
Ivan {\v{S}}upi{\'{c}} and Joseph Bowles.
\newblock ``Self-testing of quantum systems: a review''.
\newblock \href{https://dx.doi.org/10.22331/q-2020-09-30-337}{Quantum {\bf 4},
  337}~(2020).
\newblock  \href{http://arxiv.org/abs/1904.10042}{arXiv:1904.10042}.

\bibitem{schwonnek2020qkd}
Rene Schwonnek, Koon~Tong Goh, Ignatius~W. Primaatmaja, Ernest Y.~Z. Tan,
  Ramona Wolf, Valerio Scarani, and Charles C.~W. Lim.
\newblock ``Device-independent quantum key distribution with random key
  basis''.
\newblock \href{https://dx.doi.org/10.1038/s41467-021-23147-3}{Nat. Commun.
  {\bf 12}, 2880}~(2020).
\newblock  \href{http://arxiv.org/abs/2005.02691}{arXiv:2005.02691}.

\bibitem{tan2020qkd}
Ernest Y.~Z. Tan, René Schwonnek, Koon~Tong Goh, Ignatius~William Primaatmaja,
  and Charles C.~W. Lim.
\newblock ``Computing secure key rates for quantum key distribution with
  untrusted devices''.
\newblock \href{https://dx.doi.org/10.1038/s41534-021-00494-z}{npj Quantum Inf.
  {\bf 7}, 158}~(2021).
\newblock  \href{http://arxiv.org/abs/1908.11372}{arXiv:1908.11372}.

\bibitem{VW}
K.~G.~H. Vollbrecht and R.~F. Werner.
\newblock ``Entanglement measures under symmetry''.
\newblock \href{https://dx.doi.org/10.1103/PhysRevA.64.062307}{Phys. Rev. A
  {\bf 64}, 062307}~(2001).
\newblock
  \href{http://arxiv.org/abs/quant-ph/0010095}{arXiv:quant-ph/0010095}.

\bibitem{Bierhorst_2016}
Peter Bierhorst.
\newblock ``Geometric decompositions of {B}ell polytopes with practical
  applications''.
\newblock \href{https://dx.doi.org/10.1088/1751-8113/49/21/215301}{J. Phys. A
  {\bf 49}, 215301}~(2016).
\newblock  \href{http://arxiv.org/abs/1511.04127}{arXiv:1511.04127}.

\bibitem{L97}
Monique Laurent.
\newblock ``The real positive semidefinite completion problem for
  series-parallel graphs''.
\newblock \href{https://dx.doi.org/10.1016/0024-3795(95)00741-5}{Linear Algebra
  and its Applications {\bf 252}, 347--366}~(1997).

\bibitem{Lissknot}
Vaughan F.~R. Jones and J.~H. Przytycki.
\newblock ``Lissajous knots and billiard knots''.
\newblock \href{https://dx.doi.org/10.4064/-42-1-145-163}{Banach Cent. Pub.
  {\bf 42}, 145--163}~(1998).

\bibitem{soccer}
Kaie Kubjas, Pablo~A Parrilo, and Bernd Sturmfels.
\newblock ``How to flatten a soccer ball''.
\newblock In Aldo Conca, Joseph Gubeladze, and Tim R{\"o}mer, editors,
  Homological and Computational Methods in Commutative Algebra.
\newblock \href{https://dx.doi.org/10.1007/978-3-319-61943-9_9}{Volume~20 of
  INdAM Ser., pages 141--162}.
\newblock Springer~(2017).

\bibitem{wootters}
Kathleen~S. Gibbons, Matthew~J. Hoffman, and William~K. Wootters.
\newblock ``Discrete phase space based on finite fields''.
\newblock \href{https://dx.doi.org/10.1103/physreva.70.062101}{Phys. Rev. A
  {\bf 70}, 062101}~(2004).
\newblock
  \href{http://arxiv.org/abs/quant-ph/0401155}{arXiv:quant-ph/0401155}.

\bibitem{psuncert}
Reinhard~F. Werner.
\newblock ``Uncertainty relations for general phase spaces''.
\newblock \href{https://dx.doi.org/10.1007/s11467-016-0558-5}{Frontiers of
  Physics {\bf 11}, 1--10}~(2016).
\newblock  \href{http://arxiv.org/abs/1601.03843}{arXiv:arxiv:1601.03843}.

\bibitem{Prasad}
Amritanshu Prasad, Ilya Shapiro, and M.K. Vemuri.
\newblock ``Locally compact abelian groups with symplectic self-duality''.
\newblock
  \href{https://dx.doi.org/https://doi.org/10.1016/j.aim.2010.04.023}{Adv.
  Math. {\bf 225}, 2429--2454}~(2010).
\newblock  \href{http://arxiv.org/abs/0906.4397}{arXiv:0906.4397}.

\bibitem{ciripoi2018computing}
Daniel Ciripoi, Nidhi Kaihnsa, Andreas L{\"o}hne, and Bernd Sturmfels.
\newblock ``Computing convex hulls of trajectories''.
\newblock \href{https://dx.doi.org/10.33044/revuma.v60n2a22}{Rev. Un. Mat.
  Argentina {\bf 60}, 637--662}~(2019).
\newblock  \href{http://arxiv.org/abs/1810.03547}{arXiv:1810.03547}.

\bibitem{Plaumann}
Daniel Plaumann, Rainer Sinn, and Jannik~Lennart Wesner.
\newblock ``Families of faces and the normal cycle of a convex semi-algebraic
  set''.
\newblock \href{https://dx.doi.org/10.1007/s13366-022-00657-9}{Beitr. Algebra
  Geom.}~(2022).
\newblock  \href{http://arxiv.org/abs/2104.13306}{arXiv:2104.13306}.

\bibitem{M2}
Daniel~R. Grayson and Michael~E. Stillman.
\newblock ``Macaulay2, a software system for research in algebraic geometry''.
\newblock Available at \url{http://www.math.uiuc.edu/Macaulay2/}.

\bibitem{Vinzant}
John Ottem, Kristian Ranestad, Bernd Sturmfels, and Cynthia Vinzant.
\newblock ``Quartic spectrahedra''.
\newblock \href{https://dx.doi.org/10.1007/s10107-014-0844-3}{Mathematical
  Programming, Ser. B {\bf 151}, 585--612}~(2015).
\newblock  \href{http://arxiv.org/abs/1311.3675}{arXiv:1311.3675}.

\bibitem{volumes}
Ad\'an Cabello.
\newblock ``How much larger quantum correlations are than classical ones''.
\newblock \href{https://dx.doi.org/10.1103/PhysRevA.72.012113}{Phys. Rev. A
  {\bf 72}, 012113}~(2005).
\newblock
  \href{http://arxiv.org/abs/quant-ph/0409192}{arXiv:quant-ph/0409192}.

\bibitem{sampling_nonlocal}
C.~E. Gonz{\'a}lez-Guill{\'e}n, C.~H. Jim{\'e}nez, C.~Palazuelos, and
  I.~Villanueva.
\newblock ``Sampling quantum nonlocal correlations with high probability''.
\newblock \href{https://dx.doi.org/10.1007/s00220-016-2625-8}{Commun. Math.
  Phys. {\bf 344}, 141--154}~(2016).
\newblock  \href{http://arxiv.org/abs/1412.4010}{arXiv:1412.4010}.

\bibitem{volumeElliptope}
C.~R. Johnson and G.~N{\ae}vdal.
\newblock ``The probability that a (partial) matrix is positive semidefinite''.
\newblock In I.~Gohberg, R.~Mennicken, and C.~Tretter, editors, Recent Progress
  in Operator Theory.
\newblock \href{https://dx.doi.org/10.1007/978-3-0348-8793-9_10}{Pages
  171--182}.
\newblock Basel~(1998). Birkh{\"a}user Basel.

\bibitem{Schaefer}
H.~H Schaefer and M.~P Wolff.
\newblock ``Topological vector spaces''.
\newblock \href{https://dx.doi.org/10.1007/978-1-4612-1468-7}{Springer}.
  ~(1999).

\bibitem{KarolHadamard}
Wojciech Tadej and Karol \`Zyczkowski.
\newblock ``A concise guide to complex {H}adamard matrices''.
\newblock \href{https://dx.doi.org/10.1007/s11080-006-8220-2}{Open Systems \&
  Information Dynamics {\bf 13}, 133--177}~(2006).
\newblock
  \href{http://arxiv.org/abs/quant-ph/0512154}{arXiv:quant-ph/0512154}.

\bibitem{Howard}
H.~Barnum, C.P. Gaebler, and A.~Wilce.
\newblock ``Ensemble steering, weak self-duality, and the structure of
  probabilistic theories''.
\newblock \href{https://dx.doi.org/10.1007/s10701-013-9752-2}{Found. Phys {\bf
  43}, 1411–1427}~(2013).
\newblock  \href{http://arxiv.org/abs/0912.5532}{arXiv:0912.5532}.

\bibitem{yannakakis2010}
Nikos Yannakakis.
\newblock ``Stampacchia's property, self-duality and orthogonality relations''.
\newblock \href{https://dx.doi.org/10.1007/s11228-011-0175-y}{Set-Valued and
  Variational Analysis {\bf 19}, 555–567}~(2011).
\newblock  \href{http://arxiv.org/abs/1008.4958}{arXiv:1008.4958}.

\bibitem{bcr:realag}
Jacek Bochnak, Michel Coste, and Marie-Fran{\c{c}}oise Roy.
\newblock ``Real algebraic geometry''.
\newblock \href{https://dx.doi.org/10.1007/978-3-662-03718-8}{Volume~36 of A
  Series of Modern Surveys in Mathematics}.
\newblock Springer Berlin, Heidelberg. ~(2013).

\bibitem{fu2014algebraic}
Joseph H.~G. Fu.
\newblock ``Algebraic integral geometry''.
\newblock \href{https://dx.doi.org/10.1007/978-3-0348-0874-3_2}{Pages 47--112}.
\newblock Springer Basel. Basel~(2014).
\newblock  \href{http://arxiv.org/abs/1103.6256}{arXiv:1103.6256}.

\bibitem{federer1959curvature}
Herbert Federer.
\newblock ``Curvature measures''.
\newblock \href{https://dx.doi.org/10.2307/1993504}{Trans. Amer. Math. Soc.
  {\bf 93}, 418--491}~(1959).

\bibitem{wintgen1982normal}
Peter Wintgen.
\newblock ``Normal cycle and integral curvature for polyhedra in {R}iemannian
  manifolds''.
\newblock In Gy. Soos and J.~Szenthe, editors, Differential Geometry.
\newblock Volume~21.
\newblock North-Holland, Amsterdam~(1982).

\bibitem{zahle1986integral}
Martina Z{\"a}hle.
\newblock ``Integral and current representation of {F}ederer's curvature
  measures''.
\newblock \href{https://dx.doi.org/10.1007/BF01195026}{Arch. Math. {\bf 46},
  557--567}~(1986).

\bibitem{cohen2003restricted}
David Cohen-Steiner and Jean-Marie Morvan.
\newblock ``Restricted {D}elaunay triangulations and normal cycle''.
\newblock In SCG '03: Proceedings of the nineteenth annual symposium on
  Computational geometry.
\newblock \href{https://dx.doi.org/10.1145/777792.777839}{Pages 312--321}.
\newblock ~(2003).

\bibitem{roussillon2017surface}
Pierre Roussillon and Joan~Alexis Glaun{\`e}s.
\newblock ``Surface matching using normal cycles''.
\newblock In Frank Nielsen and Fr{\'e}d{\'e}ric Barbaresco, editors, Geometric
  Science of Information.
\newblock \href{https://dx.doi.org/10.1007/978-3-319-68445-1_9}{Pages 73--80}.
\newblock Cham~(2017). Springer International Publishing.

\bibitem{su2019curvature}
Kehua Su, Na~Lei, Wei Chen, Li~Cui, Hang Si, Shikui Chen, and Xianfeng Gu.
\newblock ``Curvature adaptive surface remeshing by sampling normal cycle''.
\newblock \href{https://dx.doi.org/10.1016/j.cad.2019.01.004}{Computer-Aided
  Design {\bf 111}, 1--12}~(2019).

\bibitem{CLO}
David~A. Cox, John Little, and Donal O'Shea.
\newblock ``Ideals, varieties, and algorithms''.
\newblock \href{https://dx.doi.org/10.1007/978-3-319-16721-3}{Undergraduate
  Texts in Mathematics}. Springer Cham. ~(2015).
\newblock Fourth edition.

\bibitem{Raggio}
Guido~A. Raggio.
\newblock ``A remark on {B}ell's inequality and decomposable normal states''.
\newblock \href{https://dx.doi.org/10.1007/BF00416568}{Lett. Math. Phys. {\bf
  15}, 27--29}~(1988).

\bibitem{Reality}
Marc-Olivier Renou, David Trillo, Mirjam Weilenmann, Thinh~P. Le, Armin
  Tavakoli, Nicolas Gisin, Antonio Ac{\'i}n, and Miguel Navascu{\'e}s.
\newblock ``Quantum theory based on real numbers can be experimentally
  falsified''.
\newblock \href{https://dx.doi.org/10.1038/s41586-021-04160-4}{Nature {\bf
  600}, 625--629}~(2021).
\newblock  \href{http://arxiv.org/abs/2101.10873}{arXiv:2101.10873}.

\bibitem{allpure}
Andrea Coladangelo, Koon~Tong Goh, and Valerio Scarani.
\newblock ``All pure bipartite entangled states can be self-tested''.
\newblock \href{https://dx.doi.org/10.1038/ncomms15485}{Nature Commun. {\bf 8},
  15485}~(2017).
\newblock  \href{http://arxiv.org/abs/1611.08062}{arXiv:1611.08062}.

\bibitem{BB84}
Charles~H. Bennett and Gilles Brassard.
\newblock ``Quantum cryptography: Public key distribution and coin tossing''.
\newblock
  \href{https://dx.doi.org/https://doi.org/10.1016/j.tcs.2014.05.025}{Theoret.
  Comp. Sci. {\bf 560}, 7--11}~(2014).
\newblock  \href{http://arxiv.org/abs/2003.06557}{arXiv:2003.06557}.

\bibitem{FranzWerner}
T.~Franz, F.~Furrer, and R.~F. Werner.
\newblock ``Extremal quantum correlations and cryptographic security''.
\newblock \href{https://dx.doi.org/10.1103/PhysRevLett.106.250502}{Phys. Rev.
  Lett. {\bf 106}, 250502}~(2011).
\newblock  \href{http://arxiv.org/abs/1010.1131}{arXiv:1010.1131}.

\bibitem{weakST}
J\k{e}drzej Kaniewski.
\newblock ``Weak form of self-testing''.
\newblock \href{https://dx.doi.org/10.1103/PhysRevResearch.2.033420}{Phys. Rev.
  Research {\bf 2}, 033420}~(2020).
\newblock  \href{http://arxiv.org/abs/1910.00706}{arXiv:1910.00706}.

\bibitem{privacy}
C.~H. {Bennett}, G.~{Brassard}, C.~{Crepeau}, and U.~M. {Maurer}.
\newblock ``Generalized privacy amplification''.
\newblock \href{https://dx.doi.org/10.1109/18.476316}{IEEE Transactions on
  Information Theory {\bf 41}, 1915--1923}~(1995).

\bibitem{DIQKD1}
Pavel Sekatski, Jean-Daniel Bancal, Xavier Valcarce, Ernest Y.-Z. Tan, Renato
  Renner, and Nicolas Sangouard.
\newblock ``Device-independent quantum key distribution from generalized {CHSH}
  inequalities''.
\newblock \href{https://dx.doi.org/10.22331/q-2021-04-26-444}{{Quantum} {\bf
  5}, 444}~(2021).
\newblock  \href{http://arxiv.org/abs/2009.01784}{arXiv:2009.01784}.

\bibitem{DIQKD}
Ernest Y.-Z. Tan, Pavel Sekatski, Jean-Daniel Bancal, Ren{\'{e}} Schwonnek,
  Renato Renner, Nicolas Sangouard, and Charles C.-W. Lim.
\newblock ``Improved {DIQKD} protocols with finite-size analysis''.
\newblock \href{https://dx.doi.org/10.22331/q-2022-12-22-880}{{Quantum} {\bf
  6}, 880}~(2022).
\newblock  \href{http://arxiv.org/abs/2012.08714}{arXiv:2012.08714}.

\bibitem{giustina15}
Marissa Giustina et~al.
\newblock ``Significant-loophole-free test of {B}ell's theorem with entangled
  photons''.
\newblock \href{https://dx.doi.org/10.1103/PhysRevLett.115.250401}{Phys. Rev.
  Lett. {\bf 115}, 250401}~(2015).
\newblock  \href{http://arxiv.org/abs/1511.03190}{arXiv:1511.03190}.

\bibitem{shalm15}
Lynden~K. Shalm et~al.
\newblock ``Strong loophole-free test of local realism''.
\newblock \href{https://dx.doi.org/10.1103/PhysRevLett.115.250402}{Phys. Rev.
  Lett. {\bf 115}, 250402}~(2015).
\newblock  \href{http://arxiv.org/abs/1511.03189}{arXiv:1511.03189}.

\bibitem{diqkdexp1}
D.~P Nadlinger, J.-D. Bancal, and et~al.
\newblock ``Experimental quantum key distribution certified by {B}ell's
  theorem''.
\newblock \href{https://dx.doi.org/10.1038/s41586-022-04941-5}{Nature {\bf
  607}, 682--686}~(2022).
\newblock  \href{http://arxiv.org/abs/2109.14600}{arXiv:2109.14600}.

\bibitem{diqkdexp2}
Wei Zhang, Harald Weinfurter, et~al.
\newblock ``A device-independent quantum key distribution system for distant
  users''.
\newblock \href{https://dx.doi.org/10.1038/s41586-022-04891-y}{Nature {\bf
  607}, 687--691}~(2022).
\newblock  \href{http://arxiv.org/abs/2110.00575}{arXiv:2110.00575}.

\bibitem{diqkdexp3}
Feihu Xu, Yu-Zhe Zhang, Qiang Zhang, and Jian-Wei Pan.
\newblock ``Device-independent quantum key distribution with random
  postselection''.
\newblock \href{https://dx.doi.org/10.1103/PhysRevLett.128.110506}{Phys. Rev.
  Lett. {\bf 128}, 110506}~(2022).
\newblock  \href{http://arxiv.org/abs/2110.02701}{arXiv:2110.02701}.

\bibitem{commercialQKD}
Wikipedia authors.
\newblock ``Quantum key distribution''.
\newblock  url:~\url{https://en.wikipedia.org/wiki/Quantum_key_distribution}.
\newblock (accessed:~25-October-2021).

\bibitem{stMUB}
Armin Tavakoli, Máté Farkas, Denis Rosset, Jean-Daniel Bancal, and Jedrzej
  Kaniewski.
\newblock ``Mutually unbiased bases and symmetric informationally complete
  measurements in {B}ell experiments''.
\newblock \href{https://dx.doi.org/10.1126/sciadv.abc3847}{Science Advances
  {\bf 7}, eabc3847}~(2021).
\newblock  \href{http://arxiv.org/abs/1912.03225}{arXiv:1912.03225}.

\bibitem{SW2}
Stephen~J. Summers and Reinhard~F. Werner.
\newblock ``Maximal violation of {B}ell's inequalities for algebras of
  observables in tangent spacetime regions''.
\newblock Ann. Inst. H. Poincar\'e. {\bf 49}, 215--243~(1988).

\bibitem{mermin}
N.~David Mermin.
\newblock ``Is the moon there when nobody looks? {R}eality and the quantum
  theory''.
\newblock \href{https://dx.doi.org/10.1063/1.880968}{Physics Today {\bf 38},
  38--47}~(1985).

\bibitem{janas2019}
Michael Janas, Michael~E. Cuffaro, and Michel Janssen.
\newblock ``Putting probabilities first. {H}ow {H}ilbert space generates and
  constrains them''~(2019)
  \href{http://arxiv.org/abs/1910.10688}{arXiv:1910.10688}.

\bibitem{dimwit}
Nicolas Brunner, Stefano Pironio, Antonio Ac{\'\i}n, Nicolas Gisin,
  Andr\'e~Allan M\'ethot, and Valerio Scarani.
\newblock ``Testing the dimension of {H}ilbert spaces''.
\newblock \href{https://dx.doi.org/10.1103/PhysRevLett.100.210503}{Phys. Rev.
  Lett. {\bf 100}, 210503}~(2008).
\newblock  \href{http://arxiv.org/abs/0802.0760}{arXiv:0802.0760}.

\bibitem{dimwit2016}
Yu~Cai, Jean-Daniel Bancal, Jacquiline Romero, and Valerio Scarani.
\newblock ``A new device-independent dimension witness and its experimental
  implementation''.
\newblock \href{https://dx.doi.org/10.1088/1751-8113/49/30/305301}{J. Phys. A
  {\bf 49}, 305301}~(2016).
\newblock  \href{http://arxiv.org/abs/1606.01602}{arXiv:1606.01602}.

\bibitem{dimwit2017}
Wan Cong, Yu~Cai, Jean-Daniel Bancal, and Valerio Scarani.
\newblock ``Witnessing irreducible dimension''.
\newblock \href{https://dx.doi.org/10.1103/PhysRevLett.119.080401}{Phys. Rev.
  Lett. {\bf 119}, 080401}~(2017).
\newblock  \href{http://arxiv.org/abs/1611.01258}{arXiv:1611.01258}.

\bibitem{Horodecki2qubits}
R.~Horodecki, P.~Horodecki, and M.~Horodecki.
\newblock ``Violating {B}ell inequality by mixed spin-1/2 states: necessary and
  sufficient condition''.
\newblock
  \href{https://dx.doi.org/https://doi.org/10.1016/0375-9601(95)00214-N}{Phys.
  Lett. A {\bf 200}, 340--344}~(1995).

\bibitem{gisin1991}
N.~Gisin.
\newblock ``{B}ell's inequality holds for all non-product states''.
\newblock \href{https://dx.doi.org/10.1016/0375-9601(91)90805-I}{Physics
  Letters A {\bf 154}, 201--202}~(1991).

\bibitem{wolk}
R.~Grone, C.R. Johnson, E.M. S\'a, and H.~Wolkowicz.
\newblock ``{Positive definite completions of partial Hermitian matrices}''.
\newblock
  \href{https://dx.doi.org/https://doi.org/10.1016/0024-3795(84)90207-6}{Lin.
  Alg. Appl. {\bf 58}, 109--124}~(1984).

\bibitem{Barvinok}
Alexander Barvinok.
\newblock ``A course in convexity''.
\newblock \href{https://dx.doi.org/10.1090/gsm/054}{Graduate Studies in
  Mathematics 54}. AMS. Providence~(2002).

\bibitem{dixmier}
J.~Dixmier.
\newblock ``C*-algebras''.
\newblock North-Holland mathematical library. North-Holland. ~(1982).

\bibitem{ReedSimon4}
M.~Reed and B.~Simon.
\newblock ``Methods of modern mathematical physics~{IV}: Analysis of
  operators''.
\newblock Elsevier Science. ~(1978).

\bibitem{twoproj}
Iain Raeburn and Allan~M. Sinclair.
\newblock ``The {C*-}algebra generated by two projections.''.
\newblock \href{https://dx.doi.org/10.7146/math.scand.a-12283}{Math. Scand.
  {\bf 65}, 278--290}~(1989).

\bibitem{ART}
Roy Araiza, Travis Russell, and Mark Tomforde.
\newblock ``A universal representation for quantum commuting correlations''.
\newblock \href{https://dx.doi.org/10.1007/s00023-022-01197-7}{Ann. Henri
  Poinc. {\bf 23}, 4489–4520}~(2022).
\newblock  \href{http://arxiv.org/abs/2102.05827}{arXiv:2102.05827}.

\bibitem{Pitowsky}
I.~Pitowsky.
\newblock ``Quantum probability -- quantum logic''.
\newblock \href{https://dx.doi.org/10.1007/BFb0021186}{Volume 321 of Lect.Notes
  Phys.}
\newblock Springer. ~(1989).

\bibitem{GMS}
Dan Geiger, Christopher Meek, Bernd Sturmfels, et~al.
\newblock ``On the toric algebra of graphical models''.
\newblock \href{https://dx.doi.org/10.1214/009053606000000263}{Ann. Statist.
  {\bf 34}, 1463--1492}~(2006).
\newblock  \href{http://arxiv.org/abs/math/0608054}{arXiv:math/0608054}.

\end{thebibliography}

\vspace*{\fill}
\noindent
\footnotesize
{\bf Authors' addresses:}

\smallskip

\noindent Thinh P.~Le, IQOQI Vienna
\hfill
{\tt Thinh.Le@oeaw.ac.at}

\noindent Chiara Meroni, ICERM Providence \hfill
{\tt  chiara\_meroni@brown.edu}

\noindent Bernd Sturmfels,
MPI-MiS Leipzig and UC Berkeley
\hfill {\tt bernd@mis.mpg.de}

\noindent Reinhard F.~Werner,  Leibniz Universit\"at Hannover
\hfill {\tt reinhard.werner@itp.uni-hannover.de}

\noindent Timo Ziegler,  Leibniz Universit\"at Hannover
\hfill {\tt timo.ziegler@itp.uni-hannover.de}

\end{document}